\RequirePackage{fix-cm}
\documentclass[smallextended]{svjour3}       
\smartqed  
\usepackage{graphicx}
\usepackage{natbib}
\usepackage{tabularx}
\usepackage{rotating}
\usepackage{textcomp}
\usepackage{amsmath}
\usepackage{amssymb}
\usepackage{pdfpages}
\begin{document}

\title{The InSight HP$^3$ Penetrator (Mole) on Mars: Soil Properties Derived From the 
Penetration Attempts and Related Activities.}

\titlerunning{Mole Saga Paper}

\subtitle{InSight Contribution number 223}




\author{T. Spohn \and
T.L. Hudson \and
E. Marteau \and
M. Golombek \and
M. Grott \and
T. Wippermann \and
K. S. Ali \and
C. Schmelzbach \and
S. Kedar \and
K. Hurst \and
A. Trebi-Ollennu \and
 V. Ansan \and
J. Garvin \and
 J. Knollenberg  \and
 N. M\"uller \and
 S. Piqueux \and
R. Lichtenheldt \and
C. Krause \and
C. Fantinati \and
N. Brinkman \and
D. Sollberger \and
P. Delage \and
C. Vrettos \and
S. Reershemius \and
L. Wisniewski \and
J. Grygorczuk \and
J. Robertsson, \and
P. Edme \and
F. Andersson \and
O. Kr\"omer \and
P. Lognonn\'e \and
D. Giardini \and
S. E. Smrekar \and
W. B. Banerdt}

\authorrunning{T. Spohn et al.}

\institute{T. Spohn \at
              International Space Science Institute,
              Hallerstrasse 6, 3012 Bern, Switzerland\\
             and \\
             DLR Institute of Planetary Research, 
	          Rutherfordstr. 2, 12489 Berlin, Germany 
              Tel.: +49-30-67055 300, 
              \email{tilman.spohn@dlr.de}
         \and
    M. Grott,  J. Knollenberg, N. M\"uller,   \at
              DLR Institute of Planetary Research,
	          Rutherfordstr. 2, 12489 Berlin, Germany 
    \and
    S. Reershemius, T. Wippermann  \at
              DLR Institute of Space Systems, 
	          Robert-Hooke-Str. 7, 28359 Bremen, Germany 
	       \and
    C. Fantinati, C. Krause,  \at
             DLR MUSC Space Operations and Astronaut Training,  
	          Linder H\"ohe, 51147 K\"oln, Germany 
	       \and
	R. Lichtenheldt \at
              DLR Institute of System Dynamics and Control, 
	          M\"unchener Strasse 20, 82234 Wessling, Germany 
	          \and
	K.S. Ali,  W. B. Banerdt, M.P. Golombek, T.L. Hudson, K. Hurst, S. Khedar, E. Marteau,  S. Piqueux, A. Trebi-Ollenu, S. E. Smrekar, \at
              Jet Propulsion Laboratory, California Institute of Technology,
              Oak Grove Drive,
              Pasadena, Ca 91109, USA 
	        \and
	J. Garvin \at 
	            NASA Goddard Space Flight Center, 8800 Greenbelt Road
                Greenbelt, MD 20771 USA
                \and
    J. Grygorczuk, L. Wisniewski   \at Astronika Sp. z o.o.,
                ul. Bartycka 18, 00-716 Warszawa, Poland    
               \and 
  F. Andersson, N. Brinkman, P. Edme, D. Giardini, J. Robertsson, C. Schmelzbach,  D. Sollberger  \at 
                Institute of Geophysics, Department of Earth Sciences, ETH Z\"urich, CH-8092 Z\"urich, Switzerland   
\and
V. Ansan \at
Laboratoire de Plan{\'e}tologie et G{\'e}odynamique de Nantes, Universit{\'e} de Nantes, 44322 Nantes, France 
\and
P. Delage \at
{\'E}cole nationale des ponts et chauss{\'e}es, Laboratoire Navier, Paris, France
\and
C. Vrettos \at
Department of Civil Engineering, University of Kaiserslautern, Kaiserslautern, Germany
\and
P. Lognonn{\'e} \at Institut du Physique du Globe Paris, Paris, France
\and
O. Kr\"omer \at Astrium, Bremen, Germany
}

\date{Received: date / Accepted: date}

\maketitle

\begin{abstract}
The NASA InSight Lander on Mars includes the Heat Flow and Physical Properties Package HP$^3$ to measure the surface heat flow of the planet. The package uses temperature sensors that would have been brought to the target depth of 3--5 m by a small penetrator, nicknamed the mole. The mole requiring friction on its hull to balance remaining recoil from its hammer mechanism did not penetrate to the targeted depth. Instead, by precessing about a point midway along its hull, it carved a 7 cm deep and 5-6 cm wide pit and reached a depth of initially 31 cm.  The root cause of the failure - as was determined through an extensive, almost two years long campaign  - was a lack of friction in an unexpectedly thick cohesive duricrust. During the campaign -- described in detail in this paper -- the mole penetrated further aided by friction applied using the scoop at the end of the robotic Instrument Deployment Arm and by direct support by the latter. The mole finally reached a depth of 40 cm, bringing the mole body 1--2 cm below the surface. It reversed its downward motion twice during attempts to provide friction through pressure on the regolith instead of directly with the scoop to the hull. The penetration record of the mole and its thermal sensors were used to measure thermal and mechanical soil parameters such as the  penetration resistance of the duricrust of 0.5 - 1.2 MPa and a penetration resistance of a deeper layer ($>$ 30 cm depth) of 5.3 MPa. Applying cone penetration theory, the resistance of the duricrust was used to estimate a cohesion of the latter of 4 - 25 kPa depending on the internal friction angle of the duricrust. Pushing the scoop with its blade into the surface and chopping off a piece of duricrust provided another estimate of the cohesion of 5.8 kPa. The hammerings of the mole were recorded by the seismometer SEIS and the signals could be used to derive a P-wave velocity of $114^{+40}_{-19}$ m/s and a S-wave velocity of $60^{+10}_{-7}$ m/s \citep{Brinkman2021} representative of the topmost tens of cm of the regolith. Together with a density of $1211^{+149}_{-113}$ kg/m$^3$ \citep{Grott:2021} provided by a thermal conductivity and diffusivity measurement using the mole thermal sensors, the elastic moduli could be calculated from the seismic velocities. The shear, bulk and Young's modulus were found to be $4.32 \pm 1.01$ MPa,  $9.84 \pm 6.54$ MPa, and 11.30$\pm{2.87}$\,MPa, respectively, and the Poisson ratio to be $0.31 \pm 0.15$. Using empirical correlations from terrestrial soil studies between the shear modulus and cohesion, the previous cohesion estimates were found to be consistent with the elastic moduli. The combined data were used to derive a model of the regolith that has an about 20 cm thick duricrust underneath a 1 cm thick unconsolidated layer of sand mixed with dust and above another 10 cm of unconsolidated sand. Underneath the latter, a layer more resistant to penetration and possibly consisting of debris from a small impact crater is inferred. The thermal conductivity increases from 14 mW/m K to 34 mW/m K through the 1 cm sand/dust layer, keeps the latter value in the duricrust and the sand layer underneath and then increases to 64 mW/m K in the sand/gravel layer below.         

\end{abstract}

\section{Introduction}\label{sec:introduction}

The InSight Mars lander (e.g., \cite{Banerdt2020}) is the first geophysical observatory on a planet other than the Earth and, with the exception of the geophysical instruments that the Apollo missions installed on the Moon, the only geophysical observatory on another solar system body. The goals of the InSight mission focus on the exploration of the interior structure of Mars and its evolution. An important datum for assessing planetary evolution is the present day surface heat flow from the interior as it  provides an important constraint on the thermal evolution of the planet as well as an upper bound on the bulk abundance of radiogenic elements. Therefore, the payload of InSight includes a heat flow probe in addition to an ultra-sensitive seismometer with short-period and broadband sensors, transponders to track the movement of the rotation axis of the planet, a magnetometer, and a package of atmospheric science sensors. The heat flow probe, HP$^3$, had been planned to install a string of 14 temperature sensors down to a depth of 3--5~m and measure a temperature and a thermal conductivity vs depth profile up to the target depth. The temperature sensors were imprinted on a kapton foil that was to be drawn to depth by a small penetrator, nicknamed the ``mole". The latter was equipped with temperature sensors that can be heated using a constant input power. The mole would have paused its penetration at regular depth intervals. Heating the sensors and measuring the temperature rise (24~h heating) and fall (48~h cooling) as a function of time, would have allowed a measurement of the thermal conductivity \citep{Spohn2018}. \cite{Grott:2021} have recently published a measurement of the thermal conductivity by the mole at a depth of 40~cm, its final depth after unsuccessfully trying to dig deeper. 

The design of the HP$^3$ heat flow probe was driven by the definition of heat flow $q$ given by
\begin{equation}
    q(z) = -k(z)\frac{dT}{dz}
\end{equation}
where $z$ is the depth, $k(z)$ is the depth-dependent thermal conductivity, and $T$ is temperature. 
Measuring the  heat flow from the interior requires measuring the temperature gradient at a depth where disturbances caused by diurnal, annual, and interannual surface temperature variations are small enough to allow for the targeted measurement accuracy. For HP$^3$, with $\pm 5$ mW/m$^2$ as the targeted accuracy, the minimum tip depth required was estimated to be 3 m \citep{Spohn2018}.  Practical considerations of mass and volume as well as planetary protection rules limited the maximum depth to 5 m.  The MEPAG Special Regions Science Analysis Group \citep{Rummel2014} found that the depth to buried ice in the tropics and mid-latitudes on Mars would be $>$ 5m.  The signal to noise ratio could also be improved by measuring the temperature gradient and the thermal conductivity for a significant fraction of a martian year.

After the mission had been launched on May 5\textsuperscript{th}, InSight landed on Nov 26\textsuperscript{th}, 2018 in western Elysium Planitia, Mars. The landing site is on the western side of a quasi-circular depression, interpreted to be a degraded $\sim$27~m diameter impact crater \citep{Golombek:2020a, Warner:2020, Grant:2018}, informally named Homestead hollow. The site features a smooth, sandy, granule- and pebble-rich surface and is located adjacent to slightly rockier and rougher terrain to the west (Rocky Field). Small craters ($<$10~m diameter) are common around the lander. Some of these craters have little relief and are filled with fine grained material. Farther afield, bright circular patches or hollows interpreted to be soil-filled, degraded craters are common. 

Homestead hollow has a similar morphology and soil characteristics to the degraded, sediment-filled impact craters on the Gusev cratered lava plains \citep{Golombek:2006, Grant:2006} and records degradation by eolian, impact, and lesser mass wasting processes \citep{Golombek:2020a, Grant:2004, Grant:2020, Weitz:2020}. The origin of Homestead hollow as a degraded impact crater suggests that the crater is dominantly filled with eolian sand that is $\sim$3--5~m thick in the landing ellipse, based on an initial depth/diameter ratio of 0.15 \citep{Sweeney:2018, Warner:2020, Golombek:2020b}. 


At the end of February 2019, following successful deployment of the HP$^3$ support system assembly from the lander deck to the ground, the team commanded the mole to start penetrating. It soon became clear that the mole had failed to penetrate to the target depth of 70~cm for the first hammering session.  The team tried for almost a full martian year (22~months) to diagnose the anomaly and assist the mole in penetrating deeper. The attempts were stopped in early January 2021 after it had become clear that immediate success was not to be expected and the power situation of the lander required prioritizing  other instruments on InSight.  
In the course of trying to get the mole to dig, various data sources provided constraints on the cause of the penetration anomaly and have been analyzed to give a better understanding of the properties of the martian soil at the landing site.   

In this paper, we will report in detail the operations that were performed on Mars with the mole and the robotic arm. We will further interpret the data collected from these operations in terms of mechanical and thermal properties of the regolith and its structure. In a separate paper \citep{Spohn2021} we discuss what lessons can be learned about the design and the operation of the InSight HP$^3$ mole. It is hoped that these can inform future attempts to use small penetrators,  on Mars or other extraterrestrial bodies, whether for heat flow or other scientific and exploration purposes.       

Section \ref{sec:HP3Description} describes the physical and technical properties of the mole and its support structure. Section \ref{sec:RobotArm} then describes the properties of the robotic arm and camera system that was used during the anomaly resolution attempts. Section \ref{Sec:SiteSelection} describes the site selection process and outcome, and section \ref{sec:Record} describes in detail our record of operations and observations on Mars.  
In section \ref{RegolithInteracttion1} we describe the geometry and the geological setting of the pit that the mole had carved during the first hammering sessions  using digital terrain models derived from imaging data.  In section \ref{sec:PropertiesPenetration} we derive soil mechanical and thermal properties from the mole penetration and from the interactions of the scoop at the end of the robotic arm with the regolith; thermal properties derived from mole heating experiments are also summarized. In addition, we describe the results of the interpretation of seismic signals recorded from the mole hammering by the InSight seismometer, SEIS. A synopsis section \ref{sec:Discussion} will conclude the paper.  

\section{The Mole Penetrator and its Support Structure}
\label{sec:HP3Description}

\begin{figure}
\begin{center}
  \includegraphics[width=1\textwidth]{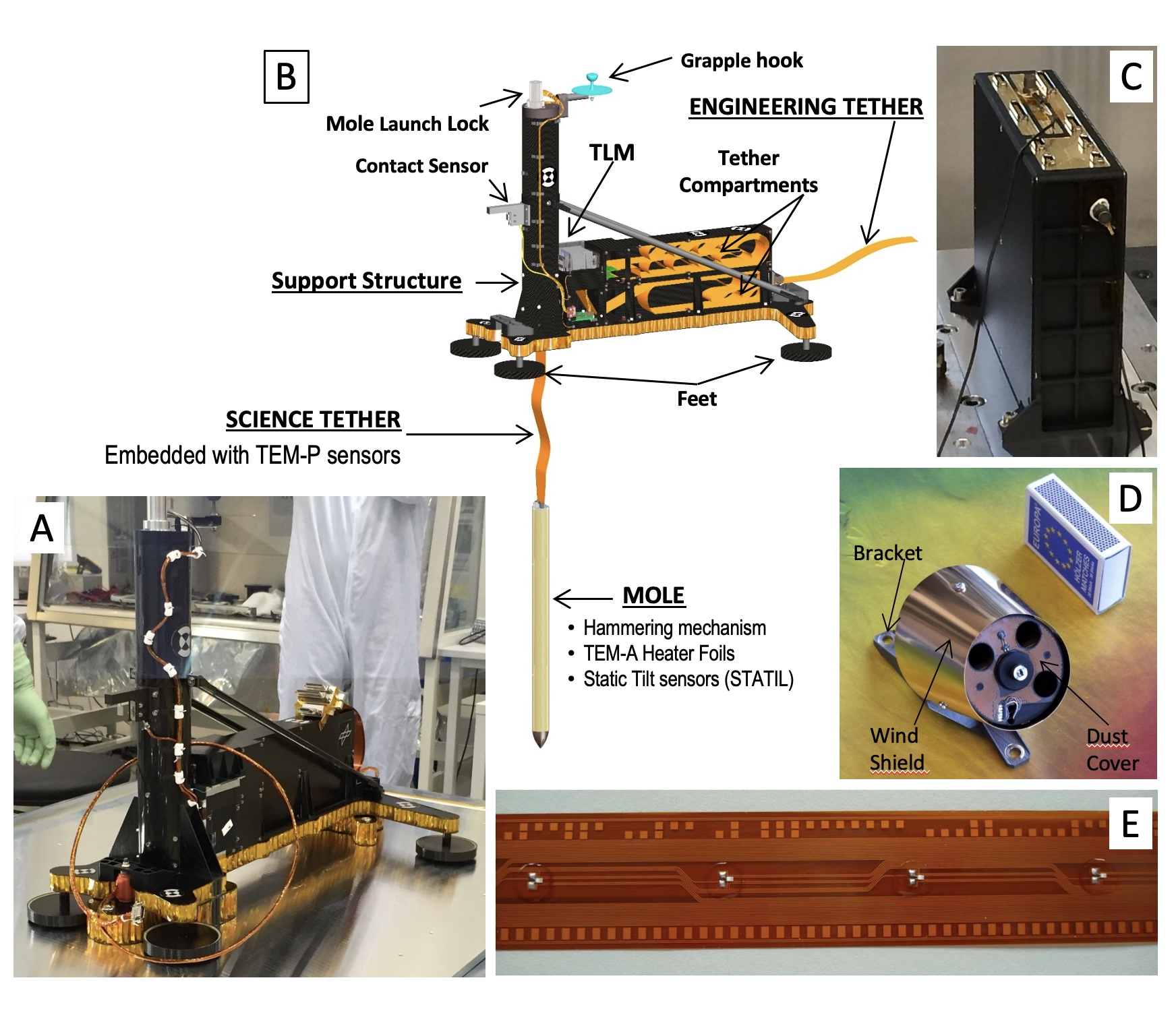}
  \caption{Elements of the Heat Flow and Physical Properties Package (HP$^3$). (A) Flight model Support System Assembly (SSA). (B) Annotated cutaway of the SSA showing Mole and Engineering Tether partly deployed. (C) Back End Electronics within the InSight lander, (D) Deck-mounted HP$^3$ Radiometer (matchbox for scale). (E) Science Tether showing embedded TEM-P sensors, relative depth markings (bottom) and Gray binary code absolute depth markings (top).  An ST prototype with closely spaced sensors is shown for illustration; the actual flight Science Tether has greater (and irregular) spacing between sensors.}
  \label{fig:Instrument overview.jpeg}   
\end{center}    
\end{figure}

The Heat Flow and Physical Properties Package HP$^3$ (see 
Fig. \ref{fig:Instrument overview.jpeg} for an annotated depiction) consists of the Back End Electronics (BEE) housed inside the InSight lander's thermal enclosure, the deck-mounted radiometer (RAD) to measure surface brightness temperature \citep{Spohn2018}, and the Support System Assembly (SSA) that is deployed to the martian surface by the InSight Instrument Deployment System (IDS) 
\citep[see section \ref{sec:RobotArm} and][]{trebi-ollennu2018}.  The Support System Assembly consists of a carbon fiber Support Structure (SS) that initially hosts the following subsystems: the Engineering Tether (ET), the Science Tether (ST), the Tether Length Measurement device (TLM), and the mole.  The ET, which is actually three separate copper/kapton ribbons bonded together, electrically connects all deployed elements of the SSA to the BEE.

The critical subsystem that enables access to sufficient depth to avoid the temperature disturbances caused by annual surface temperature variations is a self-impelling cylindrical penetrator with a length of 400~mm and a diameter of 27~mm, nicknamed the "mole". Its major components are (1) its hull, with an ogive shaped tip, foils for the thermal conductivity measurement embedded in the hull, and electrical connections to the rest of HP$^3$; (2) the internal hammer mechanism 
with a motor, drive shaft, cylindrical cam, drive springs, hammer, and brake springs; and (3) a shock-protected package of static accelerometers for measuring the mole tilt with respect to gravity (STAtic TILt sensors, or STATIL). The foils embedded within the hull are copper/kapton flexible heaters that can also be used as temperature sensors. To measure thermal conductivity, the mole is used as a modified line-heat source \citep[e.g.,][]{Carslaw1959, Bana1997} when these foils (called TEM-A, short for Thermal Excitation and Measurement--Active) are supplied with a constant power, generating heat that is then conducted to the surrounding soil. The change in mole temperature is recorded during the heating and cooling phases (24~h and 48~h, respectively) to derive thermal conductivity \citep{Grott:2021}.

The mole is connected to the HP$^3$ system by a flexible kapton/copper tether called the Science Tether (ST). This tether has two functions: First, it provides power and commands to the mole and returns data from the mole to the electronics box. Second, the ST passively measures the temperature at known points along its length.  This latter function (called TEM-P, short for Thermal Excitation and Measurement--Passive) uses 14 platinum resistance temperature sensors (PT100) embedded at unequally spaced intervals along its 5~m length, which would nominally have been pulled into the ground by the Mole. The ST has markings on either edge that enable relative and absolute measurements of tether movement.

 To determine thermal conductivity as a function of depth, and to properly separate the contributions of surface insolation and geothermal heat to the temperature data, the depth of the mole body and the vertical depth to the individual TEM-P sensors needs to be known with a precision of $\sim$1~cm. The STATIL sensors within the mole provide its angular orientation with respect to the local gravity vector with high time resolution. The Tether Length Measurement device in the SSA uses a combination of LEDs and photosensors to optically observe the markings on the Science Tether as it is extracted by the mole during penetration. The depth of the mole body and the TEM-P sensors would have been reconstructed from the STATIL and TLM readings.  

The heart of the mole is its hammer mechanism which has been described in \cite{Kroemer2019} and in \cite{Spohn2018, Spohn2021} and is only briefly described here.
This mechanism is an electro-mechanical system converting 
the energy of a compressed drive spring   
into the forward acceleration of a tungsten hammer which impacts the interior anvil surface of the mole tip. The hammer mechanism assembly (moter, gear box, drive shaft, drive spring, and hammer) is free to slide within the mole hull, bound only at its aft end by the brake spring. During a hammer 'stroke' (consisting of an initial high-force impulsive impact and several lesser impacts, collectively called 'strikes') causes the mole to move forward into the ground.  The recoil of the hammer mechanism from the strikes is largely absorbed by compression of the brake spring, with a small acceleration component transferred to the housing which must be compensated externally by friction in the soil.  


The mechanism is designed such that the forward force on impact on the tip is maximized, while the recoil transferred to the housing is minimized and stretched over a comparatively long period of time. The forward force imparted to the housing from the first hammer stroke of a healthy mole has been measured in the laboratory \citep{Wippermann2020} to be 1100$\pm$150~N.  In contrast, the recoil force transferred to the housing by the brake spring is only 5--7~N. For the mole to make overall forward progress, this small but non-zero rebound must be compensated.  Part of the compensation  results from gravity for a downward-moving  mole (the 0.85~kg mass of the mole produces 3.2 N downward force on Mars). The remaining positive rebound force of 1.8--3.8~N must be compensated by a force acting from outside the mole.  This force can be provided by friction, such as from the friction springs of the support structure (see below) or by friction from the regolith. It can also come from direct physical impediment of rebound at the back of the mole such as collapsed regolith or other solid object loaded against the back cap.  Without sufficient resistance-to-rebound, the mole will `bounce' in place and no forward progress will occur.       


\subsection{The Support Structure} \label{sec:SS}
 The design of the Support Structure (SS) was influenced by the extremely limited 
 space available on the lander deck, and by the requirement to have it stably placed on the martian surface, allowing for both assumed ranges of cm-scale surface topography and high-velocity martian wind gusts or vorticies. The maximum height of the SSA (and thus the maximum length of the mole) were further constrained by the available volume below the backshell that covered the deck until just prior to landing. These design constraints influenced the overall SSA shape; the number, size, and placement of its feet; and the placement of the Mole and tethers within it \citep[see][]{Reershemius2019}. The Support Structure's rectangular main body consists of the tether storage compartment which is split horizontally by a separation wall into a top compartment for the Science Tether and a bottom compartment for the Engineering Tether. Both tethers are stored in this compartment during cruise.

The 3.5~m long Engineering Tether, initially within the SS, is passively extracted from the aft end of the structure during deployment to the surface by the Instrument Deployment Arm (IDA, section \ref{sec:RobotArm}), ultimately laying across the deck and extending down to the deployment site. After deployment, the SS had three key functions: to support the mole within a vertical tube during the initial phase of penetration into the subsurface, to store the 5~m long Science Tether (ST) before it is extracted by the mole, and to host the TLM and other components fixed to the structure. The TLM sits between the Science Tether storage compartment and the vertical tube approximately one-third of the distance between the SS base and the top of the tube.  The ST is threaded through the TLM and then connects to the back cap of the mole.  In normal ST extraction during penetration, the markings on the side of the tether would obscure or transmit light from the TLM LEDs to opposed photosensors, providing relative and absolute measurements of extracted tether length. This design only functions when the tether is pulled through the TLM, producing a signal.

\begin{figure}
\begin{center}
  \includegraphics[width=\textwidth]{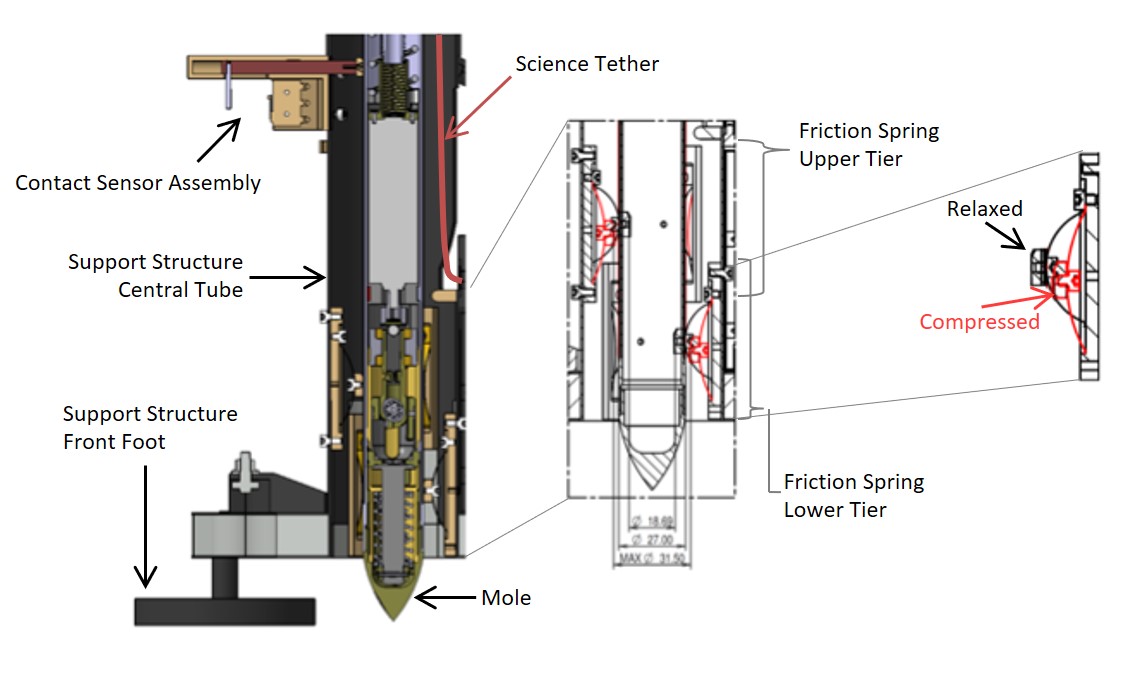}
  \caption{This partial cross-section of the forward-front portion of the Support Structure shows the mole in its position prior to penetration. The contact sensor position (top left) allows it to indicate then the mole has moved 15~cm out of the tube.  The outlines at center and right show the positions of the upper and lower friction spring tiers, and also the shape of the springs in their relaxed (black) and compressed (red) states. A portion of the Science Tether service loop can also be seen extending up from the TLM towards the back cap of the mole (not pictured).}
  \label{fig:SS_CrossSection}   
\end{center}    
\end{figure}

It is important to discuss the location of the TLM relative to the back of the mole and how it contributed to operational decisions and the unavailability of mole depth data during early penetration. The TLM's placement low on the central tube was driven in part by volumetric constraints on the lander deck, and in part by concerns about SS stability in the martian wind.  Without the mole inside, the carbon-fiber SS is very light \citep[about 2.1~kg,][]{Reershemius2019} and the 150~g TLM makes up a significant portion of the `empty' SS mass. Had the SS been constructed with the TLM placed high, near the back-end of the mole, the resulting cross section's center of pressure would have been well above the structure's center of gravity. This would have posed a tip-over risk during the post-penetration observational phase when the SS was mostly empty.  Placing the TLM low solved many issues, but created a new one: there now needed to be a 'service loop' in the Science Tether, passing up the central tube between the TLM and the attachment point at the back cap of the mole (see Fig.~\ref{fig:SS_CrossSection}). This service loop would need to be exhausted (by approximately 54~cm of mole penetration) before the ST would begin to move through the TLM and provide data.  

Recognizing this dearth of data, the SS was designed with a contact sensor to provide some intermediate indication of mole penetration before TLM data became available. Located near the vertical mid-point of the central tube, it detects mole egress by changing state when the back cap of the mole passes this point (see Fig.~\ref{fig:SS_CrossSection}). The mechanism uses a spring-loaded piston-switch connected to a tee-bar that protrudes into the central tube and loads against the outer surface of the mole. When the mole passes, the spring pushes the piston into the tube, tripping the sensor and indicating that 15~cm of the mole had moved out of the central tube.

One of the critical functions of the SS is to position and maintain the mole approximately orthogonal to the deployment surface for initial penetration, and to provide the necessary friction for the mole to make forward progress until (it was assumed) friction with the regolith would provide the necessary restoring force. The SS supplied this initial friction via a system of so-called 'friction spring' assemblies located at the bottom of the central tube. Six spring assemblies are arranged in two tiers, with the three springs in each tier spaced 120 degrees apart around the tube's interior (see Fig.~\ref{fig:SS_CrossSection}).

The bow-shape of each thin copper-beryllium leaf spring is fixed at one end, with the other end curled over and free to glide. Each spring is attached to a base that serves both as a mounting interface to the tube, and as a glide track for the free end of the spring. A contact block (also called 'gliding element') mounted on the peak of the spring touches the outer surface of the mole to provide friction. The spring assemblies are mounted with the free-gliding end oriented towards the penetration direction, allowing the mole free motion downward while resisting rebound motion. 

While static, or during forward (i.e., downward) motion, the springs apply a small amount of force to the mole. This would tend to center the mole while allowing it to progress freely through the tube. During mole rebound, which the springs were specifically designed to resist, the contact blocks' friction against the hull would transfer to the springs' fixed ends, increasing inward curvature and causing higher friction on the mole, producing a self-locking effect that resists upward motion.  Prior to landing, it was estimated that for an unconsolidated sandy regolith the friction springs would need to apply this reaction force until the mole was approximately 3/4 buried \citep{Reershemius2019}, after which it was assumed regolith friction would be sufficient to resist the rebound. 

\section{The Robotic Arm and Scoop} 
\label{sec:RobotArm}

\begin{figure}
    \centering
    \includegraphics[width=\textwidth]{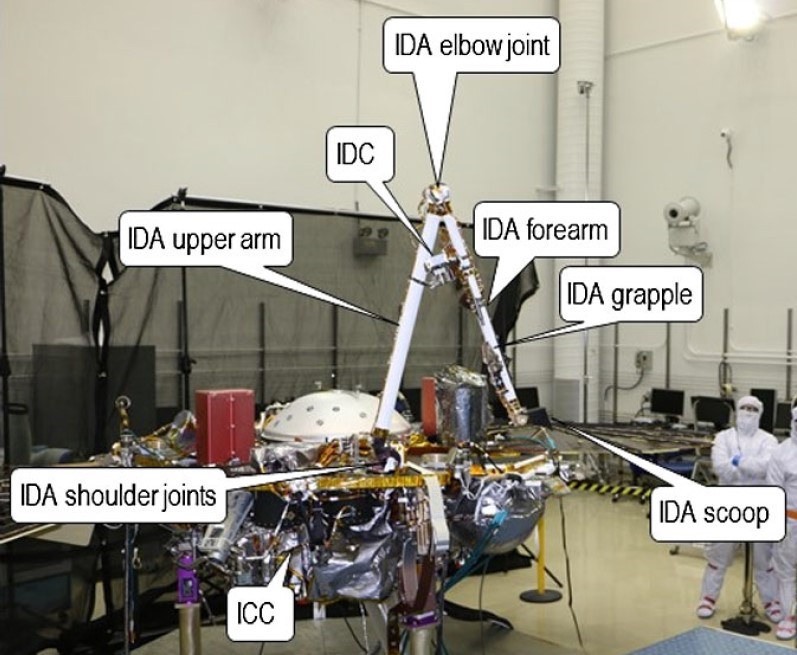}
    \caption{The Instrument Deployment System}
    \label{fig:IDS}
\end{figure}

The Instrument Deployment System (IDS) (Fig. \ref{fig:IDS}) consists of a robotic arm, two color cameras, and the motor controller electronics and software to control them.  The Instrument Deployment Arm (IDA) is a robotic arm on the InSight lander with four degrees-of-freedom.  It has a back-hoe design with a yaw joint (shoulder azimuth) at the base, and then three pitch joints, shoulder elevation, elbow, and wrist \citep{trebi-ollennu2018}.

The primary purpose of the IDA was to deploy the seismometer SEIS, its wind and thermal shield WTS, and HP$^3$, lifting them from the lander deck and placing them on the surface.  Additionally, it was intended to take images of the IDA workspace, lander, and martian atmosphere.  The images of the workspace were used to choose the deployment locations for the science instruments and for planning the motions of the IDA  (see section \ref{Sec:SiteSelection}).

\begin{figure}
    \centering
    \includegraphics[width=\textwidth]{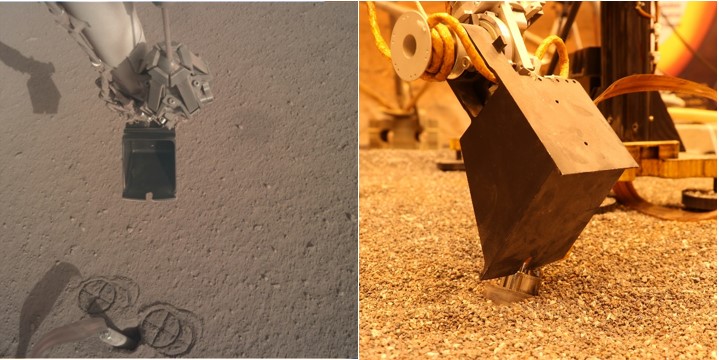}
    \caption{The IDA scoop.  In the left image, the scoop is shown above the HP3 mole in this IDC image.  The front blade is visible.  The grapple is seen stowed to the side of the IDA forearm.  In the right image, the scoop in the Earth-based testbed is shown in an “inclined push” configuration on the testbed mole back cap.  The dull blade on the outside back of the scoop is visible.}
    \label{fig:ida_scoop}
\end{figure}

The IDA has a camera mounted to the forearm and two end effectors: a scoop and a grapple.  The scoop (Fig. \ref{fig:ida_scoop}) is an open, bucket-shaped chamber with a sharp blade, about 2.29 mm thick, at the front and a second dull blade on the outside back of the bottom of the scoop.  The scoop is about 76~mm wide at the tip and 102~mm in overall length.  The grapple has five fingers which can be actively opened and shuts passively.  During the deployment of the science payloads to the martian surface, and during the re-deployment of the HP$^3$ support structure to move it off of the mole (section \ref{sec:Support System Lift}), the grapple hung from a compliant umbilical cable at the IDA wrist joint.  During IDA operations to interact with the mole and surrounding terrain, including taking images of the HP$^3$, the grapple was closed around a ball-and-cable grapple restraint mechanism, stowing it on the side of the IDA forearm (Fig \ref{fig:ida_scoop}).  The Instrument Deployment Camera (IDC) is a color camera with a 1024x1024 resolution and a 45x45 degree field of view \citep{maki2018}.  It is mounted to the forearm of the IDA, such that it can see objects suspended in the grapple and also providing a view of the scoop.  

Stereo images can be acquired by the IDC by moving the IDA either horizontally by rotating the shoulder joint or vertically by rotating the shoulder or elbow joints. However, unlike dual camera stereo systems in which both cameras have "toe in" to point to a common spot, the IDA cannot do this so the stereo images are side-by-side requiring different processing to create digital elevation models, DEMs,  \citep{Abarca2019}. The IDC has acquired a large number of color surface images, including stereo coverage at two resolutions (0.5 and 2 mm per elevation posting) of the instrument deployment workspace to select the locations to place the instruments, three complete stereo panoramas (morning, afternoon, and evening), and stereo images of the lander, its footpads, and terrain under the lander. In addition, higher resolution DEMs have been acquired using Structure from Motion (SFM) techniques in which more images with smaller offsets are acquired \citep{Garvin2019}. A common imaging sequence used for this technique involved a four by four matrix of images with small offsets followed by a single image of the entire area (4x4x1).

The Instrument Context Camera (ICC) is a color camera mounted to the underside of the lander deck.  It has a 1024x1024 resolution, and a fish-eye lens with a 124x124 degree field of view \citep{maki2018}.  It is pointed to view most of the IDA workspace.

The IDA has approximately 2~meters of reach when fully outstretched.  At the martian ground level, however, the IDA can reach 1.7~meters from the IDA base, projected to the ground, with the scoop outstretched.  The HP$^3$ was initially deployed close to the maximum reach of the robotic arm, at a distance of about 1.4~meters from the ground-level projection of the IDA frame.  This limited the ability of the IDA to perform any operations that extended more than a short distance farther than the location of the mole.  

The configuration of the IDA determines the force it can produce at the scoop.  To prevent the IDA from damaging itself when interacting with the martian terrain, the IDA flight software and command sequences limit the torques at each of the four joints to 35, 120, 65, and 10.5~Newton-meters at the shoulder azimuth, shoulder elevation, elbow, and wrist joints, respectively \citep{trebi-ollennu2018}.  At the location of the mole, in a configuration such that the flat bottom side of the scoop is pressing against the ground, this allows the IDA to produce an estimated peak force on the ground of about 46~N. 
The estimated force the IDA was capable of exerting in a lateral direction against the mole's shaft, using the shoulder azimuth joint, is about 10~N.

Hardware and software constraints limit the maximum rate at which the cameras can acquire images and save them to non-volatile memory.  If using one camera only, either IDC or ICC, the maximum acquisition rate is once every 32~seconds.  If using both cameras in an alternating fashion, consecutive IDC images will be spaced 47~seconds apart, and consecutive ICC images will also be spaced 47~seconds apart.

While the IDA was not intended for use to assist the mole's penetration, it provided the means to do so.  The IDA was used to move the HP$^3$ support structure off of the mole, revealing the underlying situation to the IDC camera.  The scoop on the IDA was later used to push against the mole itself and to push and scrape the terrain surrounding the mole.  The IDA and IDC were used to take close-up images of the mole, including during hammering attempts.


\section{HP$^3$ Instrument Deployment Site Selection}
\label{Sec:SiteSelection}
\noindent

\begin{figure}
\begin{center}
  \includegraphics{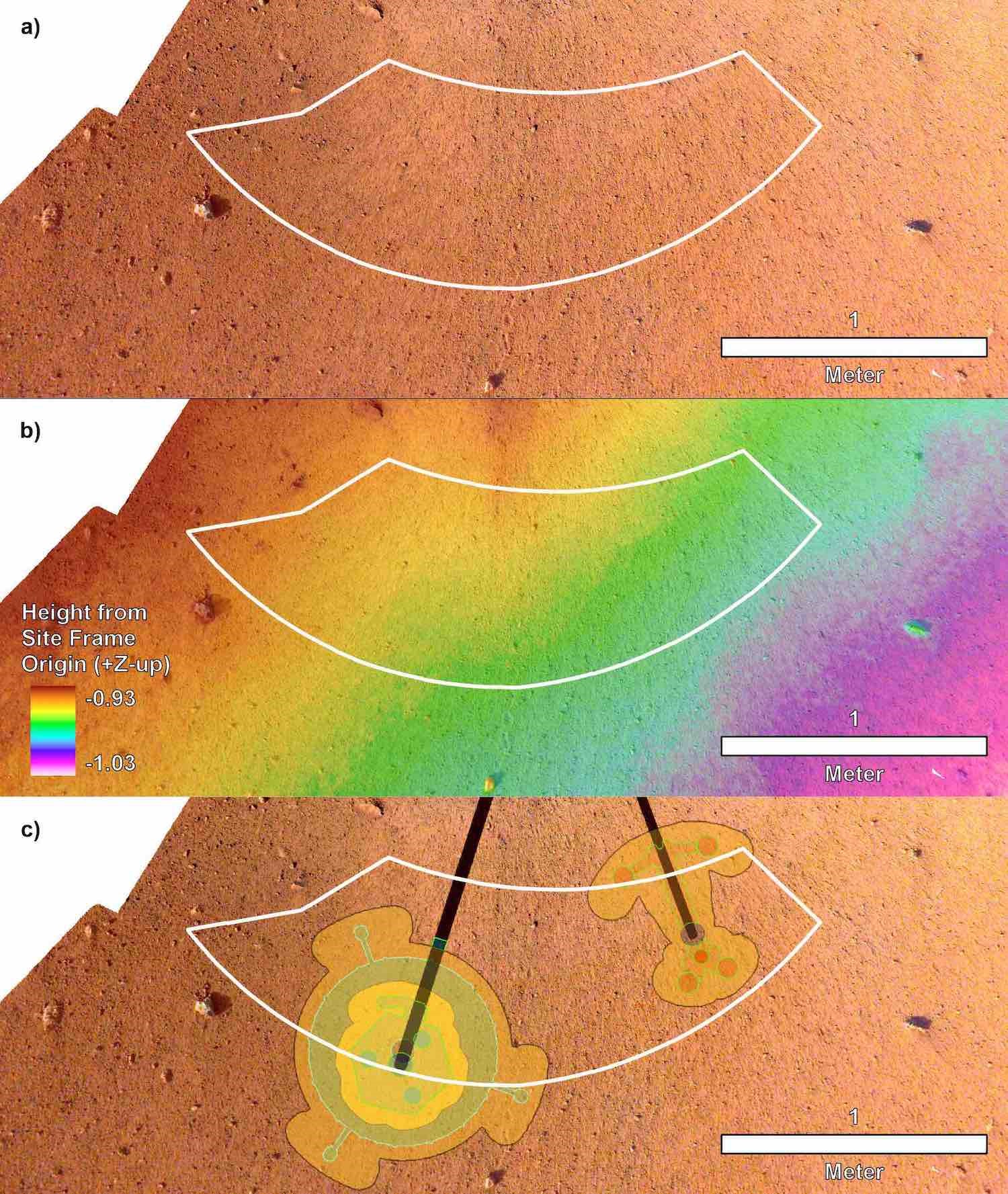}
  \caption{Image mosaic, DEM, and instrument placements selected by the ISSWG and project. (a) The first IDC image mosaic created of the workspace at 1~mm/pixel with the deployment area outlined in white. (b) High-resolution DEM produced from the second mosaic of the workspace at 1~mm per elevation posting and the deployment area outlined in white. Note that the deployment area has a total relief measured in centimeters. (c) Locations selected for the instruments with black lines to the instrument grapple points. SEIS and WTS are to the left and HP$^3$ is to the right. North is up for all. }
  \label{fig:DepeloymentAreas}   
\end{center}    
\end{figure}

The highest priority activity after landing and putting the spacecraft in a fully operational configuration was determining where to place the instruments on the surface. The Instrument Site Selection Working Group (ISSWG) determined the locations to place instruments in the workspace based on the spacecraft tilt, workspace topography, surface characteristics (soils, rocks, etc.) and instrument placement requirements. Six subgroups made up the ISSWG: 1) geologists, 2) physical property scientists, 3) arm and deployment engineers, 4) Multi-mission Image Processing Laboratory (MIPL) personnel, and instrument representatives for 5) SEIS (seismometer), and 6) HP$^3$. The workspace is in front of the spacecraft (to the south), next to where the arm is attached to the edge of the lander. The workspace extends out to roughly 2~m away from the lander and 2~m to either side in a crescent shaped area (Fig.~\ref{fig:DepeloymentAreas}). Instrument placement requirements for HP$^3$ \citep{Spohn2018} are related to surface slope, rocks, load bearing soil, tether geometry, and the desire to be away from the lander (and the seismometer) to reduce thermal interference \citep{Grott2009, Siegler2017} and are summarized in Table~\ref{tab:DeploymentReq}. Before landing, preliminary preferred instrument locations were identified as starting points for the site selection process, with both instruments as far as possible away from the lander and from each other, and with the seismometer to the west to avoid crossing tethers.

\begin{table}
 \begin{tabularx}{1.0\textwidth} {
 | >{\raggedright\arraybackslash} p{1cm}
 | >{\raggedright\arraybackslash} X 
 | >{\raggedright\arraybackslash} p{1.3 cm}  
 | >{\raggedright\arraybackslash} p{1.5 cm} |}
    \hline
    \textbf{Req ID} & \textbf{Constraint} & \textbf{Primary} & \textbf{Secondary}\\
    \hline
    HP3-01 & HP3 footplane tilt $<$15$^{\circ}$ of horizontal & HP3 & Depl/IDS Geology\\
    \hline
    HP3-02 & No rocks $>$3~cm high or relief $>$3~cm high under HP$^3$ & HP$^3$ & Depl/IDS Geology\\
    \hline
    HP3-03 & HP$3$ footpatch roughness $<\pm$1.5~cm & HP$^3$ & Depl/IDS Geology\\
    \hline
    HP3-04 & HP$^3$ placed on load-bearing soil & Geology & Depl/IDS Geology\\
    \hline
    HP3-05 & Mole egress clear of rocks $>$1.5~cm diameter & HP$^3$ & Geology\\
    \hline
    HP3-06 & No partially buried rocks near the HP$^3$ that colud be blocking the subsurface path of the mole. & Geology & HP$^3$\\
    \hline
    \hline
    HP3-07 & HP$^3$ away from lander and other sources of thermal noise (e.g., rocks) & HP$^3$ & $<none>$\\
    \hline
    HP3-08 & $<none>$ & &\\
    \hline
    HP3-09 & HP$^3$ should be $\geq$0.9~m away from SEIS and WTS shadow & HP$^3$ & SEIS\\
    \hline
    HP3-10 & Mole egress clear of rocks $>$0.5~cm diameter & HP$^3$, waived & Geology, waived\\
    \hline
    HP3-11 & HP$^3$ feet clear of stones $>$2~cm diameter (to prevent sliding) & HP$^3$ & Geology\\
    \hline
    HP3-12 & HP$^3$ on flat (enough) terrain with all 4 feet in contact with the ground, to avoid rocking during mole hammering cycles & HP$^3$ & Geology\\
    \hline
    HP3-13 & It should be checked that there are no line-of-sight obstacles obstructing the ICC field of view & Depl/IDS & HP$^3$\\
    \hline
    HP3-14 & $<none>$ & &\\
    \hline
    HP3-15 & It is desired to image all four feet of the HP$^3$ SS using the IDC & Depl/IDS, waived & HP$^3$, waived\\
    \hline
    HP3-16 & The tether should not be routed over (sharp) surface stones due to wind action over the course of the mission & Depl/IDS & Geology HP$^3$\\
    \hline
    HP3-17 & HP$^3$ and SEIS engineering tether should not touch & Depl/IDS & HP$^3$\\
    \hline
    \end{tabularx}
\caption{HP$^3$ deployment requirements (01-06) and desirements (07-17), and primary and secondary subgroups who evaluated them.}
\label{tab:DeploymentReq}
\end{table}

Within a week of landing on November 26, 2018, the IDA was deployed and began acquiring images of the surrounding terrain, spacecraft and solar panels using the IDC. The first mosaic of the workspace was available on December 10, 2018, two weeks after landing. This product was made using images acquired with all arm components above the height of the lander deck.  Orthoimage and DEM mosaics with 2~mm per elevation posting were created from stereo IDC images (Fig.~\ref{fig:DepeloymentAreas}a) \citep{Abarca2019}. The workspace revealed was particularly accommodating, with a sandy, granule and pebbly surface, few rocks, and low slopes that met all of the instrument deployment requirements over most of the deployment area. Because of this, instrument deployment locations were both near their pre-landing preferred positions. 

By pointing the arm below the lander deck and closer to the ground, a higher resolution workspace stereo mosaic was acquired by the IDC and was available on December 1\textsuperscript{st}, 2018. Individual frames had a pixel scale of 0.5~mm and the DEM from these instruments had 1~mm per elevation postings (Fig.~\ref{fig:DepeloymentAreas}b). All instrument deployment requirements at  the preliminary instrument locations were met (by a large margin) in the higher resolution data (Fig.~\ref{fig:DepeloymentAreas}c). In addition to meeting all HP$^3$ requirements, all the desirements were met except for two, which were waived because they were judged to have little impact, at the selected location (Table~\ref{tab:DeploymentReq}). At JPL, the InSight Deployment Testbed environment was 'Mars-scaped' to resemble the actual workspace on Mars. Deployment of the instruments to the selected locations was tested using an engineering model of the IDA and weight models of the instruments; these tests were successful and indicated no problems with proceeding. The instrument placement locations (Fig.~\ref{fig:DepeloymentAreas}c) were certified, approved by the instrument Principal Investigators, and selected by the project on December 17, 2018.

\section{The Mole Saga: A Record of Actions on Mars} 
\label{sec:Record}
\noindent

This section provides a chronological narrative of the actions taken on Mars from the first attempt to penetrate into the subsurface (sol~92) through the cessation of penetration anomaly response activities (sol~754).  This narrative reflects several co-evolving understandings and attitudes within the anomaly response team including: (1) a narrowing, by process of elimination, of the root causes and contributing factors to the mole’s poor penetration performance, (2) a growing understanding of the properties of the martian regolith at the landing site, (3) a progression in project risk posture with respect to assets such as the science tether, the robotic arm, and the mole itself, (4) an increase in activity complexity (e.g., from simple commands for more hammering to complex activities involving precise arm positioning and loading), (5) the success, failure, or exhaustion of a particular assistance approach, and (6) the accommodation of dwindling resources to combat the anomaly, including those available on Mars (e.g., power) and on Earth (e.g., operational personnel and schedule encumbrances). 

Consideration of the above factors lead to a particular order of operations, such as the choice to pin the mole with the scoop prior to attempting pit infill.  Methods used in later stages of the anomaly resolution, such as pushing directly on the mole back cap and scraping regolith into the pit, were evaluated early in the process but initially tabled due to a lack of understanding of the environment, confidence in operational capabilities, and/or a perception of high risk. As the team’s understanding grew and the option space shrank, some (but by no means all) of these considered methods were brought back into play.  The enumeration of all decision points, risk rankings, and descriptions of paths-not-taken is beyond the scope of this paper.  Where possible, the driving factors on the choice to pursue or abandon a given method are given in the narrative.  

One further driving consideration deserves elaboration: the scientific motivation to get the mole to its operational depth ($\geq$3~m) as fast as possible.  The mole targeted a tip depth of 3--5~m due to its requirement to emplace temperature sensors below significant influence of the annual thermal wave.  Upon arrival, however, the InSight lander introduced a step-function change in the local thermal boundary conditions by removing surface dust and decreasing the albedo \citep{Golombek:2020a}. This introduced a new shadow pattern and a new source of thermal energy.  The perturbation propagated into the regolith \citep[see][]{Grott2009, Siegler2017}, introducing a new thermal wave that would complicate the interpretation of the temperature data.  The team hoped to emplace the mole as rapidly as possible, outrunning the downward-propagating lander effect while still meeting other constraints, such as making multi-sol thermal conductivity measurements during penetration and allowing sufficient time for the thermal energy of hammering to dissipate.  Ultimately HP$^3$ lost this race, but its influence on early anomaly response decisions was significant.

\begin{table}
 \begin{tabularx}{1.0\textwidth} {
 | >{\raggedright\arraybackslash} p{2.5cm}
 | >{\raggedright\arraybackslash} p{1.3cm}
 | >{\raggedright\arraybackslash} X | }
    \hline
    \textbf{Phase} & \textbf{Sols} & \textbf{Description}\\
    \hline
    Initial Attempts (IA) & 92 – 94 & Two initial hammerings commanded w/ stop triggers of 4 and 5~hours respectively, or a 0.7~m depth reported by TLM. TLM does not report any ST extraction; STATIL reports significant tilt changes, some SS motion observed via footprints\\
    \hline
    Diagnostics \& Lift (D\&L) &	97 – 211 & 	Information gathering via imaging campaigns at various times of day and IDC positions, imaging of the SS ‘window’, and two short diagnostic hammerings.  SS is re-grappled and lifted away from the mole in three steps.\\
    \hline
    Pit Characterization (PC) &	220 – 234 &	Imaging of mole, pit, and surroundings\\
    \hline
    Regolith Interaction 1 (RI-1) &	237 – 257 &	Flat scoop pushes and chop tests attempt to collapse the pit\\
    \hline
    Pinning 1 (P1) & 291 – 318 & Mole is pinned horizontally and vertically – successful penetration of $\sim$5~cm proves there is no obstructing stone\\
    \hline
    Reversal 1 (REV1) & 322 – 325 &	Reconfiguration of the arm to protect the ST removes direct contact with the mole resulting in insufficient resistance to rebound, a mole reversal event extracts $\sim$18~cm of the mole\\
    \hline
    Pinning 2 (P2) & 329 – 380 & Mole is pinned horizontally and vertically.  Successful and fast penetration permits recovery from the reversal event to approximately the same depth as at the end of Pinning 1\\
    \hline
    Reversal 2 (REV2) &	400 – 407 & Mole is pinned with vertical preload only – another reversal event occurs, extracting $\sim$5~cm of the mole\\
    \hline
    Regolith Interaction 2 (RI-2) &	414 – 420 &	A regolith scrape test is performed, and chops are executed in an attempt to further collapse regolith into the pit\\
    \hline
    Back Cap Push – Horizontal Scoop (BCP-H) & 427 – 577 &	The mole is incrementally pushed by the scoop on its back cap, providing direct resistance to rebound and allowing the mole to descend $\sim$7~cm until the back cap is flush with the surface\\
    \hline
    Regolith Interaction 3 (RI-3) &	598 & Regolith is scraped into the pit, obscuring most of the mole\\
    \hline
    Back Cap Push – Inclined Scoop (BCP-I) & 604 – 645 & The mole is incrementally pushed using an inclined scoop, allowing the back cap to descend to $\sim$2~cm below the surface\\
    \hline
    Regolith Interaction 4 (RI-4) &	659 – 700 & More regolith is scraped into the pit; each scrape is followed by a flat-scoop tamping action\\
    \hline
    Final Free Mole Test (FFMT)	& 754 &	The scoop is positioned as for the Back Cap Push to prevent mole reversal.  500~strokes are commanded, but no forward motion is observed. \\
    \hline
    \end{tabularx}

\caption{The Mole Saga: phase names, sol intervals, and summary descriptions.}
\label{tab:Phases}
\end{table}

Table \ref{tab:Phases} names the major phases of the Mole Saga, the sols covered by each phase, and the major activities and results of the events of the phases.
Unless otherwise specified, the distance from the mole back cap to the original regolith surface, as measured along the mole body (i.e., `along-mole distance') is reported below. The depth of the mole tip measured along the mole body is obtained by subtracting the back cap distance from the length of the mole of 40 cm. Vertical tip depths underground can be determined by multiplying the latter by the cosine of the mole tilt.

\begin{figure}
\centering
  \includegraphics[width=1.45\textwidth, angle=90]{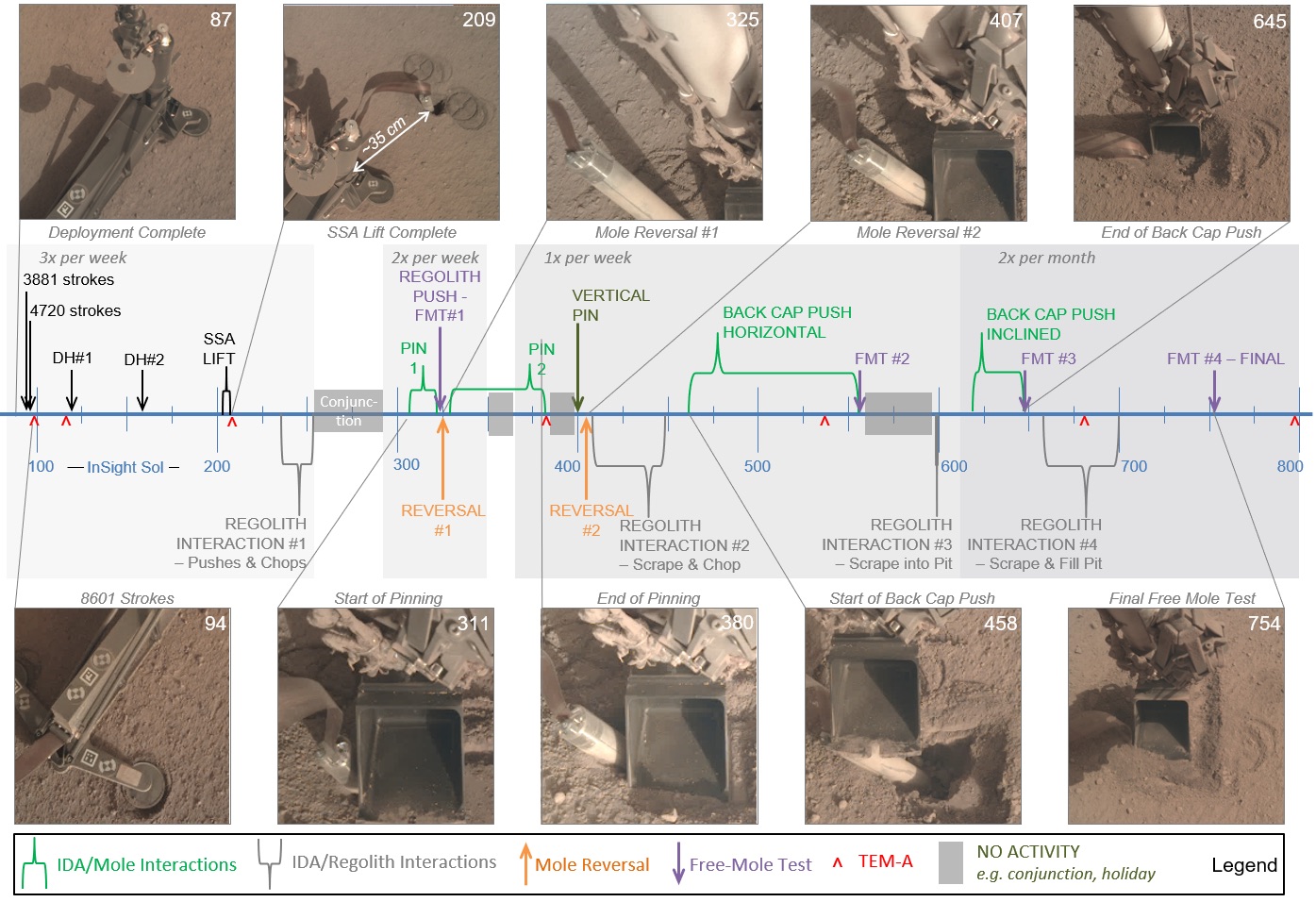}
  \caption{This linear timeline shows key periods (braces) and events (arrows and carets) of the mole penetration anomaly from the end of SSA deployment (Sol~87) to the final Free-Mole Test (Sol~764). Callout figures with sol numbers in the upper right show selected zoomed views from the IDC. Shaded background regions indicate changes in operational cadence. }
  \label{fig:MoleTimeline.jpeg}   
 
\end{figure}

\begin{figure}
\begin{center}
  \includegraphics[height=0.6\textheight, angle=90]{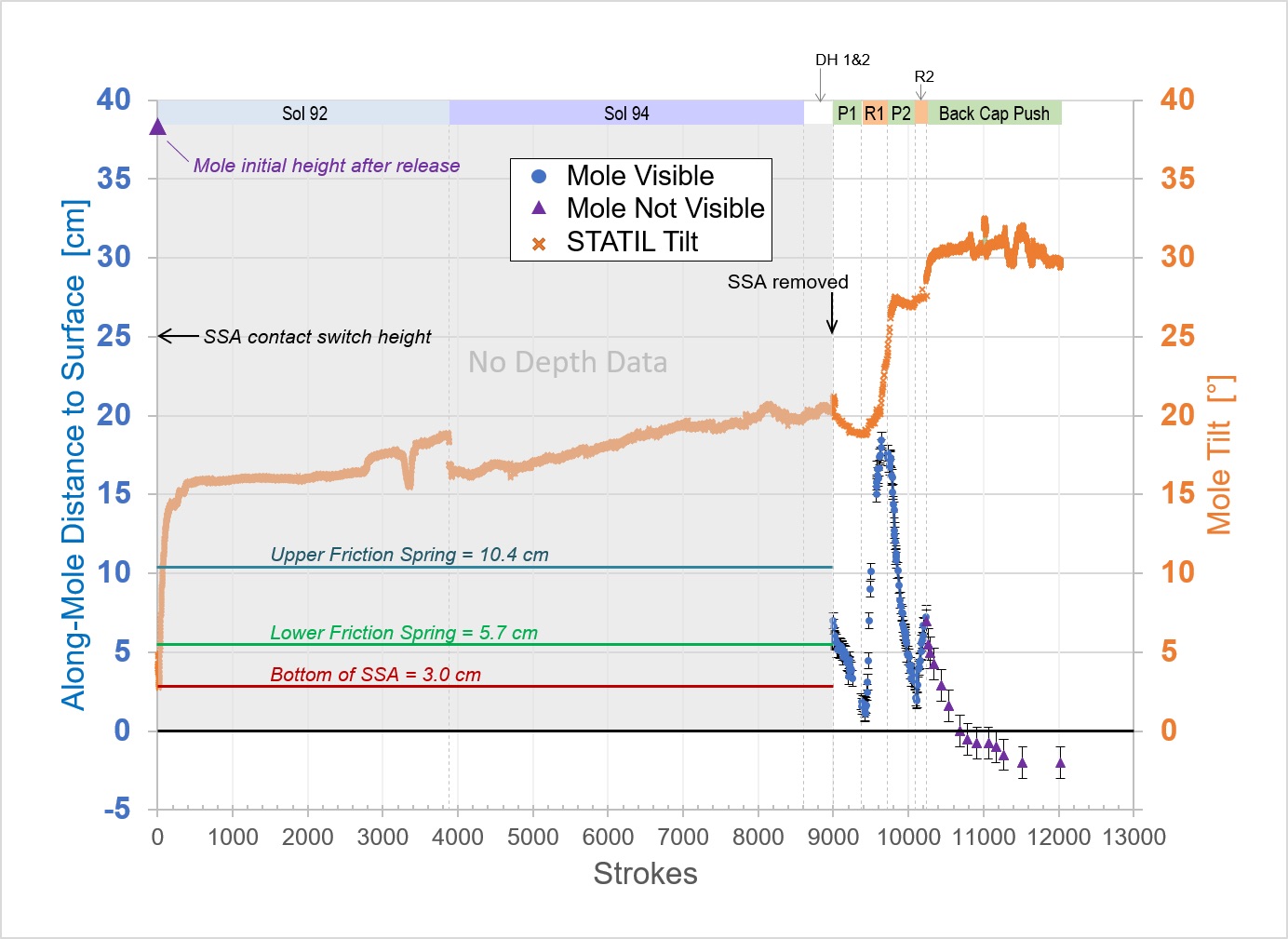}
  \caption{This plot shows (left axis, circles and triangles) the distance along-mole from the mole back cap to the original regolith surface (zero datum), and the tilt of the mole with respect to local gravity (right axis, x’s) as measured by STATIL; both axes are referenced to the total number of hammer strokes accumulated since sol 92.  Blue circles indicate along-mole distance to datum as determined from IDC images of glint features on the mole back cap.  Filled purple triangles indicate along-mole distance determined through various indirect means (e.g., SSA contact switch or IDA scoop / regolith relative position).}
  
  \label{fig:MoleDnTAll.jpeg}   
\end{center}    
\end{figure}

\begin{figure}
\begin{center}
  \includegraphics[height=0.5\textheight, angle=90]{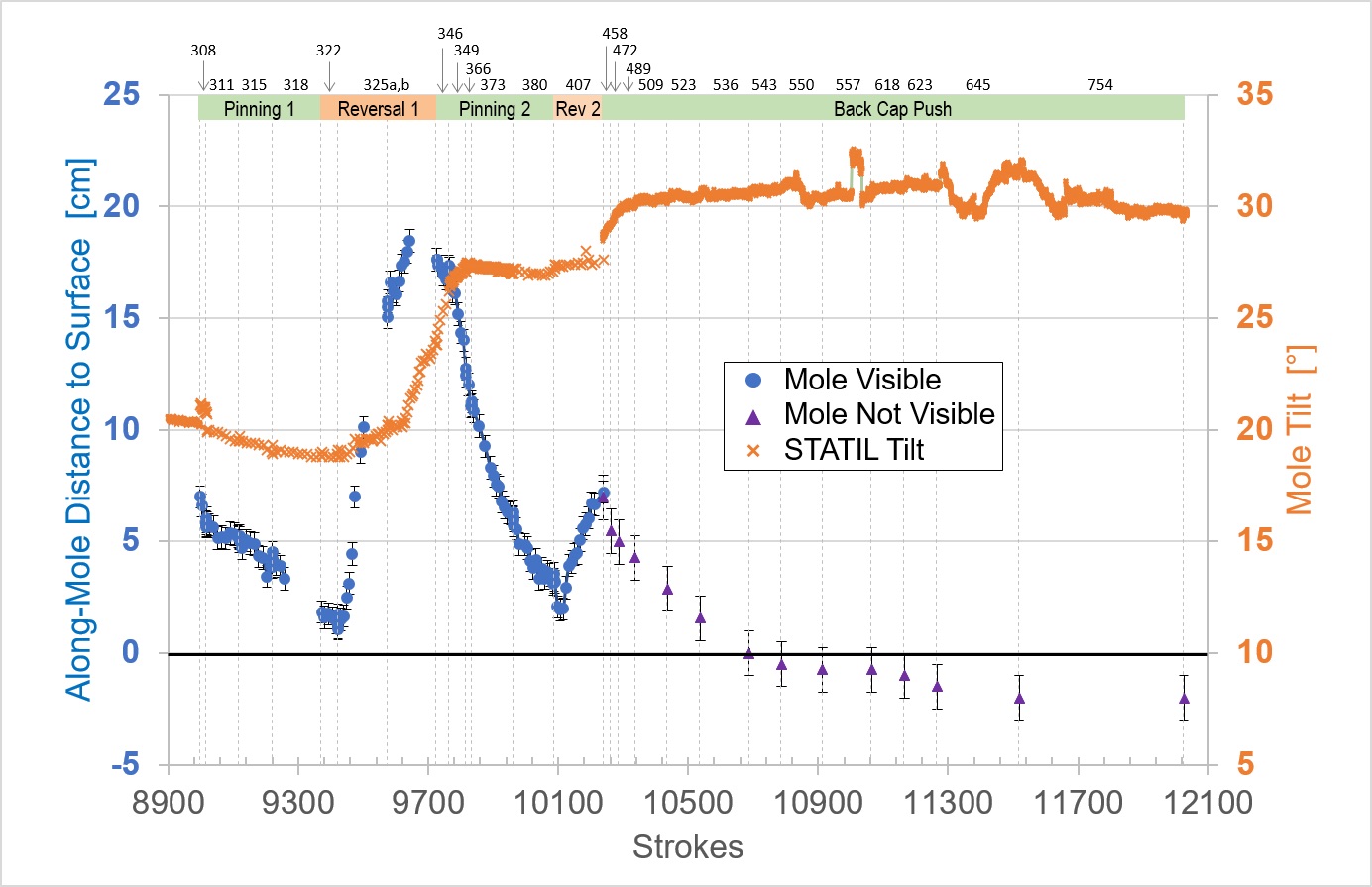}
  \caption{This plot zooms in on data shown in Fig. \ref{fig:MoleDnTAll.jpeg} beginning on sol 308 when the first pinned-mole hammering test was executed. Symbol colors and meanings are the same as in Fig. \ref{fig:MoleDnTAll.jpeg}. Note the different scales for the left-hand axis (mole depth) and right-hand axis (mole tilt from STATIL). Individual sols where hammering occurred are indicated along the top border; vertical dashed lines show the boundaries of each sol’s planned hammer strokes.  The major periods of successful mole penetration (Pinning 1, Pinning 2, and Back Cap Push) are indicated by green horizontal bars along the top, while major periods of mole reversal (Reversal 1 and Reversal 2) are in orange. }
  \label{fig:MoleDnTZoom.jpeg}   
\end{center}    
\end{figure}

This section includes three key figures for reference throughout the discussion: Fig.~\ref{fig:MoleTimeline.jpeg} shows a linear timeline of all actions taken on Mars in the 22~months (662~sols) between the first hammering attempts and the final Free Mole Test. Fig.~\ref{fig:MoleDnTAll.jpeg} plots, for the entire timeline, the mole back cap distance to the regolith (`along-mole distance' on the left-hand axis) and mole tilt (right-hand axis) as a function of hammer stroke. And Fig.~\ref{fig:MoleDnTZoom.jpeg}  presents a zoom-in on Fig.~\ref{fig:MoleDnTAll.jpeg}, highlighting the mole distance and tilt during the period when it was assisted by the robotic arm, from the first pinned hammering on sol~308 to the final Free Mole Test on sol~754.

The supplementary material contains a table with a sol-by-sol breakdown of all actions on Mars including number of commanded hammer strokes. STATIL data on mole tilt presented in the figures and discussed in the text are derived from a combination of data from the STATIL "A" and "B" channels, representing the 'x-tilt' and 'y-tilt' channels from just one of the two STATIL sensors.  

\subsection{Initial Attempts: Sols 92 \& 94}
\label{sec:Initial Attempts}
After successful deployment of the HP$^3$ SS to the martian surface on sol 76, the grapple was adjusted and released (sols~79, 81, 83), and placement position was confirmed via mosaic imaging (sol~85).   Once placement was confirmed, the mole was released from its launch lock on sol~87.  Launch lock release represented a committal of position: the mole was now free of the support structure and, though supported internally by the friction springs, could no longer be moved to any other deployment location.  Upon release, the mole was expected to drop under gravity from its locked position, allowing the tip to penetrate into the regolith by $\sim$1~cm. It is believed this occurred as expected, though no confirmation of the actual tip penetration due to this gravity drop can be given.

On Sol~92 the first hammering was commanded with a target depth of 70~cm. Due to the length of the Science Tether service loop between the TLM and the back cap of the mole ($\sim$29~cm), the ST was not expected to engage in the TLM until a the mole tip reached a depth of $\sim$54~cm.  The target depth of 70~cm was chosen to allow absolute depth markings on the science tether to be read by the TLM and so provide the stop-hammering command to the HP$^3$ electronics.  Data from penetration tests at DLR Bremen and at JPL \citep{Wippermann2020} in a variety of regolith simulants led to the expectation that the mole would reach this target depth within the first 30~minutes of hammering.  Motivated by the desire to penetrate to target depth quickly, a nominal hammering timeout period of 4~hours was set to allow for the possibility that the shallow regolith was more difficult to penetrate than in any of the terrestrial tests.  

The penetration anomaly was first recognized when data from the first hammering attempt indicated that the mole hammered for the full 4~hours (later determined using mole motor current data to be $\sim$3881~hammer strokes) and no data was reported from the TLM.  Data that were also available at this time included: (1) The mole back cap passed the contact switch (24.5~cm above the regolith surface) 4 minutes 54~seconds (77~strokes) after hammering began. (2) STATIL reported significant tilt changes, with the greatest magnitude occurring in the first 11 minutes ($\sim$170~strokes), see Fig.~\ref{fig:MoleDnTAll.jpeg}. The final reported tilt was 16$^{\circ}$ with respect to gravity. (3) At an unknown time or times during the interval, the SS moved along its long axis towards the InSight lander by $\sim$1.7~cm, as revealed by footprint markings of the initial placement site.  

Assumptions at the time, informed by terrestrial penetration experiments, suggested that over-large resistance in the regolith (either a stone or unexpectedly dense material) was a likely culprit for the poor penetration performance, since all telemetry indicated a fully functioning instrument.  A large stone entirely blocking the path of the mole was considered unlikely, though, based on pre-mission studies of the likelihood of penetration to the target depth \citep{Charalambous:2014, Golombek:2017}. So, on sol~94, a second hammering period was commanded with the same 70~cm target tip depth and a 5~hour timeout.  As with the first attempt, no readings were recorded by the TLM and the mole timed out after 5~hours (4720~strokes). 

During the second interval, the following were observed: (1) The overall tilt as reported by STATIL increased slightly, at the end fluctuating around an average of 18~degrees.  The jump in reported angle from 18~degrees to 16~degrees between sols 92 and 94 is due to a calibration shift caused by TEM-A being active during sol 92 as an auxiliary heat source for the unburied mole.  There was no actual change in mole tilt between the two sols.  
(2) At an unknown time or times during the second interval, the SS again was moved, leaving footprint impressions in the regolith.  This time the movement appeared as a rotation of $\sim$4~degrees anti-clockwise around a pivot at the rear edge of the aft right foot. The clear impressions of the SS feet edges and internal cross struts, as opposed to scrape marks, suggested that the SS was partially lifted; this is explored more in section~\ref{sec:PropertiesPenetration}.

\subsection{Diagnostics \& SS Lift: Sols 97-211}
\subsubsection{Diagnostics}
From sol 97–158, a multitude of images were taken with the ICC and the IDC (the latter with the IDA in various poses) at various times of day to try to reveal more about the mole’s attitude and what was preventing forward motion.  Images were taken near sunrise and sunset to try to observe long shadows under the SS (these were unsuccessful in providing new information).  There were some hints of the mole and the yet-to-be-discovered pit in the ICC images, but these were at the limit of the camera’s resolution and interpretation remained ambiguous until later.  Contrast-stretched images of the small ‘window’ in the back of the SS’s central tube revealed that the science tether was still inside, though it appeared to be in a skewed position. 
Two TEM-A measurements performed on sol~97 and 116 revealed that some portion of the mole was above ground and subject to diurnal temperature fluctuations.  It was thus determined that the tip of the mole was shallower than $40 \cos(18^{\circ}) = 38$~cm, i.e., the back cap was at least 2~cm above ground. 

On sols 118 and 158, two so-called ‘Diagnostic Hammerings’ were conducted of 197 and 198 strokes each (about 12.5~minutes of hammering), while the IDC was posed to take `movies’ of the SS and the Science Tether via a small window at the rear of the tubular housing of the mole.  As described in Section \ref{sec:RobotArm}, the camera software limited the maximum rate of image acquisition. In the case of sols~118 and 158, both IDC and ICC were used in parallel, limiting the cadence of IDC images to one every 47~seconds. Nonetheless, the movies so acquired revealed that the SS was jostled about during hammering, though no significant net motion such as occurred in sols~92 and 94 was observed. The window imaging did not show unambiguous motion of the science tether.

During the Diagnostic Hammerings, the SEIS instrument adopted a different digital filter configuration enabling it to listen more precisely to the individual `strikes' of the hammering mechanism. If the timing between strikes could be determined with sufficient accuracy, this could serve as a proxy for the health of the hammering mechanism (See Section \ref{sec:SEIS} below).  Since at this point it was still a possibility that the mole had been stopped by a large stone, it was hoped that analysis of the acoustic properties of the hammer strokes might also have provided clues.    

The data acquired during this period were analyzed and possibilities were debated amongst the team.  All signs appeared to point to a healthy mole, leaving three categories of root cause: (1) External Obstruction: the mole was obstructed by a large stone or an otherwise pathologically impenetrable regolith layer. (2) Internal Configuration: the mole and/or science tether was snagged or otherwise physically inhibited by the support structure. Or (3) lack of sufficient friction between the mole hull and the regolith.  Having exhausted the available sources of information, a plan was devised to remove the SS.

\subsubsection{Support System Assembly Lift}
\label{sec:Support System Lift}
The plan to remove the SS from the embedded mole had three main advantages: (1) if successful, it would eliminate the Internal Configuration category of root cause, (2) it would provide a clear view of the mole’s state and the state of the surrounding regolith, and (3) it would open up avenues for the IDA to directly or indirectly assist the mole.  Still, the lift was a risky proposition:  STATIL data and the observed SS motion suggested that the mole had rattled around inside the SS quite a bit, possibly resulting in an off-nominal tether configuration with respect to the springs. If the tether or the mole were snagged, lifting the SS could extract the mole further from the ground.  

To assess and react to these concerns, the lift was thus performed in three stages with ground-in-the-loop assessment after each stage. After the IDA grapple had been unstowed and successfully re-grappled the SS grapple hook, the SS was lifted 12~cm on sol~203. Analysis of the SS position relative to initial deployment, internal SS geometry, and mole tilt together implied that the back cap was no more than $\sim$10~cm above the regolith surface, so 12 cm was chosen as the amount of lift needed to ensure that the SS was clear of the mole, and that the narrowest (i.e., most snag-prone) portion of the science tether was below the friction springs.  Lift stage~1 was successful, there being no anomalous SS tilting or apparent mole extraction as seen in ICC images.   At the end of the lift, IDC images revealed a portion of the mole and also provided the first clear pictures of what became known as ‘The Pit.’  

The second stage of the lift, performed on sol~206, lifted the SS an additional 13~cm, for a total SS lift of 25~cm.  This amount was needed to exhaust the science tether service loop and pull a small amount of it through the TLM.  Ground-in-the-loop confirmed that 7.6~cm of tether was extracted through the TLM with no anomalous SS tilting and no apparent movement of the mole, thus indicating a successful lift.

The final stage of the lift on sol~209 lifted the SS an additional 29~cm vertically (for a total lift of 54~cm), thereby extracting sufficient science tether slack to allow the SS to be placed away from the mole.  The SS was then brought down in a step-wise fashion to a point closer to the lander, ultimately being placed with the SS’s mole egress point $\sim$35~cm from the center of the pit.  The lift operation completed successfully and the mole, pit, and science tether were now clearly visible.  This phase closed with a further TEM-A measurement on sol~211 (this was not useful for thermal conductivity, but was performed to observe the difference between a shadowed mole and one exposed to the sun and sky).

The close of this phase saw the team in possession of new data: (1) The mole height above the regolith was observed to be $\sim$7~cm (along-mole distance to the original regolith surface).  (2) The mole azimuth pointed to the southwest and its body appeared to rest against the northeast corner of the pit. (3) The pit itself had irregular yet nearly vertical walls and a depth of $\sim$5~cm or more, indicating the presence of a cohesive layer (a.k.a. duricrust) much thicker than any that had been anticipated.  This observation lent significant strength to the lack-of-friction hypothesis.  

\subsection{Pit Characterization \& Regolith Interaction 1} \label{RegolithInteracttion1}
\begin{figure}
    \centering
    \includegraphics[width=0.8\textwidth]{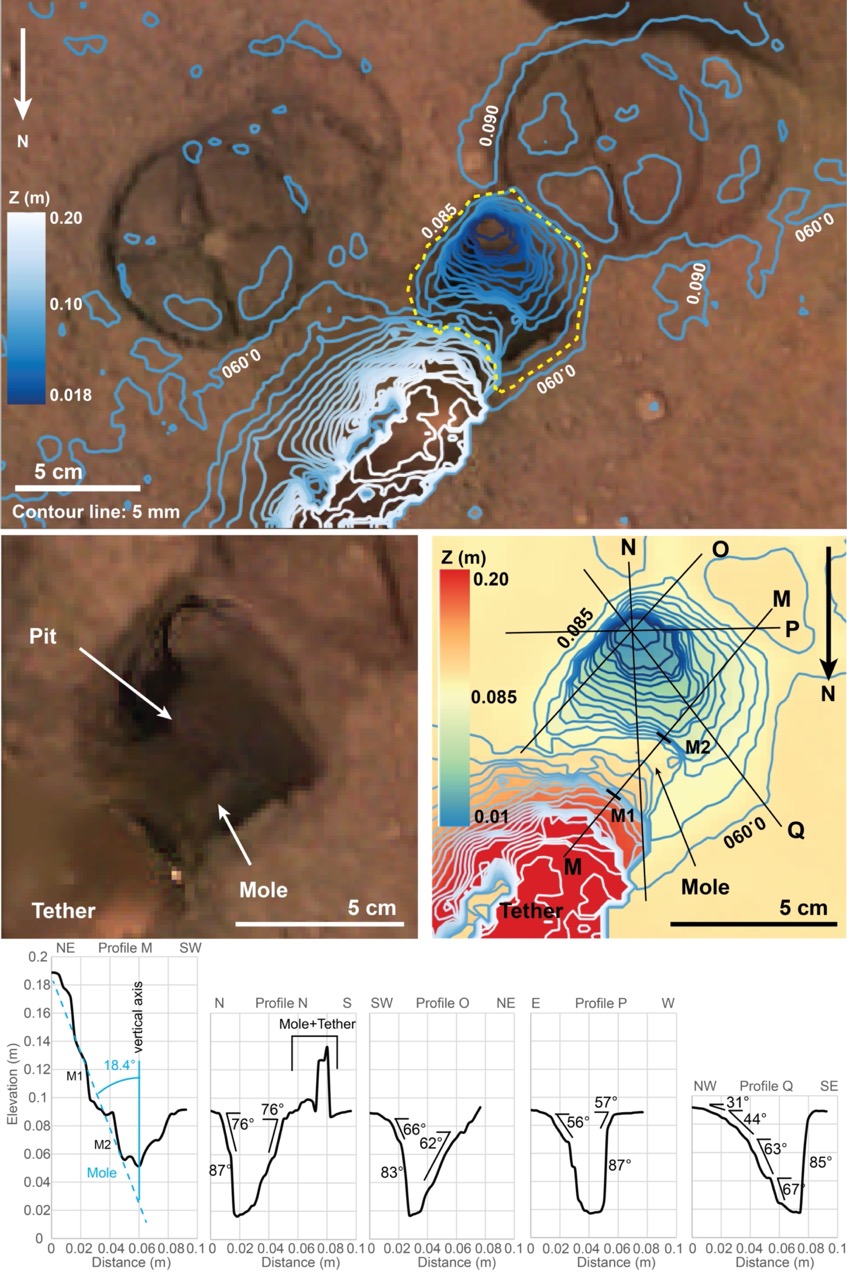} 
    \caption{Digital elevation model of the pit based on a 4x4x1 IDC imaging data set  taken on sol 230 after SSA replacement and using virtual control point methods. The top frame shows 5 mm depth interval contour lines superimposed on the orthorectified image mosaic. The rim of the pit is marked by a yellow dashed line. In addition to the pit, the imprints of the SSA feet in the fine-grained surface layer are clearly seen as well as the tether connected to the back-end of the mole. Below the top frame, from left to right, a close up orthorectified image of the pit is shown and a colour-digital elevation model (DEM) of the pit in which the reference elevation plane is ~2 cm below the deepest point of the pit. Labelled black lines correspond to the location of topographic profiles M - Q shown in the panels in the bottom row.  Profile M extends all along the mole between points M1 and M2 and up the tether. The average slope between M1 and M2 is $18.4^\circ$ which compares well with the tilt angle of the mole measured by STATIL of $18.5^\circ$.  Selected measured topographic slopes are given.
    }
    \label{The Pit}
\end{figure}

\begin{figure}
    \centering
    \includegraphics[width=1.0\textwidth]{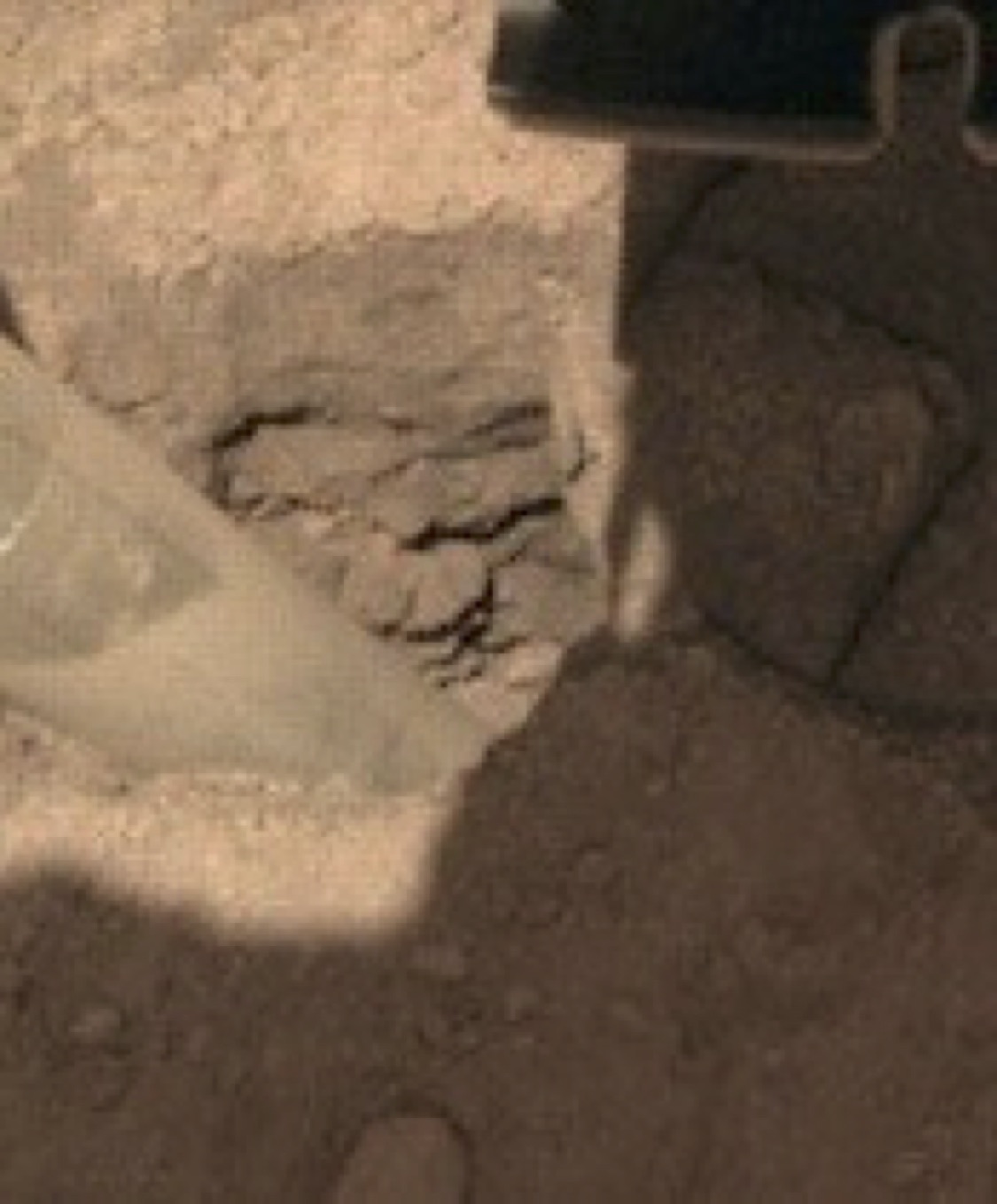} 
    \caption{ Image of hole created by the HP3 mole showing the almost vertical southerly wall of the pit and resistant layers in it. These layers have steep edges and overhangs indicating cohesion in the soil. Small rocks appear cemented in a fine-grained matrix, similar to the pits beneath the lander. Mole is in the foreground angled $\sim$15° towards the right.
    }
    \label{Mole Pit 2.jpeg}
\end{figure}

From sol~220–234, closeup and mosaic images were taken of the mole, pit, and surroundings at various times of day.  Mid-day images provided good illumination into the pit, further revealing its irregular   southerly and westerly nearly vertical walls and significant depth. Fig. \ref{The Pit} shows the results of digital elevation models of the pit and Fig. \ref{Mole Pit 2.jpeg} shows a close-up image of the pit's southerly wall with part of the mole in the foreground. The maximum depth of the pit, its volume and average depth depend on a definition of a reference surface. Using the yellow dashed line in the middle panel of Fig. \ref{The Pit} tracing the rim of the pit as reference, the maximum depth is 72~mm, the volume is $6.73 \times 10^4$~mm$^3$ and the average depth is $19.3$~mm. 
Using the 0.085~m contour line in the DEM as reference, the respective values are 69~mm, $5.4 \times 10^4$ ~mm$^3$, and $20.1$ ~mm. 
The volume estimates do not include the mole which contributes $1.1 \times 10^4$ mm$^3$ to  $1.5 \times 10^4$ ~mm$^3$  depending on the average depth of the pit. The mole blocks the view to a significant part of the bottom of the pit. It is unknown whether or not  additional volume of the pit is to be found there.  

Some of the layers within the pit have pebbles that appear cemented in a finer-grained matrix (Fig. \ref{Mole Pit 2.jpeg}). These steep, resistant layers are similar to the duricrust observed in the pits beneath the lander and the clods of material scattered during landing. The landing site shows no evidence for interactions involving significant amounts of surface or subsurface liquid water and appears to have been dry since the deposition of the Hesperian basalts \citep{Golombek:2017}. However, the layers of crust and duricust could be cemented by salts deposited by thin films of water via interactions of atmospheric water vapor and soils as suggested by chemical measurements by Viking and Mars Exploration Rover spacecraft \citep{Banin:1992, Haskin:2005, Hurowitz:2006}.

 \begin{figure}
    \centering
    \includegraphics[width=1.0\textwidth]{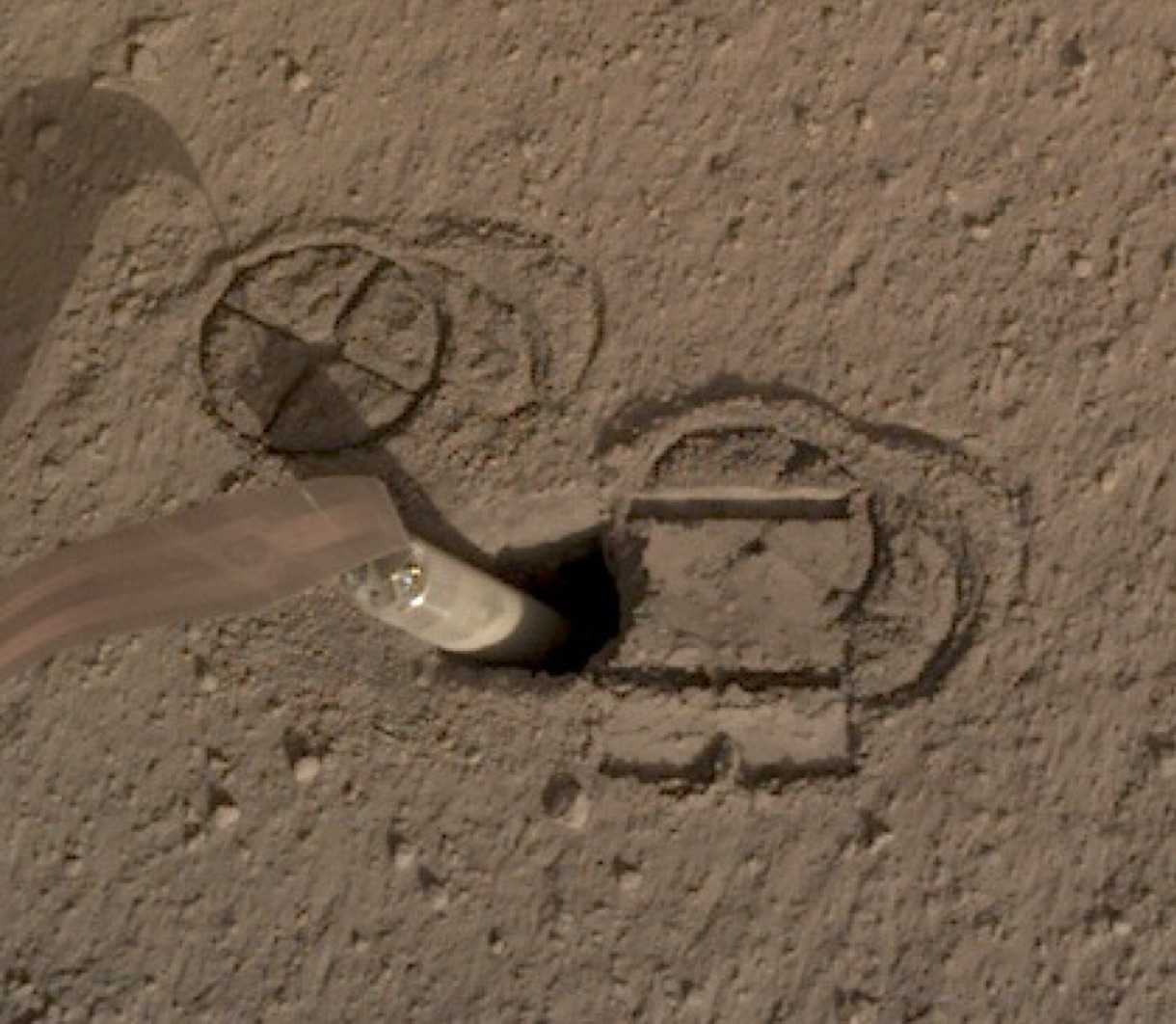} 
    \caption{ Image of mole hole and surface after interactions with the HP3 SSA feet and scoop. Circular cross patterns are imprints of the HP3 SSA feet in the soil. Smooth, reflective rectangular surface is where the flat base of the scoop (7.1 cm wide) was pressed against the soil, causing a $\sim$5 mm indentation. Horizontal troughs near the top and bottom of the scoop imprint are where the front blade of the scoop penetrated into the soil.
    }
    \label{Mole Pit 1.jpeg}
\end{figure}

By this time, the remaining root causes for the mole penetration anomaly had been reduced to two possibilities: There was an external obstruction, or there was insufficient friction provided by the surrounding regolith. Or perhaps both.  In an attempt to address the latter problem by collapsing the pit walls to provide more regolith friction, the IDA scoop was brought down to touch the regolith in two types of interactions: A flat scoop push, wherein the broad flat side of the scoop was pressed into the ground (sol~240 and 253), and a vertical chopping motion performed with the scoop's flat edge vertical and the tip pointed normal to the surface (sol~243, 250, and 253).  The scoop flat push resulted in a sharply defined area, approximately 1~cm deep, resulting from the compaction of an unconsolidated layer (see Section~\ref{sec:PropertiesScoop}). The chopping activities were largely unsuccessful at collapsing the pit, although the chop on sol~243 did break off a small wedge (1~cm x 1~cm) that fell into the pit.  Neither the chopping nor pushing attempts resulted in a significant change in pit morphology below the surface compacted layer.  The small collapsed wedge did however provide constraints on estimates of the consolidated layer strength which is evaluated in section \ref{sec:PropertiesScoop}.

\subsection{Pinning 1}
\label{subsec:pin1}
Nothing could be done about a subsurface stone or other obstacle blocking the mole.  However, the low abundance of large stones at the surface and other considerations of landing site geology (e.g., \citep{Golombek:2020a}, weakened this hypothesis.  Instead, if lack of friction were the cause, application of force on the mole from the IDA could help overcome the deficiency enough to get the mole to a depth where it could dig on its own.  By positioning the scoop adjacent to the mole and then overdriving the arm both horizontally and vertically into the mole hull (`pinning’ the mole), the frictional force between the mole and scoop and between the mole and the regolith beneath could be increased.  
It was hoped that this increased friction would exceed the 5–7~N 
rebound force threshold and allow the mole to make forward progress.

We used the seismometer as a tiltmeter to monitor the quality of the preload force exerted by the IDA. 
By low-pass filtering the seismometer response to the loading and unloading IDA motions, we could compare the relative magnitudes of the forces exerted by the IDA by comparing the tilts induced by loading, during hammering, and unloading. We found that the preload exerted by the IDA was maintained across intervals of weeks of no activity. During hammering, the preload when pushing directly on the regolith was generally progressively reduced. When pushing on the side of the mole, the preload was largely maintained, though some instances showed some decrease.

From sol~291–305, the IDA was brought into position and pinned the mole at a point $\sim$3~cm above the regolith surface.  On sol~308, 20~hammer strokes were commanded.  The resulting movie using only the IDC (image cadence: 32~seconds) showed some barely perceptible ($\sim$5~mm) motion downward, as well as mole rotation around its long axis. There was no significant change in overall tilt reported by the STATIL sensors, though they did confirm that the slight rotation of the mole during these 20~strokes was a real effect. The tendency of the mole to rotate during hammering had been observed in the laboratory.

In the period from sol~308 to 458, the back cap of the mole was visible to the IDC, allowing the distance-to-regolith to be determined from image analysis.  The technique used tracked a solar `glint feature’ at the edge of the back cap whose position was invariant to mole rotations. In this period, Mole hammerings and the associated IDC images occurred around the same time in mid-afternoon, providing a similar sun angle. This reference point provided the location of the mole in IDC image space.  The width of the mole in pixels was measured at the same longitudinal position.  This was used along with the known mole width (27.0~mm), mole tilt, and camera resolution (0.82~mrad/pixel) to derive the glint's range from the camera.  Each image was additionally co-registered to the base map of the workspace (taken on sols~16 and 243) to get an absolute, consistent reference frame for the camera model.  This adjusted camera model was then used to convert the glint location and range into XYZ coordinates, thus providing its position. Accumulated analysis errors result in an uncertainty of back cap height of $\pm$0.5~mm, shown by the error bars in Figs.~\ref{fig:MoleDnTAll.jpeg} and \ref{fig:MoleDnTZoom.jpeg}.

On sol~311, the IDA was commanded to move down by 5~mm.  This vertical re-pinning motion was accommodated by the scoop sliding along the inclined surface of the mole, thus increasing both vertical and horizontal loading.  Following this motion, 101~hammer strokes were commanded. 

The 101~strokes commanded on sol~311 resulted in a downward motion of the mole of $\sim$1~cm.  This obvious motion of the mole into the regolith was a major event. Had there been an obstructing immovable stone or impenetrable gravel layer, there would have been no forward motion.  Recall that the IDA was not pushing the mole into the regolith but only loading it from the side; all downward progress was due to mole hammering while under this limited lateral loading of 10~N at most.  Unambiguous proof was finally in hand that there was no rock blocking the mole and the External Obstruction branch of the fault tree could be eliminated.  This became clearer still when further hammering was commanded on sols~315 (101~str.) and 318 (152~str.). Sol~315 hammering was preceded by another 5~mm downward motion of the IDA, but sol~318 hammering had no preceding arm motion. Together, the pinning technique had resulted in a combined motion since sol~308 of $\sim$5~cm into the ground. Each forward motion was accompanied by further anti-clockwise rotation of the mole. At the end of sol~318, the mole back cap was $\sim$1.5~cm above the original regolith surface, see Fig.~\ref{fig:MoleDnTZoom.jpeg}.

\subsection{Reversal 1}
\label{subsec:Reversal1}
Tactics now needed to change.  The vertical motions of the scoop against the sloped side of the mole during sols~311 and 315 resulted in a horizontal as well as vertical preload (stored in the compliant composites of the IDA).  More mole motion would move the mole back cap below the scoop edge, resulting in a side-swipe of order one mole diameter that could potentially damage the science tether.  Rotation during hammering had unfortunately oriented the mole such that the tether plane was in-line with the preload direction, maximizing the potential for tether damage.

How then to continue to assist the mole while removing this risk?  Analysis suggested that some force could be transferred to the mole simply by preloading the scoop vertically onto the regolith surface immediately adjacent to it, thereby avoiding direct mole contact and removing the risk to the tether.  There were also some indications that the portion of the mole below the bottom of the pit ($\sim$30~cm of mole) might now be experiencing more friction, perhaps enough to allow it to dig freely.  Recall that the mole back cap's along-mole distance to regolith surface when the SS was removed was $\sim$7~cm, which would mean that the back cap was below the upper tier of friction springs.   At the time it was argued that perhaps the mole slipped from a regime of penetration into a regime of rebounding when the back cap passed below this upper tier. With each tier providing a step-function change to resistance to mole upward motion, it was thought that the additional 5~cm of regolith contact might have put the mole on the other side of the threshold.  There were two key flaws in this analysis.  First, the friction provided by 5 cm of additional regolith contact was significantly less than that of a friction spring tier (~27 N resisting upward motion per tier).  Second, this interpretation assumed that the support structure remained in (more-or-less) close contact with the ground throughout sol 92.  It was not appreciated at the time the degree to which the support structure had been lifted away from the regolith during the sol 92 hammering.  The geometric analysis supporting significant SS ‘ratcheting’ up an already-rebounding mole would not come until much later (See \ref{PenetrationRate}).  The now-accepted interpretation is that the mole encountered sufficient resistance to begin rebounding while still being held with both tiers of friction springs.  The upward force transferred to the SS through the springs (~54 N) was more than enough to lift the light support structure (7.4 N on Mars) away from the regolith.

As of this point it was still considered too complicated and too risky to attempt a direct push on the mole’s back cap.  Thus on sol~322, the arm was disengaged from its pinning position and pressed with a large overdrive into the regolith next to the mole.  The scoop was positioned over the pit, which meant the force could only be transferred to the mole via the pit walls to the regolith beneath.  An initial command of 50~strokes executed on sol~322.  The resultant movie showed some downward motion initially, but then mole movement became much smaller and of ambiguous direction.

With an ambition that overreached our knowledge of the situation, the team commanded on sol~325 two periods of hammering (152~strokes each) separated by a re-application of the vertical preload on the regolith. This had very unfortunate results.  The ambiguous motion seen at the end of sol~322 was seen briefly at the beginning of the first hammering (sol~325a) and then gave way to a very rapid extraction of the mole.  At an average rate of 1.0~mm/stroke, and a maximum rate of 3.2~mm/stroke, the mole withdrew from the regolith throughout the first phase of hammering.  The back cap distance to the regolith increased from 1.5~cm to $\sim$15~cm, accompanied by a tilt increase from 19$^{\circ}$ to 20$^{\circ}$.  Since there was no planned ground-in-the-loop step to recognize and react to this reversal, the reapplication of preload on the regolith occurred autonomously and the second hammering (325b) resulted in a further 3~cm of extraction, and a dramatic increase in tilt from 20$^{\circ}$ to 24$^{\circ}$, see Fig.~\ref{fig:MoleDnTZoom.jpeg}.

The story of the mole reversal rates and tilt changes during this reversal is unfortunately incomplete. A commanding error in IDC sequencing resulted in no images being taken during the latter half of either hammering phase, preventing any image analysis of mole heights during the second half of both hammerings.  The end result of these reversal events was a mole that had a severe tilt of 24$^{\circ}$ and was roughly 18~cm out of the ground with its center of mass only a few cm below the regolith surface.  See section \ref{sec:Duricrust Thickness} for a discussion of the causes of the high reversal rate and why it stopped half way through the second hammering period at 16.5~cm extraction thereby providing an estimate for the thickness of the duricrust.

\subsection{Pinning 2}
Subsequent to the mole reversal event, the IDA was retracted and used to image the mole and surroundings on sols~329 and 332. In the InSight testbed at JPL, the IDS team tested several other pinning techniques, hoping to find one less likely to knock the mole over.  Only after examining each of them was it concluded that the existing technique used in sols~298--318 was still the least risky.   Upon a careful approach to the mole with the scoop near the regolith (to reduce the lever arm to the center of mass and thus reduce the change of tipping the mole over) the mole was re-pinned horizontally on sol~339. On sol~342, vertical pinning motion of 1.5~cm was applied.  Interestingly, the repining activities had almost no effect on the mole tilt. 

On sol~346, 40~strokes were commanded that resulted in a few cm of downward motion accompanied by an increase in tilt from 24$^{\circ}$ to 26$^{\circ}$.  This tilt change motivated a vertical re-pinning on sol~349 consisting of 4~cm of commanded downward motion of the IDA scoop. This resulted in the scoop contacting the regolith and compressing the upper unconsolidated layer.   The vertical motion was followed by 50~strokes that resulted in an extremely rapid penetration of about 4~cm, ending the sol with a back cap distance-to-regolith of 14~cm and a further tilt increase to 27$^{\circ}$.  The IDA remained in this position for subsequent hammerings on sols~366 (19~str.), 373 (127~str.), and 380 (126~str.) which together resulted in further back cap distance-to-regolith reductions to 11~cm, 6~cm, and 3.5~cm, respectively.  The tilt during these sols remained constant at 27$^{\circ}$.  The maximum rate of motion during this re-penetration (see Fig.~\ref{fig:MoleDnTZoom.jpeg}) was $\sim$0.6~mm/stroke occurring over sols~349 and 366.  The hammerings on sols~373 and 380 had an average re-penetration rate of 0.3~mm/stroke. At the end of sol~380, the back cap of the mole was $\sim$3.5~cm away from the original regolith surface.  Even though still partially exposed, a TEM-A measurement was performed beginning on sol~380.

\subsection{Reversal 2}

IDA motions in Pinning~2 consisted of one horizontal motion (sol~339); augmented by vertical moves of 1.5~cm and 4~cm on sol~342 and 349, respectively.  By sol~400, it had been 20~sols since the last hammering and 51 sols since the last application of any preload.  It was feared that pre-load may have been dissipated during the previous hammerings. The concern was compounded by the scoop being in contact with the regolith since sol~349, since this may have prevented a strong scoop/mole force coupling.  To re-establish preload, the IDA was commanded on sol~400 to retract vertically (commanded 1~cm up) and re-pin using a vertical-only motion (commanded 3~cm down).  The upward motion did not bring the scoop out of contact with the regolith; rather, the motion was accommodated by the compliance in the arm links. By all expectations, this should still have retained the requisite preload to allow forward progress.  

Unfortunately, this appeared to be insufficient to resist the mole’s rebound characteristic at this depth when, on sol~407, 151 hammer strokes were commanded and resulted in a second mole reversal event.  Close examination of the IDC movie reveals that at first it appeared that penetration was continuing: the depth changed by 1~cm in the downward direction in the first $\sim$10~strokes, reaching a back cap distance-to-regolith of 2~cm. Downward motion stalled at this depth for about 20~strokes, then began to rapidly reverse at an average rate of 0.4~mm/stroke for the rest of the hammering session.  The along-mole distance between the back cap and regolith increased from 2~cm to 7~cm, with tilt remaining roughly constant at 27.5$^{\circ}$.


\subsection{Regolith Interaction 2}

During sols~414–420, the IDA was retracted and some further regolith interactions using the scoop were investigated. These consisted of a short scrape test and a further chop test. The chop test was specifically aimed at collapsing some of the pit walls, thereby increasing the amount of regolith in contact with the mole.  This was somewhat successful, resulting in a small amount of duricrust being broken off the southerly  wall of the pit on sol~420.

\subsection{Back Cap Push - Horizontal Scoop}
\label{BCP-HS}
Throughout the anomaly since the SS lift, the team had considered multiple methods by which the IDA might assist the mole \citep{Sorice2021}.  The reversal events reduced confidence that further attempts to hammer whilst pinning would be successful.  Thus the pinning method was abandoned and the team transitioned to a long campaign (sols~427–557) of pushing directly on the mole’s back cap with the scoop.  This had the advantage of supplanting friction as the main source of resistance to rebound, placing the scoop in the path of the mole’s rebound vector and directly mitigating the risk of reversal. However this delicate operation required much finer positioning than was typical for the IDA requirements and each placement was approached carefully so as to do no harm to the science tether, mole, or IDA. Since the IDA actuators could not directly follow the mole along its path into the regolith, each hammering period was followed by a repositioning of the scoop and a re-application of IDA preload.  Initially this required ground-in-the-loop after each preload and hammer action, though in later stages the preload and hammer steps were combined as confidence in the methodology grew.  

Though a long and arduous process (requiring 4.5~months to execute 9~back cap hammerings), it was successful:  (1) the mole only moved down, the scoop providing resistance to rebound sufficient both to prevent reversals and allow the mole to progress, (2) the mole moved a total of 8~cm along its axis, ending this phase with the back cap $\sim$1~cm below the original regolith surface (the maximum depth reachable by the flat scoop after the loose 1~cm surface layer had been compressed), and (3) the tilt increased only about 4 degrees from $\sim$27$^{\circ}$ to $\sim$31$^{\circ}$.  

After the 6th of the 9~back cap hammerings, on sol~536, the mole’s back cap was flush with the original (uncompressed) regolith surface.  

During this and the subsequent Inclined Scoop phase, the right side of the back cap could not be seen by the IDC, thus preventing application of the glint technique described above for determining mole height. Instead, the team measured horizontal motion of the mole using the left side of the back cap relative to the ground (outside of any area churned during hammering).  That distance was scaled based on known measurements (scoop slot and scoop width) and their apparent width in the image.  Apparent progress in ICC images (scaled based on mole shaft width) were used to confirm that these IDC depth estimates were reasonable. Overall the technique provided a mole depth uncertainty of $\pm$1.0~cm and this is reflected in the error bars for the ‘Mole Not Visible’ points in Fig.s~\ref{fig:MoleDnTAll.jpeg} and \ref{fig:MoleDnTZoom.jpeg}.


\subsection{Regolith Interaction 3}
On sol~598, a full 12~cm scrape was performed to bring more material into the pit.  This was fully successful and resulted in the mole being nearly completely covered. 


\subsection{Back Cap Push - Inclined Scoop}
The horizontal scoop used during the previous back cap push campaign could not descend further than the level of the compressed unconsolidated layer, $\sim$1~cm below the original regolith surface.  By this time, the various scrape and chop actions had widened the pit to approximately one scoop width.  This allowed the team to use an inclined scoop (30$^{\circ}$ from horizontal) to continue to preload the back cap using the scoop tip rather than its bottom edge.  Three inclined back cap push activities were commanded on sols~618 (101~str.), 632 (101~str.), and 645 (252~str.).  

These were successful in their execution, though they caused only a small increase in mole depth, with a combined 454~strokes resulting in only $\sim$1~cm greater back cap descent.  Though the mole could not be seen directly, the science tether could, and  it had enough visible features to track mole progress (or lack thereof).  Interestingly, the mole was observed to change orientation and tilt in an irregular fashion during these hammerings, with tilt fluctuating between 29.5$^{\circ}$ and 32$^{\circ}$.  The position of the science tether against the scoop was seen to migrate, and there was evidence of regolith pumping out of the pit adjacent to the mole’s position. 

The IDC movies revealed periods during these three hammerings where particles within the scoop did not move. This suggests some brief moments of ‘Free Mole’ hammering without IDA contact. Other images in the same hammerings showed substantial particle motion, suggesting that at those times the mole was attempting to reverse and was rebounding into the scoop.  This recalls the observation of sol~407 where the mole made some downward progress, stalled, then reversed.

\subsection{Regolith Interaction 4}

The final actions to help the mole were focused on increasing regolith contact, and thus potential hull friction, as much as possible.  The goal was to cover the mole with scrapes of regolith then use the arm to compact this material and pre-load the mole via the soil.  Recall that something similar was attempted on sols 322 and 325 to aid the mole when it could no longer be safely pinned.  In those previous actions the pit was empty and the force of the scoop on the surface was transferred to and dispersed in the competent duricrust layer. The desire now was to scrape regolith into the pit in several stages, tamping the pile after each one to compact and densify the material.  It was hoped this would provide a more direct load path between the scoop and the mole.

From sol~659 to 700, three 12~cm scrapes were performed, bringing material from the far side of the pit into the pit itself.  These were successful and resulted in a completely buried mole.  A TEM-A measurement was included on sol~680 to take advantage of this and provided the first high-quality thermal conductivity data. 
The scrapes provided piles of sand the angles of which are used in section \ref{sec:scrapeangle} to estimate the friction angle of the sand.

\subsection{Final Free Mole Test}
By this time, power and thermal considerations for InSight were complicating operations as dust continued to accumulate on the solar panels and Mars approached aphelion. The team had to consider the history of the mole’s penetration rates, which were quite low (typically 0.1~mm/stroke and less) in the context of the expected lifetime of InSight.  At what point could the mole make enough progress to be at an acceptable depth (3~meters) before the lander could no longer support its operation?  Reaching the target depth would not be useful if the heat from hammering could not dissipate in time to make a clean measurement. TEM-A measurements and pre-landing analysis placed this necessary cooling time at around 100~sols.  While it was not certain InSight could survive the thermal minimum of aphelion, if the mole could reach its target depth before this point it was reasoned that heat from hammering could dissipate during the aphelion lull and allow some good quality measurements of the thermal gradient when operations resumed.  

For this plan to be successful, the low rate of penetration implied multiple days of continuous hammering to get the mole to an acceptable depth in time.  Though it was hoped that the rate of penetration would increase at some depth (e.g., assuming that the mole passed through a densified layer), this could not be counted on.  In order to make this constraint fit within the worsening power and thermal situation, it was decided amongst the team that after the scrapes and tamps of the previous period were completed, there would be one final Free Mole Test.  In this test, the IDA scoop in a horizontal orientation would be maximally preloaded onto the regolith filling the pit above and around the mole.  The mole would then be commanded to hammer 500~strokes, the high number being chosen such that the result, whatever it was, would be unambiguous.  

Operational constraints and winter holidays pushed the final Free Mole Test to January 9th.  Then, on sol~754, 500~strokes were performed.  No further downward motion was detected by observing the science tether, although a substantial amount of lateral tether motion was observed.  The mole tilt varied irregularly between 29.5$^{\circ}$ and 32$^{\circ}$, some regolith poured out of the pit from below onto the surface next to the tether, and regolith particles on the IDA and in the scoop were seen to move erratically.  This latter evidence suggests the mole was attempting to reverse and rebounding into the scoop, similar to what was seen during the back cap push activities with an inclined scoop.  Thus it was determined that the final Free Mole Test was not successful and further attempts to assist the mole to achieve greater depth were abandoned.  

A final IDA retraction and mosaic was performed on sol~775, and a TEM-A test with a fully buried mole was again performed on sol~795.
\\
\\

\section{Soil Mechanical and Thermal Properties Derived from Actions and Measurements During the Mole Recovery Activities}
\label{sec:PropertiesPenetration}

In this section we will interpret the data collected during the almost 2 years of operating the mole and the IDA to support the mole on Mars. We will discuss the dimensions and the fill of the pit to derive density and porosity ratios between the duricrust and the sand underneath the crust. We will further discuss the penetration rate and derive values for the penetration resistance and an estimate of the thickness of the duricrust. The results of scoop--soil interactions will be used to calculate the cohesion of the duricrust and the sand and estimate the internal friction angle for the latter. The thermal measurements with the HP$^3$ radiometer and the TEM-A hardware have been used to estimate the thermal conductivity and the density up to 40 cm depth. Seismic velocities, elastic moduli and the Poisson's ratio have been determined using the recordings of the seismometer SEIS. In the section following thereafter we will combine the results in a synopsis and present a model of the top 40 cm of the martian soil at the site of the HP$^3$ mole pit. 

\subsection{Pit formation and soil porosity}
\label{sec:soil porosity}

\begin{figure}
    \centering
    \includegraphics[width=0.9\textwidth]{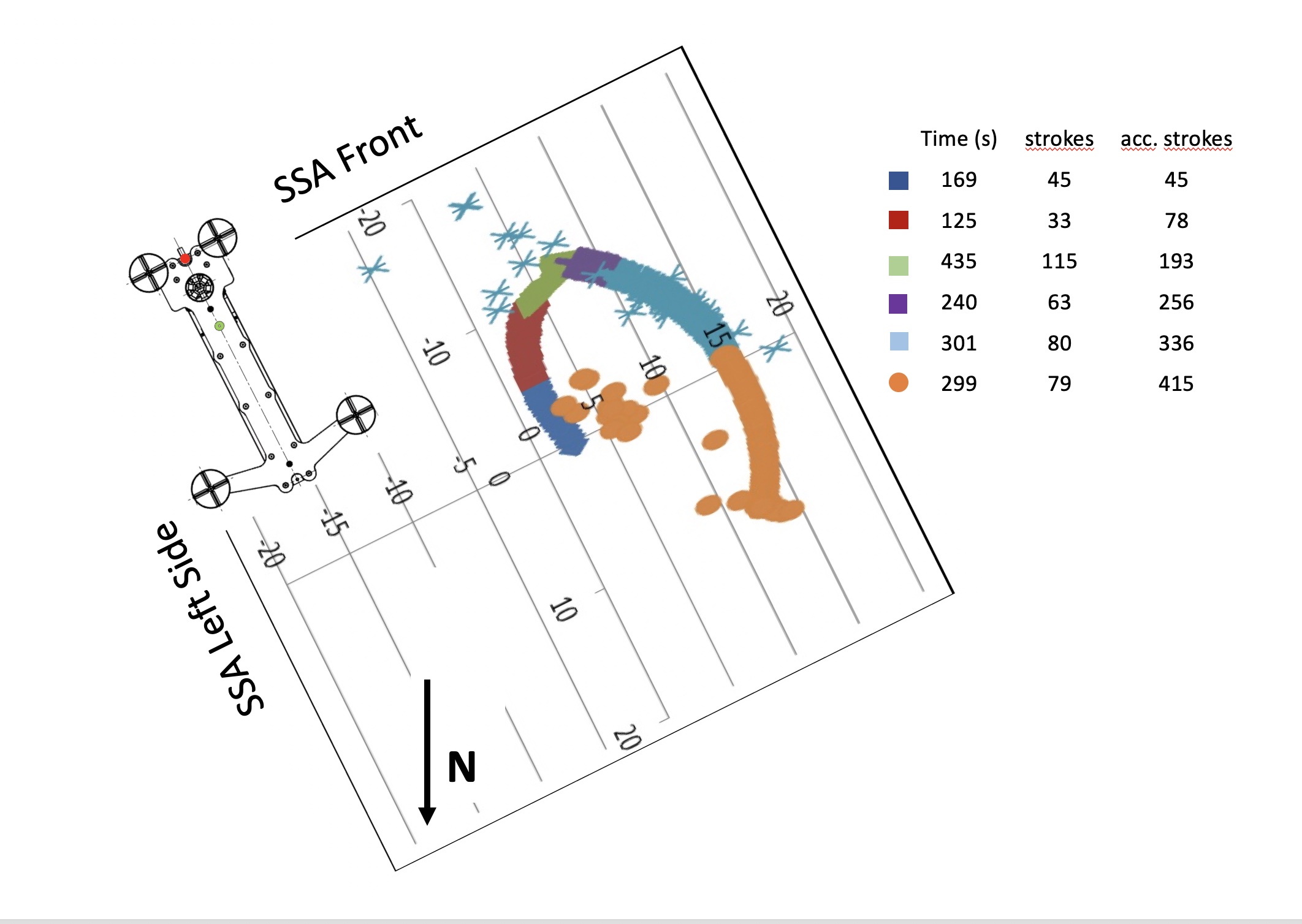}
    \includegraphics[width=0.9\textwidth]{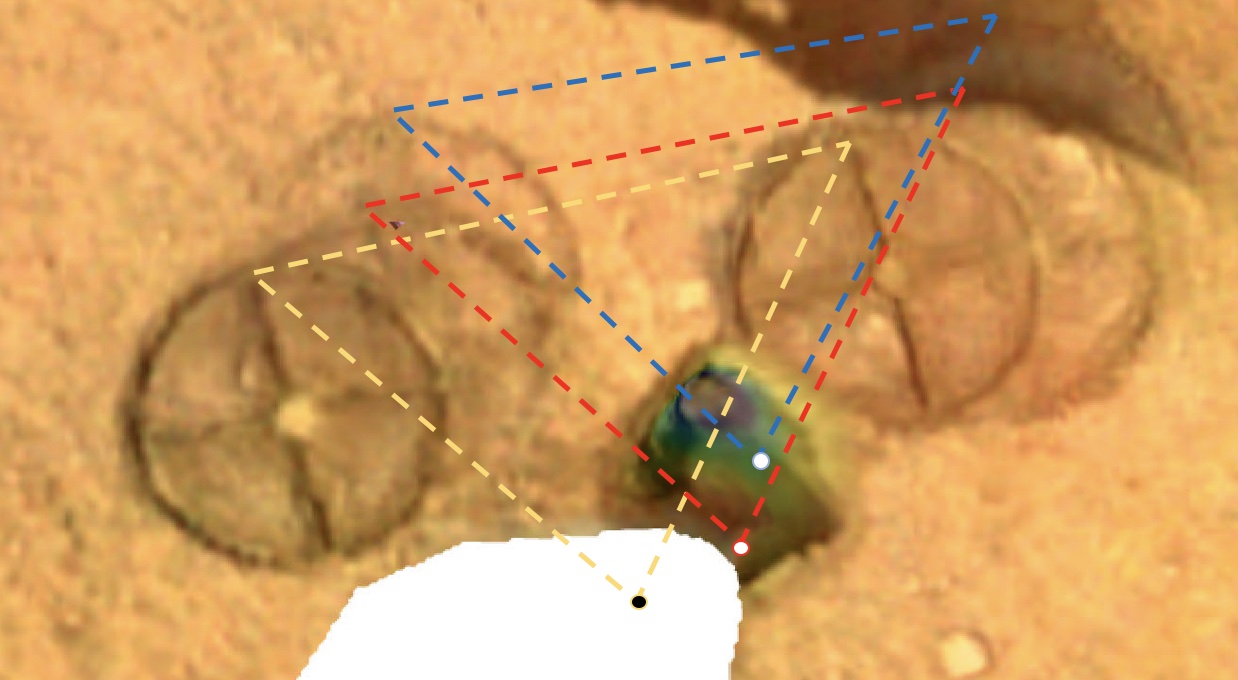}
    \caption{Top: Reading of the x-y sensors of STATIL during the first 325 hammer strokes on Sol 92 in degrees. The recordings are ambiguous with respect to rotation of the mole and x-y motion of the tip. The recordings are consistent, however, with a south and west movement of the tip and followed by a northward rotation of the mole as suggested by the position and the attitude of the mole and the twisted orientation of the tether in images taken at Sol 230. 
    Bottom: Reconstructed path of the back-cap from the footprints of the feet using the known dimensions of the support structure. While STATIL data indicate a movement of the tip southward and then westward, the back-cap moved northeastward and then eastward}
    \label{mole tip}
\end{figure}

We begin by discussing  the formation of the pit during the  first two hammerings on Sols 92 and 94 (compare Tab. \ref{tab:Phases} and Fig. \ref{fig:MoleDnTAll.jpeg}) and its depth. 
The pit has been described in section \ref{RegolithInteracttion1} and is shown in  Figs. \ref{The Pit} through \ref{Mole Pit 1.jpeg}. It is about $2.4-2.8$ mole diameters (65 - 76 mm) wide and $20 \pm 1$ mm on average deep with a maximum depth of  72 mm. Its dimensions and and shape and the position of the mole in the pit pointing towards a southwesterly direction at a tilt of $18.5^\circ$ suggest that it has been carved by the mole through a precession-like movement about a point roughly midway on its vertical axis. We have attempted to follow the mole movement by tracking  the path of its tip and of its back-end in Fig. \ref{mole tip}.  Unfortunately, the STATIL data are ambiguous with respect to mole tip movement and mole rotation. The top panel shows the angular distance in degrees that the mole tip would have moved if the STATIL readings could all be interpreted as tip movement. Accordingly, the tip moved first in a southerly direction before turning west and possibly back. The light blue and the orange dots after stroke 256, however, likely indicate a rotation of the mole about its axis which is consistent with the position and tilt of the mole in the pit and the observed twist of the science tether. In the bottom panel we are mapping the motion of the back cap by using characteristic markings on the imprint in the sand of the support structure feet. The lower apex of the triangles mark the midpoint of the back-cap as is known from the dimensions of the support structure. The blue triangle indicates  the first position after deployment followed by the red and the yellow triangle. Accordingly, the back-cap moved northeast first and then turned east. Taken together these data support a half circle precession movement of the mole.

\begin{figure}
\begin{center}
  \includegraphics[width=1\textwidth]{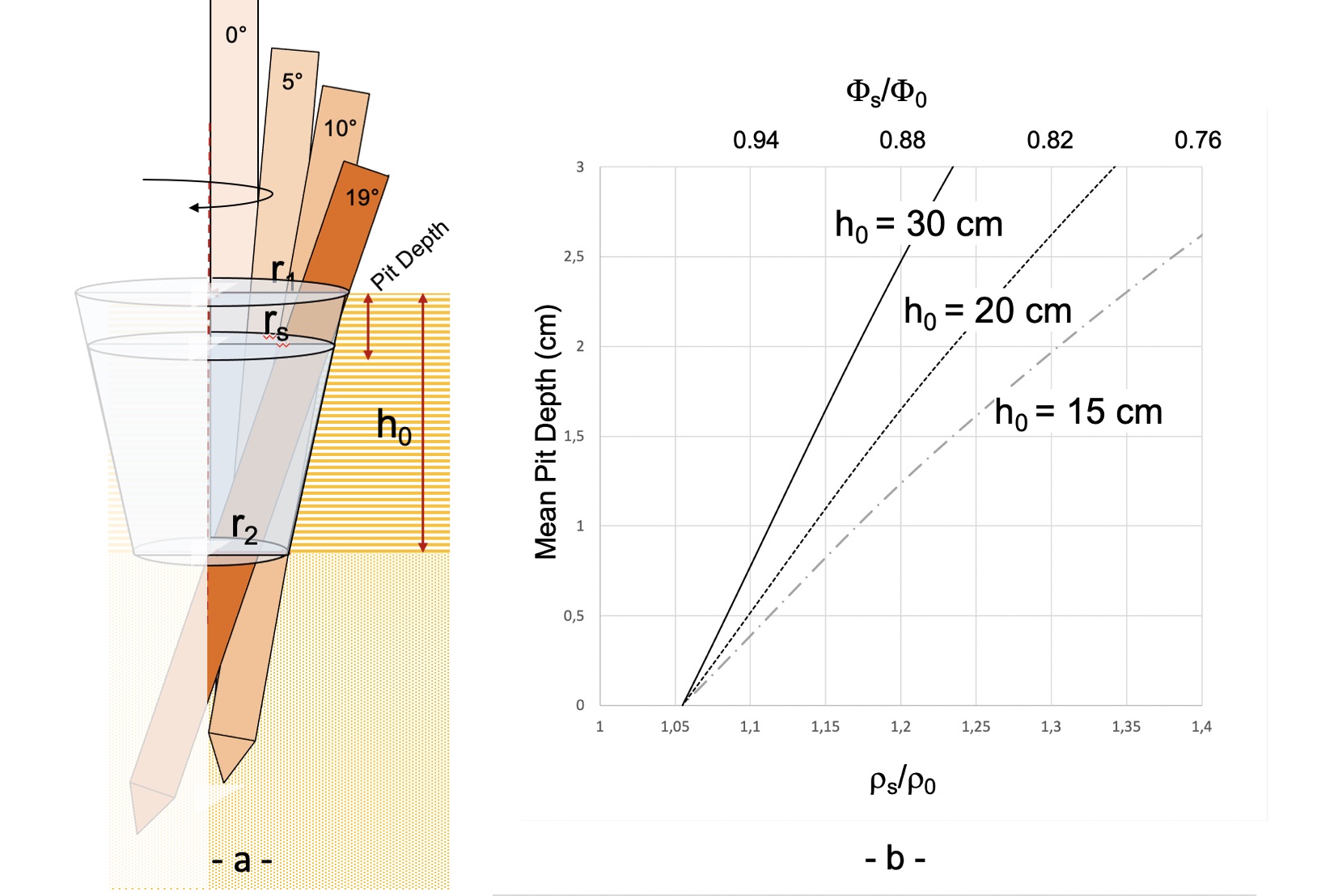}
  \caption{Model to estimate the depth of the pit carved by the mole during the first two hammering sessions. A simple geometrical model is considered, with the pit represented by half a frustrum of a cone of height $h_0$ and with the mole increasing tilt to $18.5^\circ$.  Panel a shows a sketch of the model.  Panel b gives the mean pit depth as a function of $h_0$ and the ratio of the bulk densities between the duricrust  and the sand. The top horizontal axis gives the porosity assuming a particle density   of 3200 $kg/m^3$.}
  \label{fig:pit fill}   
\end{center}    
\end{figure}

Some explanations of the formation of the pit suggest that the original fill was drained to hollows existing at depth before the pit was formed. Images of the pit wall suggest that there may be hollows in the duricrust but there is no way to prove or disprove their existence. An alternative proposal is based on the high porosity of the crust concluded from the interpretation of the TEM-A data that suggest a bulk density of $1211^{+149}_{-113}$ kg/m$^3$ and a bulk porosity of $63^{+9}_{-4}$ \% \citep{Grott:2021}. Assuming that the mole upon penetrating destroyed the fabric of the duricrust and reduced the porosity of the material, it is proposed that the pit formed as a consequence of the mole grinding duricrust to sand. In Fig. \ref{fig:pit fill} we test this hypothesis by calculating the depth of the pit from the volume of the pit fill with sand from duricrust. The ratio between  the bulk density of the sand $\rho_s$ and the crust $\rho_0$ is a model parameter as well as the thickness of the duricrust $h_0$.

We consider a simple geometrical model of the pit and the mole movement. We approximate the volume of the crust worked by the mole by half a frustum of a cone of height $h_0$ and top radius $r_1 \approx 6 r_m$, where $r_m$ is the radius of the mole. The bottom radius $r_2 = r_1 - h_0 \, tan \, \alpha$, with $\alpha$ chosen to be 10°. r$_s$ is the radius of the level of the pit fill and is calculated from the model by varying duricrust thickness and $\rho_s/\rho_0$. The pit depth is then calculated from r$_s$. Panel b gives the pit depth as a function of $h_0$ and the ratio of the bulk densities between the duricrust  and the sand. Panel b of Fig. \ref{fig:pit fill} shows the mean depth to the bottom of the pit calculated for three assumed thicknesses of the duricrust and as a function of the ratio between the densities $\rho_s/\rho_0$. The figure also shows the ratio between the porosities assuming a particle density   of 3200 kg/m$^3$, which we assume not to change during the process of pit formation. 

It can be seen that the observed average depth of the pit of about 2 cm can be explained with a thickness of the duricrust of more than about 15~cm. We will argue in section \ref{sec:Duricrust Thickness} for a thickness of the duricrust of about 19~cm. The ratio of the  densities does not need to be more than 1.15 to 1.3. \cite{Morgan:2018} give a representative value for martian sand of 1300 - 1350 kg/m$^3$, suggesting a duricrust bulk density of around 1100 kg/m$^3$.  




\subsection{Rate of Mole Penetration and Soil Penetration Resistance}
\label{PenetrationRateResistance}
\subsubsection{Observed Rate of Penetration}
\label{PenetrationRate}
The rate of penetration of the mole has varied substantially over the limited depth reached (compare Fig. \ref{fig:MoleDnTZoom.jpeg}). Unfortunately, the accuracy of the estimates of the tip depth is limited in that range since the TLM was designed to engage only after the mole tip reached a depth of 54~cm as discussed in sections \ref{sec:SS} and \ref{sec:Initial Attempts}.  There are only a few constraints on the initial penetration rate during sol~92 and 94, and unfortunately a large number of unconstrained behaviors complicate interpretation of the rate of mole progress.  

For the initial penetration rate on Sol 92, the available constraints are (1) the time the mole back cap passed the contact switch, (2) the STATIL tilt data, and (3) the internal geometry of the SS.  The back cap passed the contact switch after 77~hammer strokes ($\sim$293~seconds). The switch is located in the tube such that when the back cap of the 40~cm mole passes it, 17.8~cm of the mole is below the bottom of the tube, and 14.8~cm of the mole is below the SS feet. Had the mole been oriented vertically, we would conclude that $\sim15 cm$ of the mole had penetrated into the regolith. The STATIL data, however, tell a very different story.  

During the first 400~strokes the mole tilt changed rapidly. Initial tilt reported by STATIL was $\sim$4$^{\circ}$, about two degrees less vertical than the local ground slope of $\sim$2$^{\circ}$.  Within the first minute ($\sim$16~strokes), this reduced to 2.5$^{\circ}$, then began to rise again, plateauing at $\sim$14$^{\circ}$ after about 11~minutes ($\sim$stroke~170).  A slight break in the time-varying tilt curve around 4.8~minutes (stroke~77) is consistent with the moment of contact switch passage. Further STATIL data from this interval and sol~94 are well recorded, but unfortunately provide no further constraints on mole motion without significant assumptions.

Limiting the discussion to just the moment of contact switch passage, STATIL reported a mole tilt of 11$^{\circ}$, implying 9 to 13$^{\circ}$ of mole tilt with respect to the SS vertical tube. However, the vertical tube of the SS cannot accommodate such a large relative mole tilt when the back cap is level with the contact switch; rather the maximum allowable mole-vs-SS tilt is only 5$^{\circ}$, assuming a full compression of the friction springs. To resolve this geometric conundrum, we invoke the following: (1) The SS is very light weighing only 7.4~N on Mars, which is about equal to the  rebound force resisted by the SS friction springs. (2) The SS moved at a time or times unknown during the sol~92 (and possibly sol~94) hammerings, as revealed by the footprint markings in the upper soil layer in e.g., Fig. \ref{mole tip}. (3)  The front footprint impressions show little to no scuffing, especially after sol~94 where the internal cross-beams of the feet have left clear markings in the soil.  This suggests the SS was lifted away from the regolith and then replaced in a new position, rather than just being dragged or pushed horizontally. (4) The SS was seen to jump or jostle about during the Diagnostic Hammerings on sol~118 and 158. And finally (5) as revealed when the SS was lifted away, the mole azimuth had its tip pointing towards the southwest, meaning the rebound vector pointed northeast.  The SS motions of y-axis translation and aft-right-foot rotation, seen after the first and second hammering intervals, respectively, are roughly consistent with the mole pushing against the SS during mole rebound.  Thus we conclude that the mole rebound acted to lift the front feet of the light support structure away from the soil at some time or times during sol~92 (and possibly also sol~94).  Unfortunately, mid-hammering movies were not captured during the 4~hr hammering interval, and there is no tilt data for the SS itself. 

\begin{figure}
    \centering
    \includegraphics[width=0.7\textwidth]{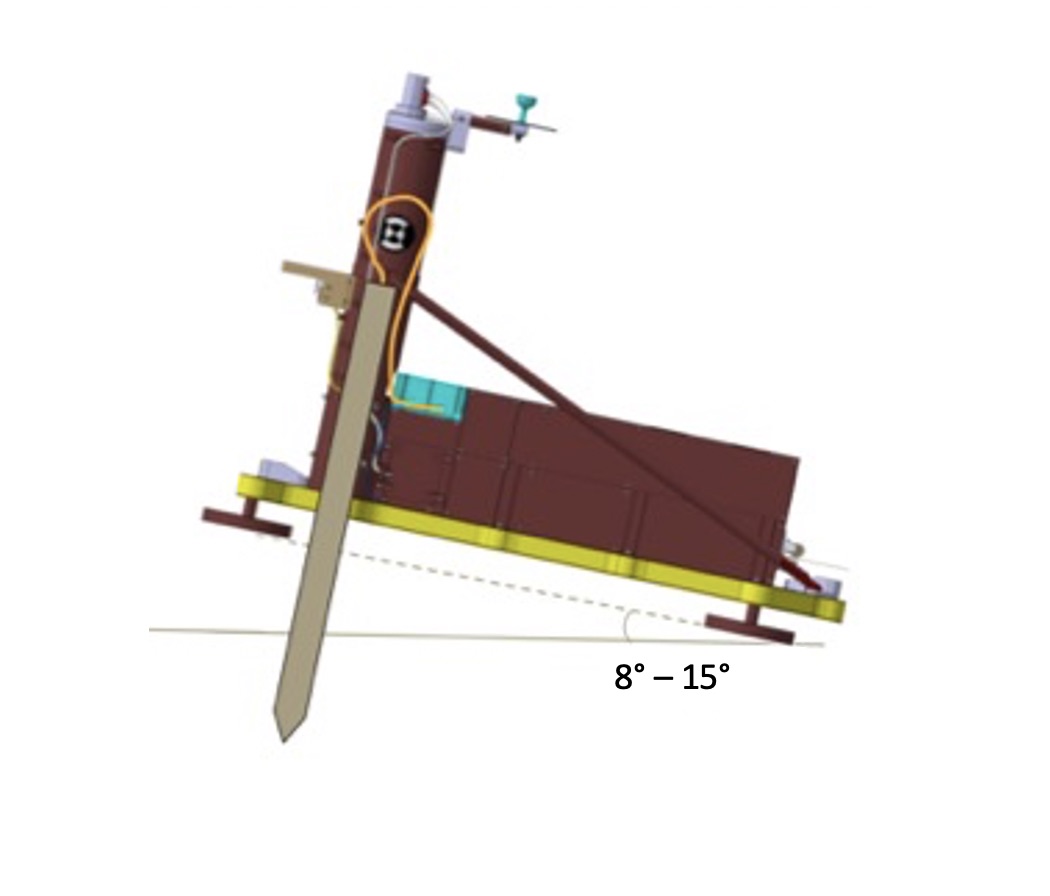}
    \caption{Illustration of the support structure lift and rotation about an axis through its back feet as suggested by the interpretation of the STATIL data recorded on sol 92 and discussed in section \ref{PenetrationRate}}.
    \label{fig:Inclined SSA}
\end{figure}

Using the STATIL reported tilt of the mole ($11^{\circ}$), the length of the mole and position of the contact switch in the SS, the maximum allowable mole-vs-SS tilt (5$^{\circ}$), and the local ground slope (2$^{\circ}$), we can bound the geometry of the system at the moment of contact switch passage.  The smallest possible lift that is consistent with all the data implies an SS tilted up by 8$^{\circ}$ relative to the ground around a pivot point at the back of the aft feet (see Fig. \ref{fig:Inclined SSA}). The mole, extending 14.8~cm below the bottom of the SS feet, would then have a tip depth of $\sim$9~cm. The largest possible SS lift implies an SS tilted up by 15$^{\circ}$ relative to the ground around the same pivot. The mole, extending the same 14.8~cm below the bottom of the SS feet, would then have a tip depth of $\sim$4~cm. These then place limits on the average penetration rate for the first 77 strokes of  0.5-1.2~mm/stroke.  
By the end of each multi-hour session, the SS had settled back to the surface.  With the back cap necessarily below the contact switch and all four SS feet on the ground (confirmed by post-hammering images on sol~92), the mole tip was therefore deeper than $\sim$15~cm at the end of sol~92.

There is a richness of data from the individual STATIL sensors reported during both of the long hammering intervals (Fig. \ref{fig:MoleDnTAll.jpeg}).  After contact switch passage on sol~92, the tilt briefly plateaued at $\sim$14$^{\circ}$ then increased by $\sim$1$^{\circ}$ over a two-minute interval beginning at 16~minutes and a further $\sim$1$^{\circ}$ over a three-minute interval beginning at 21~minutes.  STATIL then reported long intervals of constant tilt with irregular changes of $\pm$1$^{\circ}$ during the remaining 3~hours and 39~minutes of the sol~92 hammering interval, resulting in a final tilt of $\sim$18$^{\circ}$.  Unfortunately, this data is fundamentally ambiguous with respect to mole rotation or changes in mole tilt azimuth.  Without direct observations of the mole during this time, no further constraints can be placed on mole forward motion. But in any case, if the mole continued to hammer at the rates estimated above, it would have reached the tip depth of $31cm$ after 175 to 520 additional hammer strokes or after 11-32 minutes.



Better time- and depth-resolved estimates of the penetration rates are available after Sol~308 where IDC images could be used to compute the height of the back-cap above ground. The error of the depth determination from imaging data is estimated to be $\pm0.5$~cm for cases where the glint feature on the back cap was visible, and $\pm1.0$~cm for cases where the scoop fully or partially blocked view of the back cap.  Considering only the downward motion of the mole reported by stroke in Fig.~\ref{fig:MoleDnTZoom.jpeg}, the average rate of penetration from sol~311 to sol~322 was 0.11~mm/stroke.  Faster rates were observed during recovery from Reversal~1, with a maximum `re-penetration' rate of 0.6~mm/stroke during sols~349, 366, and the beginning of 373.  The later portion of sol ~373 and sol~380 averaged $\sim$0.3~mm/stroke.  Likewise, recovery from Reversal~2 was initially rapid ($\sim$0.6~mm/stroke on sol~458) but quickly decreased to an average rate of $\sim$0.13~mm/stroke over sols~472-536.  All subsequent penetration was $<$0.1~mm/stroke with very low penetration rates of $<$0.05 mm/stroke from Sol 543 on. On Sols 557 and 754, the mole did not penetrate at all.  


\subsubsection{Comparison with Test Data and Models - Soil Penetration Resistance}
\label{PenetrationResistance}
Laboratory tests of the mole \citep{Wippermann2020} in cohesionless quartz-sand (WF 34) and mildly cohesive, high friction angle sands (Syar, MSS-D) at DLR Bremen and at JPL in Pasadena have measured significantly higher penetration rates than on Mars at similarly shallow depths (compare Tables \ref{tab:penetration rates-2} and \ref{tab:penetration rates}). The term "Syar" was used by the team as a shorthand for a commercially available crushed basalt sand\footnote{ "Syar Industries" is the trademarked name of a building material supply company in northern California.  Crushed basalt from the Lake Herman Volcanics Group was procured by the InSight project from Syar Industries as a bulk regolith simulant.  The term 'Syar' was used by the team to distinguish this simulant from other basaltic materials used in testing.}. Syar sand has sharped-edged grains with sizes ranging up to 100 $\mu$. A mix of 80 weight-\% sand and 20 weight-\% Syar dust was used at DLR. \cite{Wippermann2020} list a friction angle for the Syar mix used of 54.8° \citep[see also][]{Spohn2021}. MSS-D is Mars Simulant Sand-D. Mechanical properties of MSS-D 
have been described in \citet{Delage2017} and are listed in \cite{Spohn2021} who give a friction angles for MSS-D of 37°.  

We list in table \ref{tab:penetration rates-2} the penetration rates  that were recorded for the smallest tip depths during the tests. Because the gravity on Earth is roughly three times higher than on Mars, the soil overburden pressure was proportionally higher in the terrestrial laboratory.  Some data are available for the decrease of the penetration rate through the first 30 $cm$ (Tab. \ref{tab:penetration rates}) in Syar and MSS-D sands. With the exception of a test in pure MSS-D sand (second line from the top in Tab. \ref{tab:penetration rates}), the ratios are smaller than the minimum estimate of the decrease of the penetration rate on Mars. The latter test  stands out because of  a high initial  penetration rate, $R_0$.

\begin{table}[ht]
    \centering
    \begin{tabular}{l|c|c|c|c|c|c}
    \hline
        Simulant & Mole & Stroke & Test ID & Tip & Penetration & Penetration \\
         & Model &Energy&  & Depth&Rate &Resistance\\
         & &(J) &  &(mm)&(mm/stroke) &(kPa)\\
        \hline
        Syar                 $ $  & PPFM  & 0.85 & DPT 4 & 400 & 1.34 &  \\
        Quartz sand                 & PPFM  & 0.85 & DPT 1 & 300 & 2.03 &  \\
        WF 34                   & PPFM  & 0.70 & DPT 6 & 300 & 2.33 &  \\
                                & PFE-1 & 0.85 & DPT 1 & 400 & 1.73 &  \\
                                & PFE-1 & 0.85 & DPT 2 & 400 & 3.67 & 200 \\
                                & PFE-1 & 0.85 & DPT 3 & 400 & 2.65 & 250 \\
                                & PFE-1 & 0.85 & DPT 4 & 400 & 2.92 &  \\
                                & PFE-2 & 0.70 & DPT 1 & 400 & 2.37 &  \\
                                & PFE-2 & 0.70 & DPT 2 & 630 & 2.91 &  \\
                                & PFE-3 & 0.70 & DPT 1 & 720 & 2.10 &  \\
                                & QM    & 0.70 & DPT 1 & 650 & 2.56 &  \\
         \hline

    \end{tabular}
    \caption{Comparison of penetration rates at depths between 300 and 720 mm in the Deep Penetration Tests with Syar sand and with quartz-sand WF 34 described in more detail in \citet{Wippermann2020}. Most readings were taken at 400 mm tip depth.  The mole models listed are the Preliminary Proto Flight Model PPFM, three Proto Flight Equivalent Models PFE-1 through -3, and the Qualification Model QM. Some models differ in their stroke energy but the  rates have been adjusted for a stroke energy of 0.7 J. The measured penetration rates vary by a factor of about 2 but even the penetration rate in Syar sand is about an  order of magnitude higher than estimated for  a similar depth of 310 mm on Mars. For DPT-2 and DPT-3 with PFE-1 penetration resistances of the sand were measured before and after the tests using a commercial hydraulic cone penetrometer HYSON 100 kN – LW of the manufacturer A.P. van den Berg. At the depth of interest here, the resistance values did not change much between pre- and post-test recordings}
    \label{tab:penetration rates-2}
\end{table}

\begin{table}[ht]
    \centering
    \begin{tabular}{l|c|c|c|c}
            \hline
        Simulant & $N_{St}$ & $R_{0}$ (mm/stroke) & $R_{30}$ (mm/stroke) & $R_{0}/R_{30}$\\
        \hline
        Syar $ $ (with stones) & 31 & 4.839 & 1.183 & 4.1\\
        MSS-D & 14 & 10.71 & 1.408 & 7.6 \\
        MSS-D (with stones) & 23 & 6.521 & 3.877 & 1.6 \\
        MSS-D (compacted) & 64 & 2.312 & 0.469 & 4.9 \\
        MSS-D (compacted with stones) & 57  & 2.632 & 0.618 & 4.2\\
        Mars & - & 0.5 - 1.2 & 0.11 & 5-12\\ 
       \hline  
        
    \end{tabular}
    \caption{Comparison of the change of penetration rate through the first 30 cm observed in laboratory experiments and on Mars. $N_{St}$ is the number of hammer strokes to reach 30 cm tip depth. $R_0$ is the penetration rate at the surface and $R_{30}$ the rate at 30 cm tip depth. The fourth column gives the ratio between the two rates. The penetration rate on Mars decreased at least as much as in compacted sands or more. The number of hammer strokes needed to reach 30 cm depth on Mars is not known, unfortunately}
    \label{tab:penetration rates}
\end{table}
Many models have been published aiming at predicting the rate of penetration of penetrometers in sand. These include analytical theories such as \cite{Rahim2004} based on the cavity expansion theory of \cite{Salgado1997} as well as numerical models based on e.g., Dynamic Cone Penetration Theory \citep{Poganski2017} and Discrete Element Modeling (e.g., Lichtenheld and Kr\"omer, 2016; Zhang et al., 2019; ).  In general, they find the penetration resistance $\sigma_{P}$ for a penetrator of a given stroke energy $E$ to be inversely proportional to the penetration rate  

\begin{equation}
    R \cdot \sigma_{P} = \frac{\epsilon E}{A}
    \label{equ:resistance}
\end{equation}
where  $R$ is the penetration rate, $\epsilon$ is the efficiency of the mole converting its stroke energy $E$ into deformation energy, and $A$ is the cross-section area of the penetrator. The stroke energy of the mole flight model is known to be $0.7 J$ and its cross-section area can be calculated from its radius of 1.35 cm. Accordingly,
\begin{equation}
    R \cdot \sigma_{P} \approx \epsilon \times 1.22 \: \mathrm{kPa \cdot m/stroke}.
    \label{equ:resistance_1}
\end{equation}
Some pre-flight mole models had a higher stroke energy of 0.85 J. The penetration rates given in Table \ref{tab:penetration rates-2} have been corrected for the difference and referenced to a stroke energy of 0.7J. 
 
 The efficiency of energy conversion $\epsilon$ is more difficult to estimate and the result is more uncertain. \cite{Rahim2004} use 0.75 while \cite{Zhang2019} find 0.5 as a typical value. The penetration resistance was measured before and after the tests with a commercial penetrometer and compared with the penetration rate during two of the laboratory tests in quartz-sand described in \cite{Wippermann2020} with the Proto Flight Equivalent Model PFE 1 (compare Table \ref{tab:penetration rates-2}).   These data allow an estimate of $\epsilon$ at least for these tests as \citet{Zhang2019} find the static resistance (the resistance to a slowly penetrating penetrometer) to be very close to the dynamic resistance for penetration resistances smaller than 10 MPa.  By comparing the penetration rates with the penetration resistances up to 3~m tip depth, we find $\epsilon = 0.47 \pm 0.05$.  It should be noted, however, that in both tests, the penetration rates at depths $> 3$ m decreased substantially although the measured penetration resistance did not increase accordingly. The reason for the decrease in penetration rates for quartz-sand and the difference to the test with Syar sand has been explained by \cite{Wippermann2020} as being due to the effects of friction on the tether.

Adopting the mole efficiency from the laboratory experiments in quartz-sand, the estimated initial penetration rate on Mars of 0.5 - 1.2 mm/stroke implies a penetration resistance of  $0.48 - 1.2$ ~MPa. 
Penetration beyond the tip depth of 31cm, where the penetration rate was observed to be $\leq$ 0.11 mm/stroke and where the rebound was balanced with the help of the robotic arm through back-cap push (section \ref{BCP-HS}), suggests a penetration resistance of $5.3$~MPa. This value would have increased even further as the mole got deeper and the penetration rate kept  decreasing.

\citet{Zhang2019} citing evidence from Discrete Element Modeling as well as from field measurements report that the penetration resistance increases with the square of the relative density and linearly with overburden pressure.  \cite{Rahim2004} find the penetration resistance to depend on initial porosity and internal friction angle of the granular material while cohesion was found to be of smaller importance. For small internal friction angles of 20° or less, the dependence on initial porosity is small with resistance increasing by a few percent when porosity is decreased from e.g., 0.6 to 0.3. Resistance will increase by a factor of 5.5, though, in that same porosity range for a friction angle of 40°. Data collected by \cite{Golombek2008} and \cite{Herkenhoff2008} show that martian soils have friction angles between 30° and 40° with dust having friction angles as low as 20°. 

Observing that the mole tip was at a depth of $4$ to $9$ cm after the initial relatively rapid penetration phase and at about $31$ cm depth when the penetration resumed helped by the robotic arm, we have an increase of overburden pressure at the tip depth by a factor of 3 to 7.5, assuming that the density in the column is constant and that the pressure in the duricrust indeed increased linearly with overburden thickness. That alone could explain part of the decrease in penetration rate from $4-9$ cm to $31$ cm tip depth if the duricrust extends that deep. If the duricrust is thinner and underlain by sand of low or no cohesion with a postulated low-pressure penetration resistance closer to some hundred kPa, a significant increase by a factor of up to 3  or more in relative density would  be required. It is possible, if not likely, that the soil at that depth was compacted during the first 8600 strokes hammered during Sols 92 and 94. Vibration generated by the hammer strokes could also have been a factor in compacting the soil. It should be noted, however, that the seismic energy in the hammer signals recorded by SEIS during hammering is less than a percent of the stroke energy when geometrically projected to the mole tip as the source area (compare section \ref{sec:seismic energy} below). Penetration models usually find a compacted region in front of the penetrator with a thickness of a few times the radius of the penetrator. The mole penetrated roughly 7.5 radii aided by the robotic arm beneath the depth of interest without an increase in the rate having been observed. On the contrary, the rate rather decreased further, from 0.6 mm/stroke for 20 strokes on Sol 308 to 0.06 - 0.15~mm/stroke between Sols 311-322 and less than 0.05 mm/stroke from Sol 543 on.  

A simple explanation for the low penetration rate from 31 cm tip depth on is that the mole had entered into a  more resistant layer than sand that got more resistant with depth, e.g.\ a layer of gravel or a layer of small stones embedded in sand. An early test with a breadboard model (the MM-mole model) has been reported in \cite{Wippermann2020}. That mole penetrated a mono-layer of Columbia river basalt stones of 5-15 cm size. The rate estimated from the data  was about 0.03 mm/stroke but that mole had a smaller spring energy than the flight model. The TEM-A thermal conductivity measurements \citep{Grott:2021},  do not show any evidence for layering or an increase of conductivity with depth as might be expected if an  highly compacted sand region extended from $30$ to $37$ cm depth. A layer of gravel or small rocks is thought to be consistent with the data (compare section \ref{sec:PropertiesThermal}) and could have a density of 1600 kg/m$^3$.  

\begin{figure}
\begin{center}
  \includegraphics[width=1\textwidth]{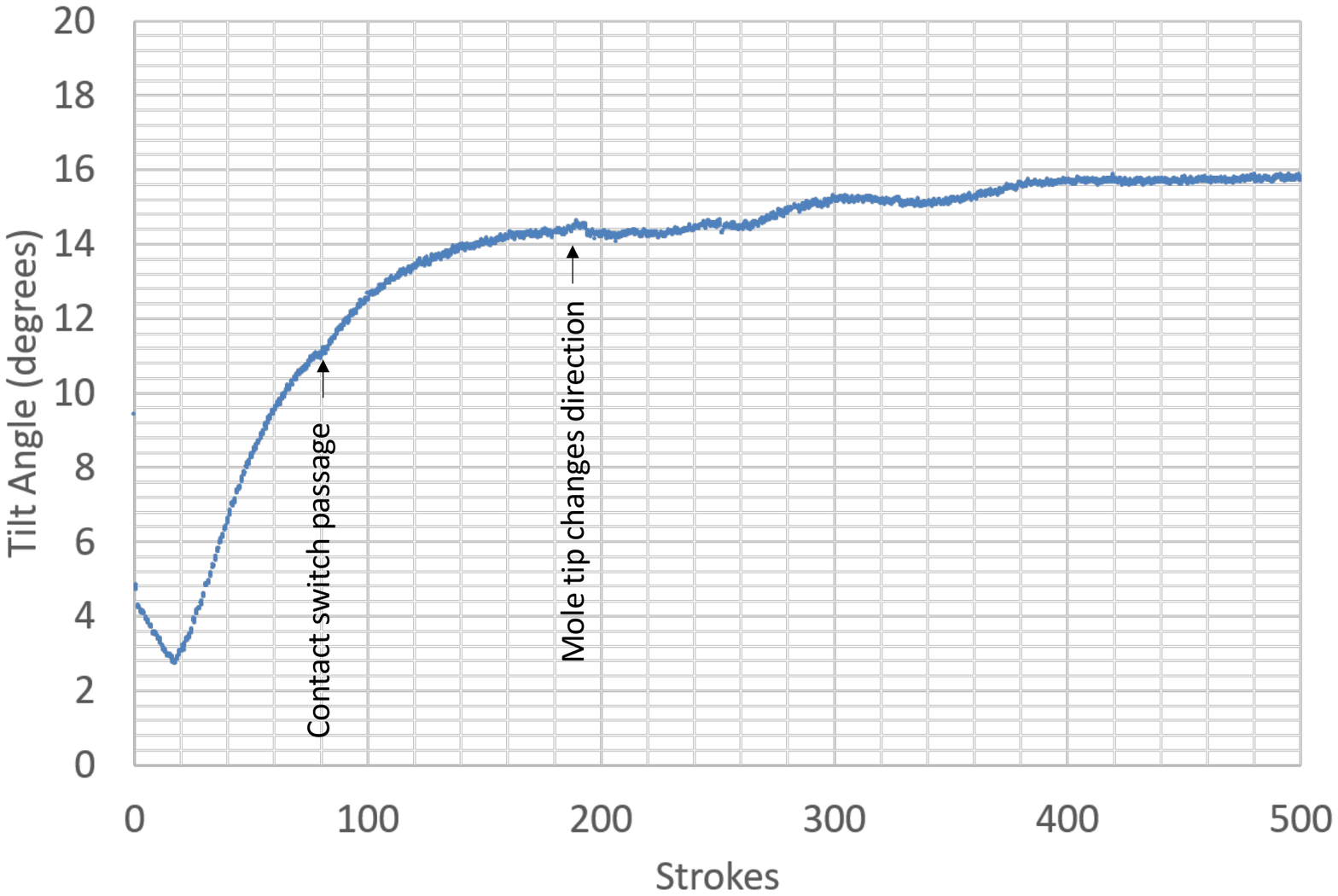}
  \caption{Statil recording of the first 500 strokes on sol 92. Marked are the mole  passage of the contact switch as sensed by STATIL and the stroke when the mole tip suddenly changed direction from mostly southward to mostly northward (compare Fig. \ref{mole tip}). Note that because of the ambiguity in the STATIL data the inferred change in direction of tip movement may also be a sudden change in sense of rotation. Also note the wavy character of the increase in tilt angle after stroke 200}
  \label{fig:First 500 strokes}   
\end{center}    
\end{figure}

Accepting the hypothesis of a mechanically layered structure with a duricrust overlying sand followed by a more resistant layer, the question poses itself of whether anything can be said about the intermediate sand layer. Unfortunately, there are no unambiguous data available to constrain its properties. However, the STATIL data (compare Fig. \ref{fig:First 500 strokes})  can be used to see whether there are any features that could be interpreted as indicating a transition in mole behaviour. Such a feature might be caused by a transition in soil properties and might be  in addition to the kink in the data that clearly marks the mole's passage of the contact switch at 77 strokes. Assume first that the mole penetrated to 9 cm during the first 77 strokes at the implied rate of 1.2 mm/stroke. At that rate it would have reached 31 cm tip depth after 183 additional strokes, 260 strokes altogether. There are some wiggles in the data around that stroke number but nothing very characteristic. Assume next that the mole penetrated to 4 cm initially. Then - using a similar argument with 0.5 mm/stroke - we get to 31 cm tip depth after a total of 77+540 about 620 strokes. No characteristic features to be seen there.  If instead the characteristic feature at 190 strokes is assumed to mark the encounter with the resistant layer and the following features as indication of trying to penetrate this layer, a rate of 2 mm/stroke is calculated for the penetration from a tip depth of 9 cm to 31 cm with 113 strokes. This is about the average rate of those in quartz-sand in the Deep Penetration Tests  (compare table \ref{tab:penetration rates-2}). With 4 cm tip depth after 77 strokes, we get a similar rate of 2.4 mm/stroke. We further note that the 190 strokes mark is the point where the mole tip in Fig. \ref{mole tip} abruptly changes direction from southerly to northerly directions. 

It is reasonable to assume that the mole penetrated faster after it got through the duricrust into sand and before encountering a more resistant layer at 31 cm. After all, the support structure re-settled to the ground during that time. The mole reaching 31 cm depth could have started to precess and widen the hole, thereby increasing its inclination to 16°.

Overall,  the penetration resistance on Mars was found to be significantly larger than what was observed in the test bed on Earth. Possible explanations include a finite compressive strength of the duricrust provided by cohesion which the sands in the terrestrial test bed did not have. After all, during the first 77 hammer strokes the mole lifted the support structure weighing 7.4 N by several centimeters. The still lower penetration rate at depths below $31$ cm is difficult to explain with sand even if compacted as the thermal data recorded  by the TEM-A sensors suggest a high porosity even for that layer.

  \begin{figure}
    \centering
    \includegraphics[width=1\textwidth]{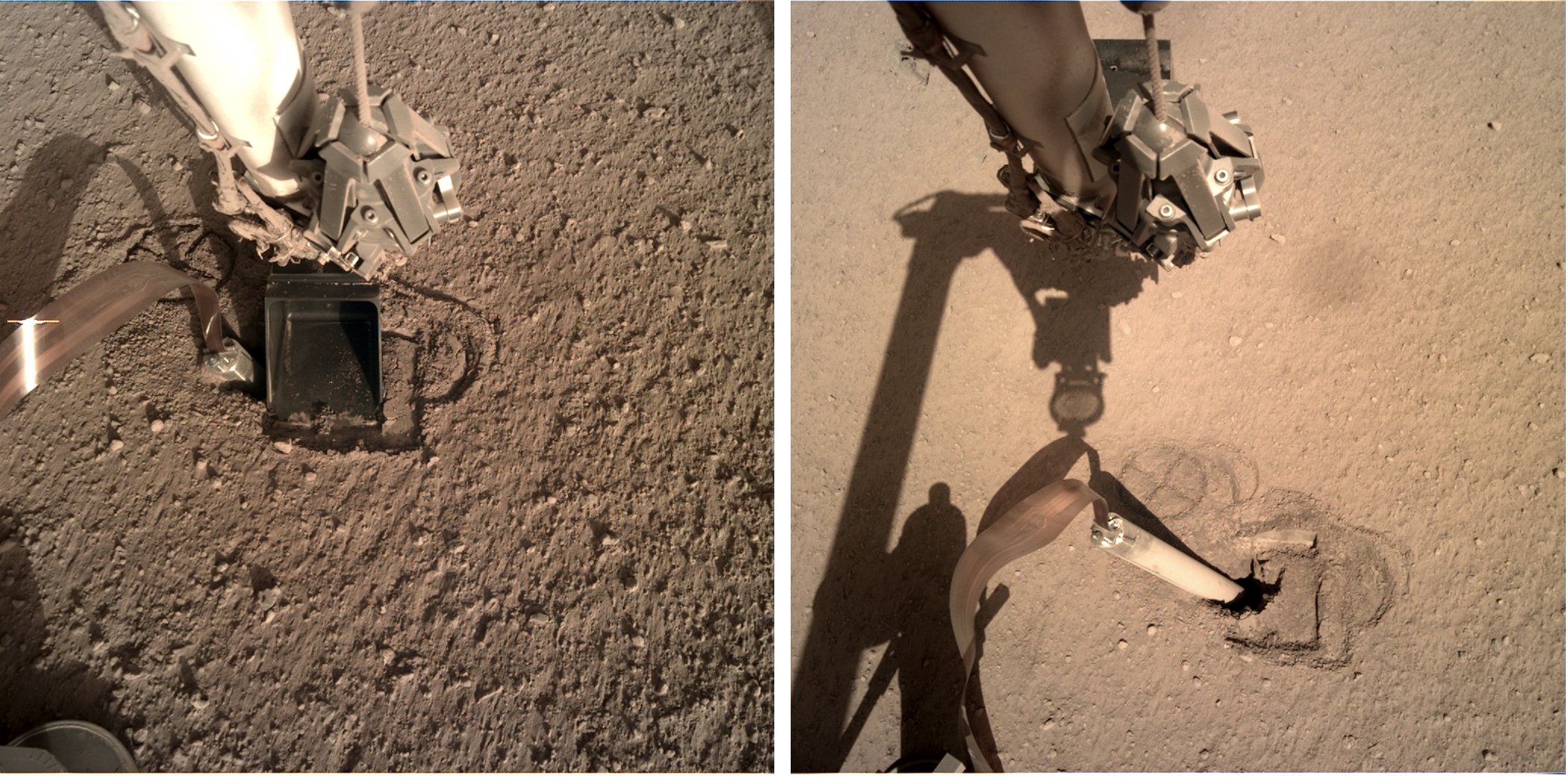}
    \caption{Configuration of the scoop during regolith push. The left panel shows the scoop pressed onto the surface next to the mole for the intended regolith push on Sol 322. The right panel shows the mole pit, the backed-out mole and the scoop indentation after lifting the scoop on Sol 333.}
    \label{fig:regolith push}
\end{figure}

\subsection{Estimate the thickness  of the duricrust from  mole backing out on Sol 325 after regolith push}
\label{sec:Duricrust Thickness}

As we have reported in section \ref{subsec:Reversal1}, the mole backed out of the ground by a total of 17.4 cm on sol 325. Not being able to pin the mole any further,  the scoop had been pressed  onto the surface next to the mole to provide vertical stress that would increase friction on the mole hull. The vertical force of the scoop was estimated to be 50 N, equivalent to a vertical stress immediately underneath the 7.6 x 10.6 cm scoop of 6.6 kPa. Stress propagation in an elastic half space suggests that this stress is concentrated underneath the load but decreases to about one tenth at a depth of two scoop widths. The scoop was thus placed above the mole (Fig. \ref{fig:regolith push}) although part of the the scoop would then be above the  pit through which vertical stress could not be transferred. Not knowing the thickness and the stiffness of the duricrust, it was not clear how well stress could be transferred to the mole hull. Simple calculations for an elastic  half-space loaded by the scoop suggested that the necessary friction could be provided.    
A small number of strokes was commanded on sol 322 and after the mole continued to penetrate for another centimeter at a rate of 0.15 mm/stroke, a hammering session of 2 times 150 strokes was planned. 

The mole backed out right from the beginning of the first 150 strokes set on sol 325 at an average rate of -0.9 mm/stroke. 
When the mole was 18.5 cm along-mole-distance out of the ground (compare Fig. \ref{fig:MoleDnTZoom.jpeg}), the upward motion stopped and the mole moved down for 8 mm between the last two frames taken, which is just about the error margin of $\pm$ 5 mm of the mole distance measurement. The two frames were 83 strokes apart. Unfortunately, we do not have more frames in between the two.  All in all, the mole had moved a total of 17.4 cm, from 1.1 cm to 18.5 cm along mole distance. During the time, the mole tilt increased about 18° to 27° (Fig. \ref{fig:MoleDnTZoom.jpeg}). Because of the mole tilt, the vertical distance was 15.5 cm. 

The mole reversing its direction of motion has been observed in the laboratory at martian atmosphere pressure in high-friction-angle  sand such as Syar. A proposal to explain the backward motion assumes that the mole is embedded in sand underneath the duricrust and in sand that has accumulated by its penetration through duricrust as described in section \ref{sec:soil porosity}. When the mole bounces in place some sand may flow in front of the tip during an upward motion thus raising the floor underneath the bouncing mole. The rate of accumulation of sand would be about equal to the rate of upward motion ignoring some compaction of sand when the mole falls back onto the sand. With an average upward motion of about 1 mm/stroke, a millimeter of sand would have to fill in underneath the mole per stroke. The mole would have stopped its upward motion when its tip reached the bottom of the duricrust when the lateral flow of sand stopped. If this simple model is correct, then the distance of the upward motion provides an estimate of the thickness of the duricrust. From the length of the mole out of the ground of 18.5 cm, its total length of 40 cm and the tip angle of 27°, we get an estimate of the thickness of the duricrust of 19.2 cm.      




\subsection{Soil mechanical parameters derived from scoop – soil interactions}
\label{sec:PropertiesScoop}

\subsubsection{Cohesion Estimate from Regolith Interaction 1} \label{sec:CohesionEstimate}

The pit that formed around the HP$^3$ mole offers a unique opportunity to combine slope stability analysis with measurements of IDA forces at the scoop and images to estimate the mechanical properties of the martian soil. The stability of the pit is examined with a three-dimensional Finite Element Method (FEM) calculation using the PLAXIS 3D program. In considering the problem of slope stability, we assume that the material is homogeneous and that a Mohr-Coulomb failure criterium is satisfied along the failure plane, i.e., that the regolith shear strength $\tau$ is defined by: $\tau = c + \sigma$ tan $\phi$, where $\sigma$ is the normal stress on the potential failure plane, $\phi$ is the internal friction angle, and $c$ is the cohesion.

\begin{figure}
    \centering
    \includegraphics[width=0.8\textwidth]{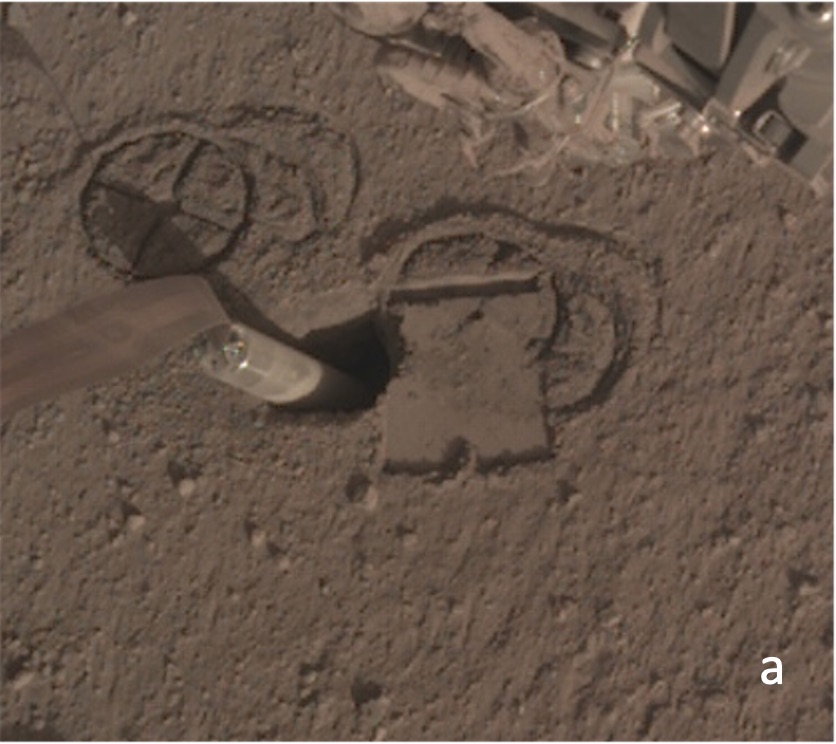}
    \includegraphics[width=0.8\textwidth]{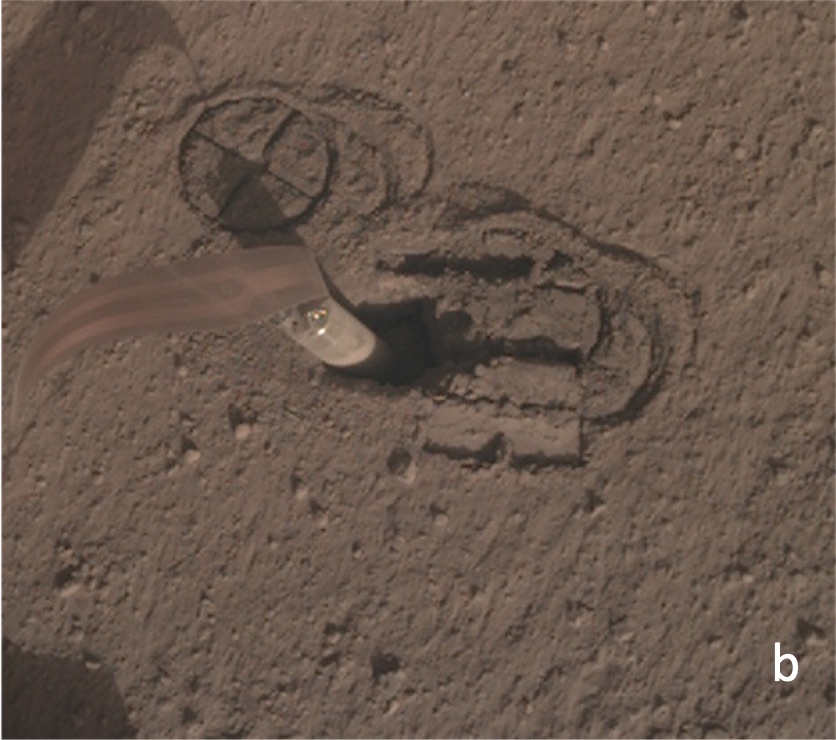}
    \caption{Images of IDA scoop interactions with the surface material near the HP3 mole pit. (a) After a flat push on Sol 240. (b) After a tip push on Sol 250.}
    \label{fig:Sol240_Sol250}
\end{figure}

As outlined in Section \ref{RegolithInteracttion1}, on Sol 240, the flat part of the IDA scoop was used to apply a preload at the edge of the HP$^3$ mole pit in an attempt to cause failure of the western wall. The IDA algorithm used to compute the force at the end-effector \citep{trebi-ollennu2018} determined that the force applied by the scoop was $F_z$ = 29 N in the vertical direction and $F_r$ = 15 N in the radial direction. Interestingly, this ratio corresponds to a friction angle of about 30°, showing that the scoop is not far from sliding along the regolith surface, with sliding probably impeded by a notch at the bottom of the scoop.
Such force did not cause slope failure (Fig. \ref{fig:Sol240_Sol250}a). It is worth noting that, without a slope failure, the slope stability analysis provides a lower bound estimate of the cohesion. The force $F_r$, which acts away from the lander, does not affect the stability and only the vertical force $F_z$ is considered in the analysis.

The topographic map of the HP$^3$ pit presented in Fig. \ref{The Pit} is used to measure the slope inclination angle $\beta$ = 85°, height h = 0.07 m, and width w = 0.045 m. Minimum estimates of the cohesion are calculated for a presumed internal friction angle $\phi$ of 30° and a bulk density $\rho$ of 1200 kg/m$^3$. The results indicate that a minimum cohesion c of 0.4 kPa is required for the slope to be marginally stable when a force $F_z$ is applied by the flat part of the IDA scoop.

Subsequently, on Sol 250, the tip of the scoop was pushed into the soil near the HP$^3$ mole pit. The force applied by the tip of the scoop was $F_z$ = 45 N. The IDA interaction with the surface resulted in the failure of a soil wedge (Fig. \ref{fig:Sol240_Sol250}b). High-fidelity Digital Elevation Models obtained using Structure-from-Motion (SfM) computation \citep{Garvin2019} provided information on the geometry of the failure wedge, i.e., a failure angle of 35° and a height of 0.013 m. If we use this finding in conjunction with the slope stability model, and take into account the lower bound estimate of the cohesion obtained from the flat push on Sol 240, we obtain a cohesion c of 5.8 kPa for a value of the internal friction angle $\phi$ of 30° and bulk density $\rho$ of 1200 kg/m$^3$. The effect of uncertainty in the model input parameters was explored using a sensitivity analysis. Results show that the cross-sectional area on which the force is applied has the most effect on the cohesion value, while variations in internal friction angle, slope height, soil bulk density and vertical force applied do not greatly influence the cohesion estimate.

These cohesion values are consistent with a steep-sided open pit, the wall slopes created by the IDA scrapes, and are similar to relatively strong, blocky, indurated soil at Viking Lander 2 \citep{Moore1987}. 

\subsubsection{Scrape angles from regolith interaction 4}\label{sec:scrapeangle}

\begin{figure}
   \centering
    \includegraphics[height=0.3\textheight, angle=90]{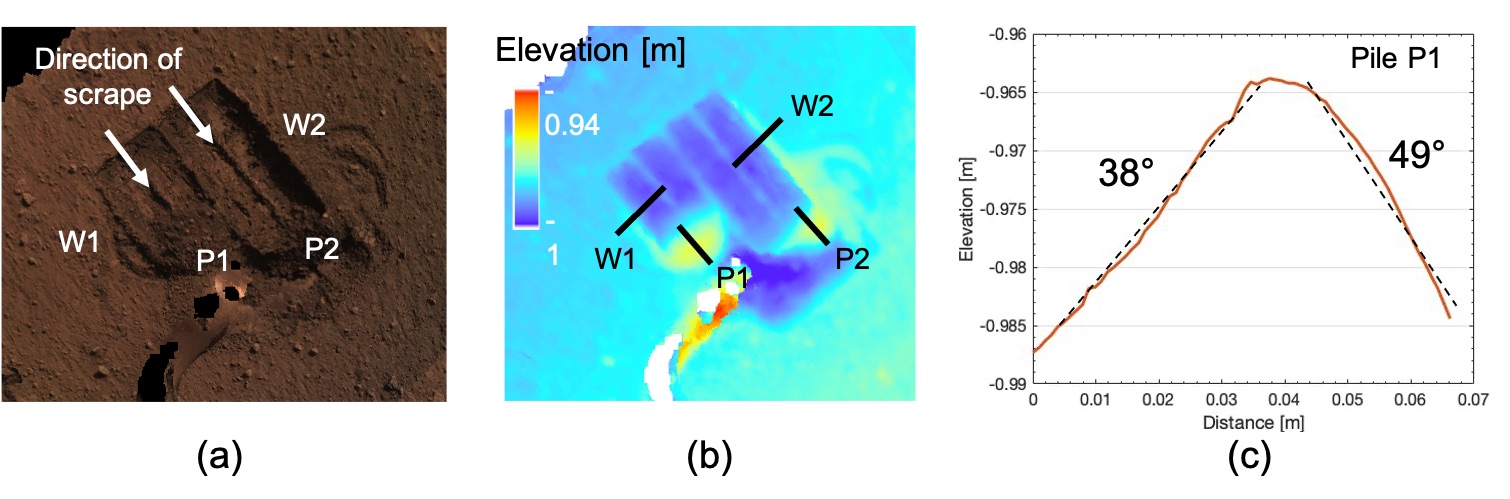} 
    \caption{Digital elevation model of the pit based on the stereo pair taken on sol 673 after IDA scraping. (a) Orthoimage showing the piles P1 and P2 and walls W1 and W2 left after the scoop scraped the regolith. (b) Digital elevation model  (c) Elevation profile for pile P1.} 
    \label{Scrape_sol673}
\end{figure}

On sol 673, two overlapping 12-cm long scrapes were commanded to bring regolith from the far side of the pit into the pit. The scrapes created two piles close to the mole, referred to P1 and P2 in Fig. \ref{Scrape_sol673}a, and walls parallel to the direction of the scoop's scrapes, denoted by W1 and W2 in Fig. \ref{Scrape_sol673}a. As a result of the IDA scraping actions, the piles P1 and P2 are created by bulldozing mounts of grains over the relatively flat ground surface. It is worth noting that mounds obtained by scraping typically yield different geometries and properties than piles formed by pouring the material from a given height, from which the angle of repose is typically measured \citep{BEAKAWIALHASHEMI2018,Chik05,ASTMC1444}. The scoop scraping action disturbs the regolith and likely breaks the cohesive bonds between the grains.

Elevation profiles are extracted from the sol 673 digital elevation model (Fig. \ref{Scrape_sol673}b) to measure the slopes of the piles and walls. We find that the slopes of the bulldozed mounds of regolith can be as high as 38-39° on the upstream side, where the grains have been pushed by the scoop. Once the scoop loses contact with the regolith, some of the grains likely come down along the created slope. On the downstream side, where the grains have been pushed away by the scoop with no back sliding once the contact with the scoop was lost, the slopes of the piles are between 49 to 53°. The walls scraped by the vertical sides of the scoop have slope values of 78° and 70°, for the wall W1 and W2, respectively.

The slopes of the piles are larger than that of a pile poured from a given height defined, in non-cohesive granular materials, by an angle of repose that is close to the friction angle at the density of the pile (the larger the falling height, the higher the density and friction angle). Part of the difference in slope angle may be due to the difference in setting up the pile. In their investigation of soil tool interactions in extraterrestrial conditions, \citet{Jiang2017} conducted some 2D DEM modelling of a heap, made up of circular particles of 1.2 mm average diameter, pushed by a vertical blade. Compared to the 30° angle of repose of their material (a value that they found independent of the gravity field), they observed that the downstream slope of the pushed heap was 45°, a value comparable to that observed in pile P1. They also found that this angle increased to 47° when accounting for the cohesion resulting from the significant van der Waals inter-grains attractive forces under the high vacuum conditions (10$^{-7}$ Pa) of the Moon. Indeed, the tests conducted on a lunar simulant by Bromwell (1966) and Nelson (1967) under high vacuum and high temperature (394 K) showed an increase in friction angle of 13° and an increase in cohesion of 1.1 kPa. In this regard, it would be interesting to investigate the possible increase in van der Waals inter-grain forces under Mars atmospheric conditions, so as to better estimate the possible resulting cohesion.

The slopes of the walls W1 and W2 left by the scoop's scrapes are significantly larger than the slopes of the piles and slope failure was not observed on these walls. This result can be interpreted as a result of the presence of some cohesive forces in the undisturbed regolith within the mounds. 




\subsection{Seismic observations of the hammering}
\label{sec:SEIS}
InSight's seismic experiment for interior structure (SEIS) seismometer assembly was installed on Mars to monitor the martian seismicity and image the deep interior of the planet with seismological methods \citep[e.g.,][]{Lognonne2019,Staehler2021}. Due to the limited reach of the IDA, SEIS was placed on the ground about 1.2\,m away from HP$^3$. Induced ground displacement and mechanical vibrations from all HP$^3$ related operations such as the mole hammering and IDA activities to assist the hammering generated seismic (elastic) signals that were recorded by SEIS. These seismic signals were used to support the mole anomaly recovery activities. For example, the seismometer was used as a tiltmeter to monitor the quality of the preload force exerted on the mole by the IDA (see Section \ref{subsec:pin1}). 

In addition, the seismic recordings of mole hammering provide a unique opportunity to study the near-surface structure and elastic parameters at the InSigth landing site. Knowing these parameters is relevant, for example, to understand the coupling of the seismometer to the ground, to infer on the local geological structure, composition and history at the landing site, and to collect information on the martian regolith for future missions \citep{kedar2017,Golombek2018}. Listening to the HP$^3$ hammering marks, to the best of our knowledge, the first controlled-source seismic experiment ever conducted on another planet \citep{Brinkman2019}. This experiment on Mars can be seen as a continuation of successful seismic experiments on the Moon \citep{Cooper1974,Sollberger2016} and on comet 67P/Churyumov-Gerasimenko \citep{Knapmeyer2016}. Interestingly enough, the experiments on comet 67P/Churyumov-Gerasimenko bear some similarity to the mole hammering on Mars in terms of source type and scale.

Seismic studies of the subsurface at the InSight landing site covering the topmost 5\,m initially planned to be reached by the mole and the section below include first traveltime analyses of the first HP$^3$ hammering sessions \citep{Lognonne2020}, compliance inversions \citep{kenda2020subsurface,Lognonne2020}, and an ambient vibrations Rayleigh wave ellipticity study \citep{hobiger2021}. These initial seismic investigations suggest a shallow low velocity layer (P-wave velocities $<$ 300 m/s; S-wave velocities $<$ 150 m/s) that cannot be thicker than 1 to 1.5 m. Below 1 to 2 m depth, the fine-grained regolith seems to be mixed with blocky ejecta resulting in increased bulk seismic velocities (P-wave velocities $>$ 700~m/s; S-wave velocities $>$ 400~m/s) as indicated by both the Rayleigh wave analysis and compliance inversions. Below this transition zone, a sequence of high and low velocity layers found by the Rayleigh wave ellipticity inversion for the topmost 200 m is interpreted as a sequence of lava flows inter-fingered with a sedimentary unit. \cite{Manga2021} have concluded that the seismic velocities in the top 10 km underneath InSight are too low to suggest an ice-saturated cryosphere. 

\subsubsection{Preparing SEIS for listening to HP$^3$ hammering}
Studying the near-surface using the HP$^3$ seismic signals did not address any of the primary InSight mission goals. Furthermore, exploiting the HP$^3$ seismic signals was not conceived before key decisions on the system design were already taken. Therefore, a series of ad-hoc adaptations had to be implemented to realise this opportunistic experiment and extensive feasibility tests had to be performed.

The SEIS sensor assembly was deployed by the IDA on the ground on sol 22. SEIS consists of six seismic sensors, covering a nominal seismic bandwidth from 0.01 to 50\,Hz: a three-component  very broad band seismometer (VBB) and a three-component short period seismometer (SP), all mounted on a three-legged leveling system. After the sensor assembly was deployed, a  wind and thermal shield was placed over SEIS to provide a first level of environmental noise protection. 

SEIS is operated using the so-called E-Box housing all acquisition and control electronics. Programmable digital finite impulse response (FIR) filters, for example, are used to low-pass filter the seismic data before down-sampling in preparation of transmission to Earth \citep{Zweifel2021}. Changing these FIR filters during hammering proved to be critical for successful recording of the hammering signals \citep{Sollberger2021}.

\subsubsection{Pre-mission preparation activities}
The preparation for the recording of HP$^3$ seismic signals began early in the mission \citep{kedar2017}. In a first phase, the mole seismic source time function was measured and used to generate sets of time series simulating the signals recorded by SEIS from a mole that penetrates through a simple layered model. It was concluded that in spite of the low nominal resolution of the SEIS data relative to the hammer source duration, it would be feasible to retrieve key elastic parameters such as the seismic P-wave and S-wave velocities of the near-surface, including possibly detecting sharp interfaces up to several meters beneath the InSight lander \citep{Golombek2018,Brinkman2019}.  

Once the scientific value of listening to the hammering was demonstrated, an extensive field analogue experiment was carried out in the Mojave Desert, California.  The site was selected since it provided a sharp contact between sedimentary and an igneous rock layer, similar to the landing site at Elysium Planitia on Mars. A seismic survey was conducted to characterize the site stratigraphy. A mole engineering model together with broadband seismometers were installed in a similar geometry to the HP$^3$-SEIS configurations planned for Mars.  The seismic signals were recorded at 1,000 Hz sampling frequency, and then down-sampled to 100 Hz sampling frequency to simulate SEIS highest nominal resolution setting. 

We demonstrated that under those conditions the seismic velocity in the soft sediment can be determined with high fidelity using the HP$^3$ STATIL time tags and the HP$^3$ mole depth.  Yet, it was determined that due to the relatively low temporal resolution of SEIS, due to reverberation of the HP$^3$ mole and the fact that the HP$^3$ seismic source shows a double pulse 0.06 s apart \citep{kedar2017} caused by double strikes of the hammer mechanism (section \ref{sec:HP3Description}) that determining the depth of the sediment-rock interface would be challenging.  The field experiment highlighted the need to accurately synchronize the SEIS and HP$^3$ clocks to take full advantage of the STATIL hammer stroke time tag and to implement strategies to maximize the temporal resolution.

\subsubsection{Regolith properties from HP$^3$-SEIS}
\begin{figure}
    \centering
    \includegraphics[width=0.7\textheight,angle=90,origin=c]{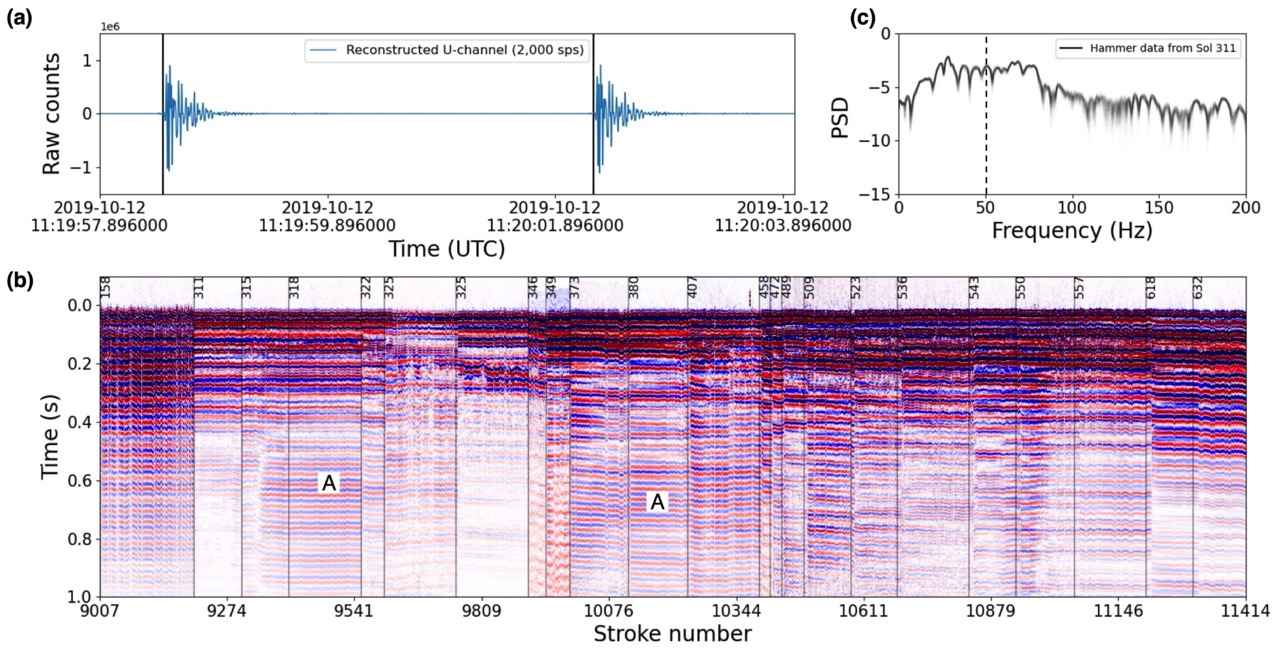} 
    \caption{Seismic data collected during HP$^3$ mole hammering. The SEIS SP data were recorded with an adapted acquisition procedure that allowed reconstructing the broadband waveforms. (a) Waveforms of two subsequent mole strokes separated by around 3.7\,s recorded during the diagnostic hammering on sol 158. Vertical black lines mark the hammering time. (b) All broadband east-component SP data recorded for hammering sessions between sol 158 and 632. Vertical bars show the beginnings of the sessions marked with the corresponding sol. Time $t = 0$\,s corresponds to the mole hammering time. The seismic signal show clear first arriving energy being interpreted as the P-wave arrival. The mole hammer strokes also excite the 25-Hz resonance denoted with A, which is assumed to originate from vibrations of the SEIS housing and/or leveling system. (c) Power spectral density computed for all data recorded during the hammering on sol 311. Note how the frequency bandwidth of the hammer signal exceeds the Nyquist frequency of 50\,Hz of the nominal SEIS acquisition (marked by the dashed line) highlighting the value of the reconstruction method by \citet{Sollberger2021}.}
    \label{hp3seis}
\end{figure}

Analysing the seismic signals traveling between the mole and SEIS allows inferring the regolith elastic parameters governing seismic wave propagation \citep{Lognonne2020,Brinkman2021}. Because of the short travel path of around 1.2\,m, the traveltimes of the seismic waves were expected to be on the order of several milliseconds only and therefore shorter than the SEIS sampling interval of 10 ms (governed by the sampling frequency of 100\,Hz). In order to reach the necessary high temporal resolution and to record the broad-band hammering signals, we developed a recording and data-processing strategy to overcome the nominal sampling limitations \citep{Sollberger2021}. Firstly, the anti-aliasing FIR filters to prepare the seismic data for down-sampling to the highest nominal sampling frequency of 100 Hz were turned-off during seismic acquisition of most hammering sessions. This resulted in the seismic signals being aliased, containing energy in the frequency range 0--250\,Hz but (multiply) folded around the nominal Nyquist frequency of 50\,Hz. Based on the assumption that the HP$^3$ hammering signals are highly repeatable (see Fig.~\ref{hp3seis}a for two example hammer strokes measured on Mars), the original seismic waveforms can be reconstructed at a high virtual sampling rate using a sparseness-promoting algorithm \citep{Sollberger2021,SEISData2020}. 

To compute seismic velocities, it is important that the mole stroke triggering times are accurately known to compute the absolute traveltimes between the mole and SEIS. Because HP$^3$, SEIS, and the lander operate with independently running clocks, a high-precision clock-correlation procedure had to be designed and implemented.

We analysed around 2,000 traveltime picks extracted from the waveform data displayed in Fig.~\ref{hp3seis} recorded with the high-resolution SEIS settings (hammering sessions between sol 311 and 632). Based on these traveltimes, we estimated a bulk P- (compressional) and S- (shear) wave velocity of $114^{+40}_{-19}$\,m/s and $60^{+10}_{-7}$\,m/s, respectively, for the regolith volume between the mole and SEIS \citep{Brinkman2021}. Assuming a density of 1200\,kg/m$^3$, the velocity estimates translate into bulk, shear, and Young's moduli as well as Poisson's ratio of 9.84$\pm{6.54}$\,MPa, 4.32$\pm{1.01}$\,MPa, 11.30$\pm{2.87}$\,MPa, and 0.31$\pm{0.15}$, respectively. When interpreting these estimates, one should keep in mind that they were derived from elastic waves with a dominant frequency content of around 40 to 80\,Hz (see Fig.~\ref{hp3seis}c). 

The observed seismic velocities are interpreted as bulk averages for shallowest few tens of centimeters. The velocity values appear low compared to lab measurements for unconsolidated dry quartz sand on Earth. However, extrapolations of lab measurements on martian regolith soil simulants to the low gravity (low overburden pressure) conditions at the surface of Mars result in very similar P-wave velocities of 100-120\,m/s \citep{Delage2017,Morgan2018}. 

\subsubsection{Radiated seismic energy from HP$^3$ hammering}
\label{sec:seismic energy}
An estimate of the radiated seismic energy, $E_R$, can be made by integrating over the elastic energy flux from each hammer stroke.  Using the formulation of \cite{Shearer_2019}
\begin{equation}
    E_R = \rho \int_S \int_{-\infty}^{\infty} \alpha (\dot{u}_\alpha^2 + \dot{v}_\alpha^2 + \dot{w}_\alpha^2) dt dS = 4\pi \rho \alpha r^2 I_P
\end{equation}
with
\begin{equation}
  I_P \equiv \int_{-\infty}^{\infty} (\dot{u}_\alpha^2 + \dot{v}_\alpha^2 + \dot{w}_\alpha^2) dt  \quad ,
\end{equation}
where $\rho$ marks density assumed to be $\sim$1200 kg/m$^3$, $\alpha$ is the measured P-wave velocity, $\dot{u}_\alpha$, $\dot{v}_\alpha$, and $\dot{w}_\alpha$ are the three SEIS measured components of the ground velocity during a hammer stroke, and $r$ the distance from HP$^3$ to SEIS.  It is further assumed that the $\sim$25 Hz reverberations observed during each hammer stroke (see Fig.~\ref{hp3seis}b) were excited by the source (as opposed to being excited by the wind), and that the energy measured at SEIS is predominantly P-wave energy radiated spherically from a point source.  Averaging over multiple hammer stroke recordings from the hammering session conducted on sol 158, we obtain $E_R \sim$ $1.3 \times 10^{-3}$ Joules per hammer stroke which may be compared with the mole stroke energy of 0.7 J. Thus, the seismic energy in the hammer signals recorded by SEIS during hammering is less than a percent of the stroke energy when geometrically projected back to the mole tip as the source area. This suggests that only a small part of the mole stroke energy is partitioned into vibrational energy.

\subsection{Implications from thermal measurements}
\label{sec:PropertiesThermal}

\begin{figure}
    \centering
    \includegraphics[width=1\linewidth]{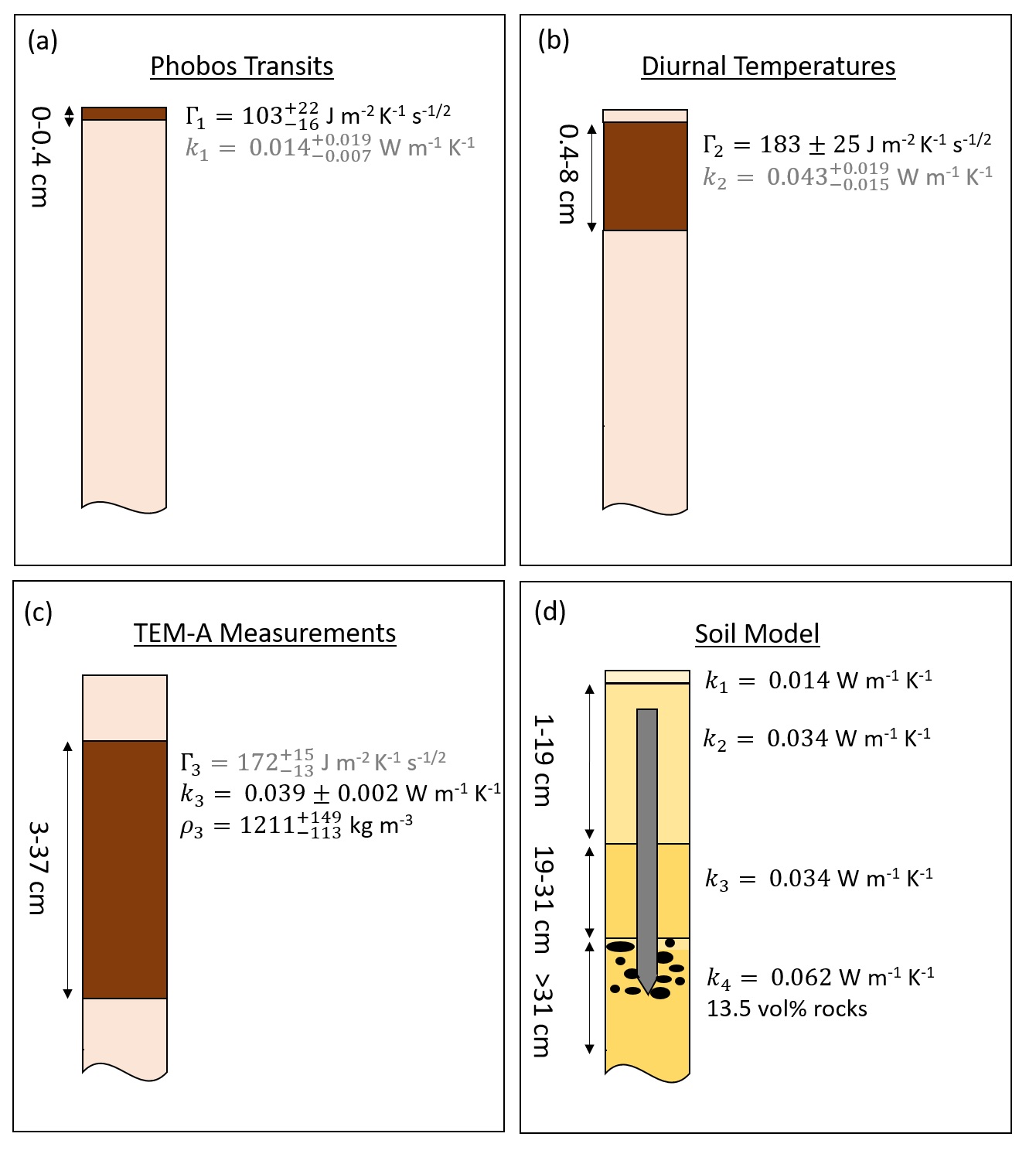} 
    \caption{(a)-(c) Summary of regolith physical properties derived from HP$^3$ RAD and active heating experiments using TEM-A. The sensing depths of the different methods are indicated.  Quantities that are immediately calculated  from the data such as thermal inertia in panels a and b and thermal conductivity and density in panel c are given in black. Values inferred from the data are given in gray. To convert thermal conductivity to thermal inertia, a soil heat capacity of 630 J kg$^{-1}$ K$^{-1}$ has been assumed \citep{Morgan:2018}. (d) Soil thermal model compatible with all observations assuming four regolith layers: A top unconsolidated sand/dust layer, a duricrust, an unconsolidated sand layer, as well as a layer including small rocks or gravel. Thermal conductivity of the rocks was assumed to be 3 W m$^{-1}$ K$^{-1}$.}
    \label{Fig:ThermalConductivity}
\end{figure}

The thermal properties of the soil around the lander have been probed by the HP$^3$ radiometer, which observes two spots north of the lander \citep{Spohn2018,Mueller:2020}, as well as the thermal sensors inside the HP$^3$ mole termed TEM-A (Thermal Excitation and Measurement - Active \citep{Spohn2018,Grott:2019}). These measurements are sensitive to different depth ranges and results of the investigations are summarized in Fig. \ref{Fig:ThermalConductivity}. 

The surface response to insolation changes is diagnostic of the surface thermal inertia, which is defined as 
\begin{eqnarray}
    \Gamma = \sqrt{k\rho c_p}
    \label{Eq:ThermalInertia}
\end{eqnarray}
where $k$ is thermal conductivity, $\rho$ is density, and $c_p$ is specific heat capacity. For fast changing illumination conditions, measurements are sensitive to shallow depths, while long-term periodic changes probe deeper soil layers. So far, measurements of the surface temperature response to transits of the martian moon Phobos \citep{Mueller:2021} as well as measurements of the temperature response to diurnal insolation changes \citep{Piqueux:2021} have been performed.  

The transit measurements are sensitive to the depth range between 0 and 0.4 cm, and a best fitting thermal inertia of $103^{+22}_{-16}$ J m$^{-2}$ K$^{-1}$ s$^{-1/2}$ has been determined for that layer (compare Fig. \ref{Fig:ThermalConductivity}(a), where the layer is indicated in brown). These data are complemented by the analysis of diurnal temperature changes, which are sensitive to about 8 cm depth. For this layer, a best fitting thermal inertia of $183 \pm 25$ J m$^{-2}$ K$^{-1}$ s$^{-1/2}$ has been determined (Fig.\ref{Fig:ThermalConductivity}(b)). Finally, thermal properties of the soil have been probed by direct thermal conductivity measurements using the mole as a modified line heat source \citep{Grott:2021}, and a thermal conductivity of $0.039 \pm 0.002$ W m$^{-1}$ K$^{-1}$ has been determined for the  3 to 37 cm depth range (Fig.\ref{Fig:ThermalConductivity}c).

To compare the radiometer and TEM-A measurements to one another, thermal inertia can be converted to thermal conductivity using Eq. \ref{Eq:ThermalInertia} if some assumptions regarding soil density and heat capacity are made. Here we use a heat capacity of $c_p=630$ J kg$^{-1}$ K$^{-1}$ as appropriate for basaltic sand at 220 K \citep{Morgan:2018} and a soil bulk density of 1211 kg m$^{-3}$ as derived from the active heating experiments \citep{Grott:2021}. Resulting derived values for thermal inertia (Fig. \ref{Fig:ThermalConductivity}c) and thermal conductivity (Fig. \ref{Fig:ThermalConductivity}a-b) are given in gray. Except for the uppermost unconsolidated layer, estimates for the soil thermal conductivity fall within the range expected for uncemented martian soils, which is 0.02 to 0.1 W m$^{-1}$ K$^{-1}$ \citep{Grott:2007}. 

All measurements of thermal properties indicate that the soil at the landing site is a poor thermal conductor, and the derived soil thermal conductivities place strong constraints on the allowable degree of soil cementation. Only minor amounts of cement are consistent with the derived low thermal conductivities \citep{Piqueux:2009,Piqueux:2009b}. Further study is needed to see how well these constraints can be reconciled with the observed  cohesion in the duricrust. 

\begin{figure}
    \centering
    \includegraphics[width=0.5\linewidth]{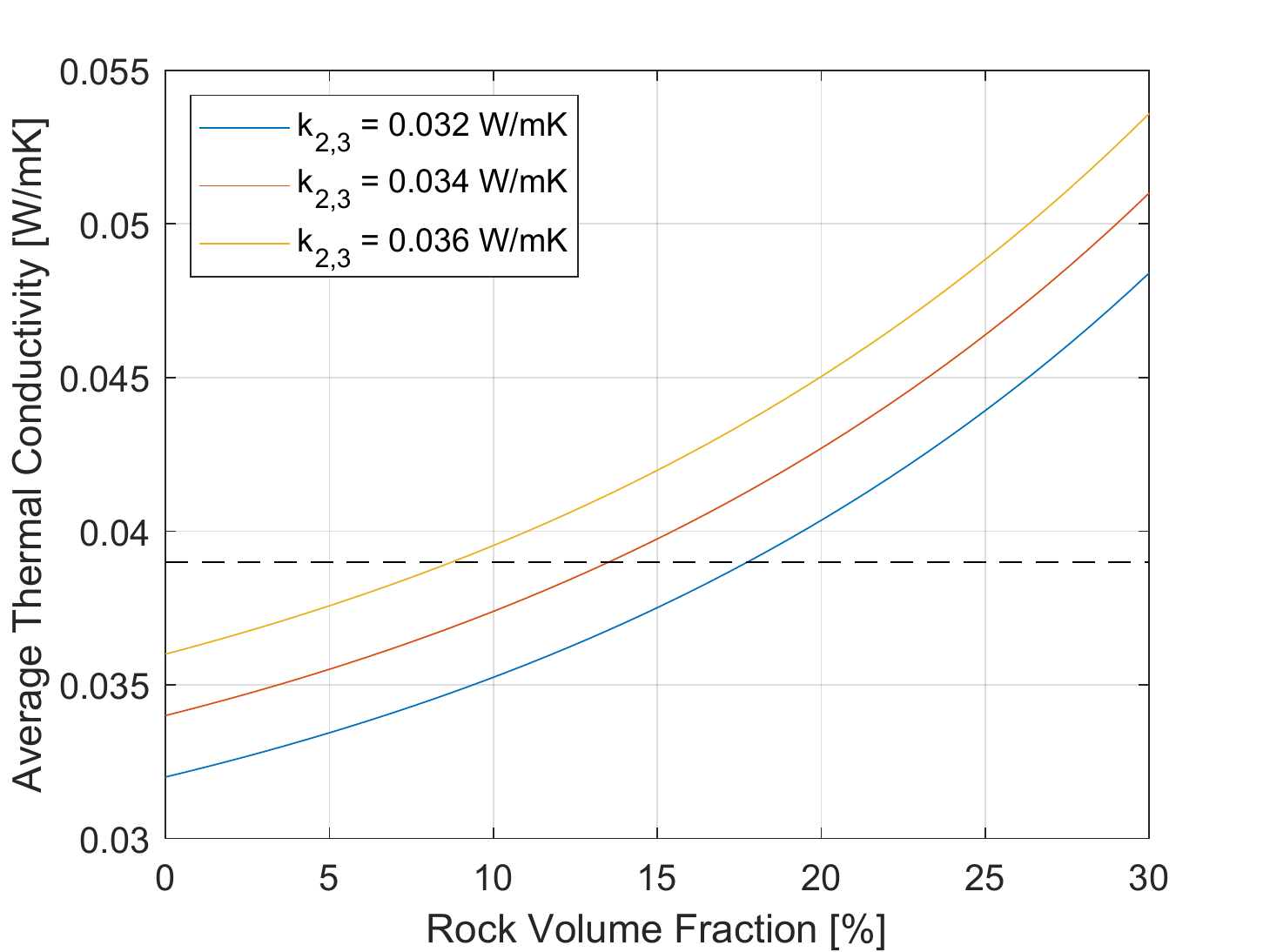} 
    \caption{Average thermal conductivity in the 3 to 37 cm depth range as a function of the volume fraction of stones in a hypothetical gravel layer located below 31 cm depth (compare Fig. \ref{Fig:ThermalConductivity}). Results are shown for three different thermal conductivities $k_{2,3}$ of the uppermost duricrust and intermediate sand layer, respectively. The average thermal conductivity of the entire soil column as measured using TEM-A is indicated by the horizontal dashed line. For 100 µm diameter particles, $k = 0.032$ to $0.036$ W m$^{-1}$ K$^{-1}$ \citep{Presley:1997}. Thus the volume fraction of rocks is limited to be smaller than 18\%.}
    \label{Fig:RockAbundance}
\end{figure}

While thermal soil properties determined for the different depth ranges do not indicate layering below the uppermost 0.4 cm, layering can also not be ruled out because the different measurements yield average values in their respective depth ranges. In particular, TEM-A yielded the average thermal conductivity between 3 and 37 cm depth. In order to explore the range of admissible soil properties, we consider a four layer soil model as shown in Fig. \ref{Fig:ThermalConductivity}(d), consisting of a 1 cm thick unconsolidated layer of sand mixed with dust through which the thermal conductivity would increase, a duricrust layer between 1 and 19 cm, a layer of unconsolidated sand between 19 and 31 cm, and a gravel layer below 31 cm. Assuming minimal cementation and a minimum particle size of 100 µm compatible with mobilization by winds \citep{Kok:2012}, the duricrust and the layer of unconsolidated sand can be assigned a minimum thermal conductivity of $k_{2,3}=0.034$ W m$^{-1}$ K$^{-1}$. 

To calculate the rock abundance compatible with the TEM-A results, we use mixing models for layering parallel to the direction of heat flow during the measurements \citep{Beardsmore:2001} and determine the maximum thermal conductivity admissible in the gravel layer. We then use mixing laws for randomly mixed material \citep{Beardsmore:2001} to estimate the rock abundance in the gravel layer itself assuming a rock thermal conductivity of 3 W m$^{-1}$ K$^{-1}$. Results of the calculations are shown in Fig. \ref{Fig:RockAbundance}, where the average thermal conductivity is given as a function of the volume fraction of rocks in the gravel layer for three different thermal conductivities $k_{2,3}$ of the duricrust and unconsolidated layer. The average thermal conductivity as determined using TEM-A is indicated by the horizontal dashed line. Results indicate that $\sim$15 vol\% of rocks in the gravel layer would still be compatible with the TEM-A results, and this scenario is summarized in Fig. \ref{Fig:ThermalConductivity}d.  

\section{Synopsis}
\label{sec:Discussion}


The HP$^3$ was a bold experiment, attempting to reach unprecedented depth in the martian regolith with a very compact, low power, and low mass mechanism. Its science goal was to measure the martian surface heat flow, a quantity that has been often modeled \citep[e.g.,][]{Schubert:90, Plesa:2018} and that provides an important constraint for the energy budget of the martian interior, its thermal and dynamic history and its composition \citep[e.g.,][]{Spohn2018, Smrekar:2018}. Most recently \cite{Khan:2021} have attempted to use the recordings of SEIS to invert for the lithosphere temperature gradient and estimated the heat flow after assuming a value for the thermal conductivity. Their value of about 20 mW/m$^2$ is consistent with most recent thermal models of Mars. 

HP$^3$ proved to have lower performance margins than originally planned and the system encountered an environment more difficult  than expected. In a separate paper, \cite{Spohn2021} have discussed lessons learned for the design and the operation of an HP$^3$-type heat flow probe. They conclude that a more massive design may have been able to meet the challenges, but at the expense of more mass and likely at greater cost.  A further dimension in which the effort was challenged was the operations schedule: there was pressure to achieve operation depth ahead of the shadowing thermal wave, and deployment delays eroded the time available. This in turn motivated more aggressive (i.e., longer) hammering commands during the initial penetration sols, which may have been detrimental.  More generous margins in any of these dimensions may have allowed success, but the competed mission framework does not foster large margins. 

Although HP$^3$ did not meet its primary science goal, the two years of carefully operating the mole and the robotic arm provided a wealth of data on the martian soil that were not available before. The primary data that HP$^3$ acquired were the radiometer data \citep{Piqueux:2021, Mueller:2021} and the measurements of the thermal conductivity using the TEM-A sensors on the mole \citep{Grott:2021}. Power permitting, these data will be continued to be acquired to study the time variability of the thermal inertia and the soil thermal conductivity, the latter to include a study of the effect of the gas pressure in the porous regolith. 
\begin{figure}
    \centering
\includegraphics[width=1.5\textwidth, angle=90]{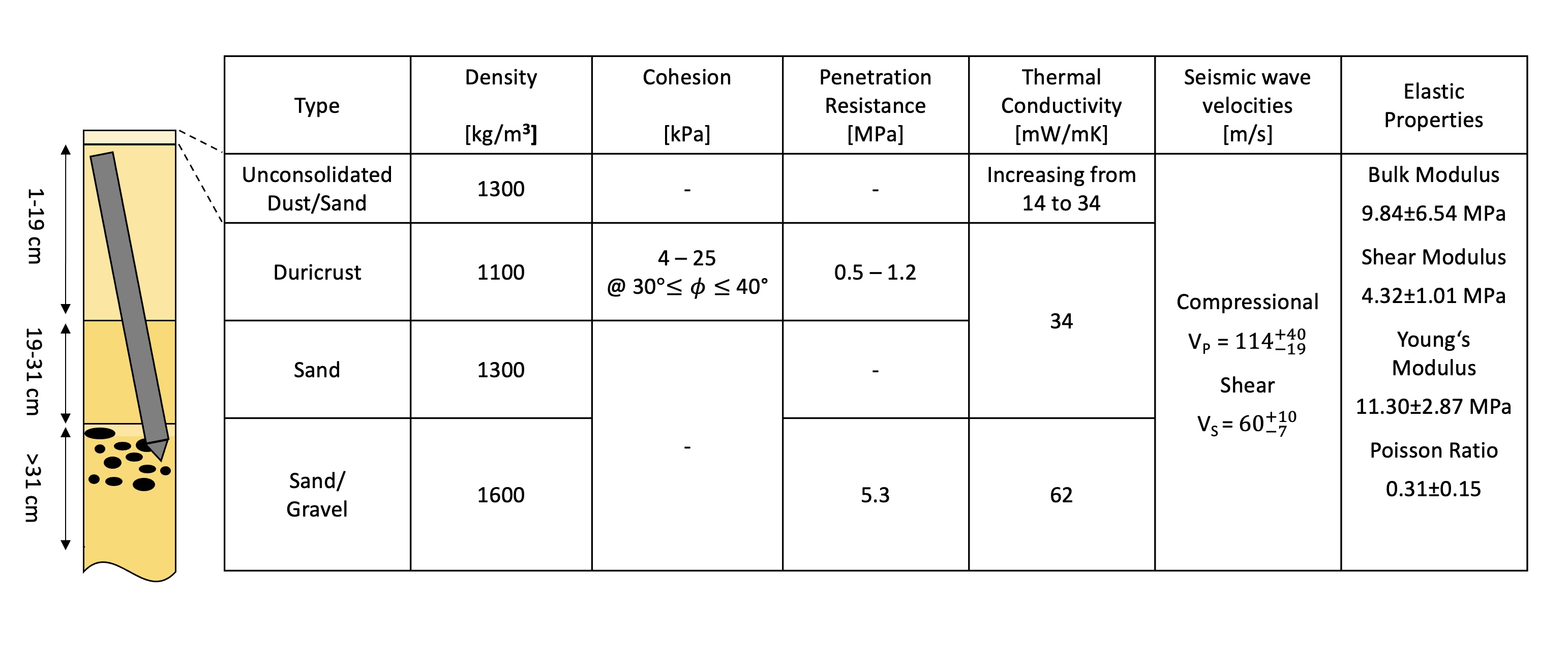}
\caption{Model of the martian soil at the HP$^3$ mole pit. The assumed range of internal friction angle $\phi$ for the listed cohesion value range is indicated }
    \label{Fig:Synopsis}
\end{figure}

In its attempt to penetrate, the mole acted as a penetrometer as is used in civil engineering and geology to study the properties of soils \cite[e.g.,][]{Terzaghi:1947,  Verruijt:2018}. Unfortunately, the data were not acquired through a carefully planned soil mechanics experiment and the data coverage was not ideal. Still, it is believed that combining the data acquired from  the mole and scoop operations, the thermal data from the radiometer and TEM-A, and the data from SEIS recording the mole hammering can be combined to form a consistent record of the soil properties as shown in Fig. \ref{Fig:Synopsis}. 

Working from the top surface to depth, we find a sand layer possibly mixed with dust of about 1 cm thickness to form the top layer. This layer has been observed in the images taken by the robotic arm instrument deployment camera IDC and has been indented by the feet of the SS and by the scoop, when the latter was pressed onto the surface by the robotic arm. The mechanical strength of the layer is weak, at least much weaker than the duricrust below and was compressed by the 3.8 kPa force per unit area exerted by the arm. The sand has been scraped to form ridges and a slope analysis as reported in section \ref{sec:scrapeangle} showed slopes of 49-53° and 70° and 78° of the fore and side walls of the scrape, respectively. It is conceivable that electrostatic forces cause these high values (in comparison with estimates of the  internal friction angle of around 30°). These forces may also cause some weak cohesion of the sand. Thermal data reported in section \ref{sec:PropertiesThermal} suggest a low value of thermal inertia of about 103 J m$^{-2}$ K$^{-1}$ s$^{-1/2}$ and - accordingly - a small thermal conductivity of 0.014 W m$^{-1}$ K$^{-1}$ increasing through the layer to 0.034 W m$^{-1}$ K$^{-1}$ and suggesting high porosity. 

Underneath the sand layer, the mole found a duricrust that provided significant resistance to mole penetration of about 0.5 to 1.2 MPa, depending on uncertainties in the timing of the penetration progress. The penetration resistance even at small depth can be explained by a significant cohesion. Slope stability analysis as reported in section \ref{sec:CohesionEstimate} after pressing the blade of the scoop into the duricrust and causing part of it to fail suggest a cohesion of 5.8 kPa. 

Using the analytical theory of \cite{Terzaghi:1947}, as reported in \cite{Poganski2017} the penetration resistance can be related to the cohesion and the angle of internal friction $\phi$. At small depth, the relation is 

\begin{equation}
  \sigma_{P} \approx \, c \, (\chi - 1) \,  \frac{(1 + sin \, \phi)\, \chi - 1}{\chi - 1}\, cot \, \phi
    \label{equ:resistance2}
\end{equation}

with 

\begin{equation}
  \chi \equiv \frac{1 + sin \, \phi}{1 - sin \, \phi}e^{\pi \, tan \,\phi}    \label{equ:resistance3}
\end{equation}

Assuming an angle of internal friction of 30° - 40° and a penetration resistance of 0.5 to 1.2 MPa as inferred from the penetration rate during the first 77 strokes (section \ref{PenetrationResistance}), equation \ref{equ:resistance2} suggests a cohesion of about 4 - 25 kPa, consistent with the estimate from slope stability analysis, given the uncertainties in the problem. A comparison with the latter estimate tends to favour a penetration rate near the higher end of the values given in section \ref{PenetrationRate} and near the lower end of the penetration resistance values given in section \ref{PenetrationResistance}. It should be noted, however, that even a cohesion of 25 kPa is small in comparison with cohesion values from terrestrial soils. \cite{Grott:2021} have argued that the thermal conductivity of 0.034 W/m K inferred from the TEM-A data would be difficult to reconcile with extensive bridges of cement between sand grains. The actual cohesion of the duricrust may possibly be acquired with thin layers of cementation that would still be consistent with a small thermal conductivity. Electrostatic attraction and interlocking of grains may also contribute to the cohesion but cannot easily be argued to be particularly relevant for the duricrust layer.   

The thickness of the duricrust could not be directly measured. The digital elevation data for the pit discussed in section \ref{RegolithInteracttion1} and the images taken by the IDC suggest a thickness of at least 7 cm while we have argued from the distance of the mole backing out for a thickness of about 20 cm in section \ref{sec:Duricrust Thickness}. Because the inversion of the TEM-A data do not suggest a layering of the soil between 3 cm and 31 cm depth a thicker duricrust cannot be ruled out. 

Inversion of the TEM-A data suggest that the layer between 3 cm and 31 cm depth has a density of $1211^{+149}_{-113}$ kg/m$^3$ consistent with a porosity of $63^{+9}_{-4}$ \%. In section \ref{sec:soil porosity} we explain the formation of the pit as a result of a precession of the mole and by the grinding the duricrust to sand that filled the pit. We conclude that the ratio of the densities between the duricrust and the sand would depend on the thickness of the former. For our preferred thickness of about 20 cm, the ratio would be about 1.2. We can use the constraints from the TEM-A data to estimate the density of the duricrust and the sand.  With  $1211^{+149}_{-113}$ kg/m$^3$ a consistent pair of values would be 1100 kg/m$^3$ for the duricrust and 1300 kg/m$^3$ for the sand, respectively. \cite{Morgan:2018} give a representative value of martian sand of 1300 - 1350 kg/m$^3$. A layer of sand may extend from the bottom of the duricrust to a depth of about 31 cm where the mole hit a more resistant layer.  The mole likely penetrated through the sand layer at a rate of around 2 mm/stroke. This would be a penetration rate typical of the HP$^3$ mole in cohesionless sand, similar in rheological properties to quartz-sand. The duricrust could have formed from this sand by aqueous reactions \citep[e.g.,][]{Banin:1992, Haskin:2005, Hurowitz:2006}.
 
After penetrating to a depth of 31 cm (or 26 cm measured along the mole) the mole encountered a layer into which it penetrated about 6 cm vertically aided by pinning and back cap pushing at a rate of only 0.1 mm/stroke or even less. A rate, that is approximately by an order of magnitude smaller than the rate estimated for the duricrust. The small  penetration rate suggests a penetration resistance of  5.3 MPa. We can only speculate about the nature of this layer. A significant reduction of penetration rate has been observed in the laboratory when penetration through a layer of gravel was attempted \citep{Wippermann2020}. The size of the gravel stones was about the size of the mole diameter, up to a small multiple thereof. Such a layer of gravel may be present as  buried ejecta from an  impact crater \citep{Golombek:2020a}. It is also conceivable although less likely that the mole has compacted the sand in front of the tip and provided sufficient resistance. While the penetration resistance undoubtedly increases with increasing compaction it is questionable whether the mole could have sufficiently compacted a layer of four times its radius. Thermal modelling shows that the TEM-A data would be consistent with a layer of almost twice the conductivity of the duricrust with a rock fraction of 15 vol-\%.  Taking a particle density of martian crust basalt of 3200 kg/m$^3$ and a bulk density of the sand of 1300 kg/m$^3$ we estimate a bulk density of the gravel layer of 1600 kg/m$^3$.

Recordings of the hammer signals of the mole have been used to estimate a seismic bulk P- and S-wave velocity of $114^{+40}_{-19}$\,m/s and $60^{+10}_{-7}$\,m/s, respectively, as well as the elastic moduli such as a shear modulus of 4.32$\pm{1.01}$\,MPa  as discussed in section \ref{sec:SEIS}. Civil engineers and soil scientists have attempted to relate the shear modulus of soils to their shear strengths and find empirical relations of the form
\begin{equation}
    G \approx A \, S^\beta
    \label{Shearmodvstrength}
\end{equation}
where $G$ is the shear modulus, $A$ and $\beta$ are empirical constants and $S$ is the shear strength which we identify at low confining pressure with the cohesion \citep[e.g.,][]{Hardin:1972}. \cite{Hara:1974} have collected data on 25 terrestrial sites and find $\beta$ to be close to 1 while the values for $A=G/S$ at $\beta \approx 1$ cluster around 500, but range up to 1600. It should be noted, however, that these empirical values are for the undrained - that is water saturated- shear strength. Given $G = 4.32\pm{1.01}$\,MPa and assuming $A = 500$, we obtain shear strength values between 6.62 and 10.7\,kPa deduced from the seismic data. These strength values are in good agreement with the cohesion estimates from the slope stability analysis of 5.8\,kPa (see Section \ref{sec:CohesionEstimate}) and from the penetration resistance estimates of 4 - 25\,kPa. 

The proposed layering is consistent with the geology of the InSight landing site \citep{Golombek:2020a}. The soils observed at the landing site are generally similar to soils at other landing sites on Mars \citep{Christensen:1992, Herkenhoff2008, Golombek2008} and their origin via impact and eolian processes is likely similar to the Spirit landing site \citep{Golombek:2020b}. 

\par

\begin{acknowledgements}
The authors thank the operations team at JPL and DLR for their continued support. Part of this work was performed at the Jet Propulsion Laboratory, California Institute of Technology under a contract with NASA. US government support acknowledged. Support of this work by DLR, the Swiss Space Office and CNES is thankfully acknowledged as well as funding from ETH Research Grant ETH-06 17-02. Lab. Navier, Ecole des Ponts (UMR CNRS 8205, Paris) received from CNES a Grant N°6397 (2021).

\noindent The seismic waveform data are available from NASA PDS (National Aeronautics and Space Administration Planetary Data System, https://pds.nasa.gov; InSight Mars SEIS Data Service, 2019; https://doi.org/10.18715/SEIS.INSIGHT.XB\_2016). The high-rate seismic data from HP$^3$ hammering were obtained using a reconstruction algorithm and are made available in a public repository at https://doi.org/10.5281/zenodo.4001920.

\end{acknowledgements}

%
\section*{Conflict of interest}

The authors declare that they have no conflict of interest.

\bibliographystyle{spbasic} 
\bibliography{Mole_Saga_Paper_SSR_con}



\section*{Supplementary material}

**Supplementary Material**

| Sol | Date | Phase | IDA | IDC | ICC | SEIS | STATIL | TEM-A | HAM | Description |
|---|---|---|---|---|---|---|---|---|---|---|
| | **2019** | | | | | | | | | |
| 92 | Feb 28 | IA | x | x | x | x | x | | 3881 | First penetration attempt, stereo pair (16:30) |
| 94 | Mar 02 | | x | x | x | x | x | | 4720 | Second penetration attempt, stereo pair (17:00) |
| 97 | Mar 04 | D&L | | | | | | x | | Conductivity measurement, 24 hours @ 1.0 W |
| 98 | Mar 05 | | x | x | x | | | | | 3x6 mosaic (16:00), Sunset long-shadow, 'mole exit' pose 1 |
| 99 | Mar 06 | | x | x | x | | | | | Sunrise long-shadow, 'mole exit' pose 1 |
| 103 | Mar 10 | | x | x | x | | | | | SS window imaging, 'window' pose 1 |
| 106 | Mar 13 | | | x | | | x | | | Deck mosaic (debris check); Tilt measurement |
| 107 | Mar 14 | | x | x | x | | | | | Sunrise long-shadow, 'mole exit' pose 2 |
| 111 | Mar 18 | | x | x | x | | | | | Sunset long-shadow, 'mole exit' pose 2 |
| 116 | Mar 23 | | | | | | | x | | Conductivity measurement, 24 hours @ 2.0 W |
| 118 | Mar 25 | | x | x | x | x | x | | 197 | Diagnostic Hammering 1 – IDC 'overhead' pose |
| 127 | Apr 03 | | | x | x | | | | | SS window imaging, 'window' pose 2 |
| 128 | Apr 04 | | x | x | x | | | | | 2x2x1 mosaic (16:30) |
| 146 | Apr 23 | | x | x | x | x | x | | - | *Diagnostic Hammering – did not execute* |
| 158 | May 05 | | x | x | x | x | x | | 198 | Diagnostic Hammering 2, 'window' pose 2 |
| 203 | Jun 22 | | x | x | x | | x | | | 12 cm lift to expose mole; no TLM reading (as expected) |
| 206 | Jun 25 | | x | x | x | | x | | | 13 cm lift (to 25 cm) to engage TLM |
| 209 | Jun 29 | | x | x | x | | x | | | 29 cm lift (to 54 cm) to extract ST and place SS |
| 211 | Jul 01 | | | | | | | x | | Conductivity measurement, 24 hours @ 2.0 W |
| 227 | Jul 17 | PC | x | x | x | | | | | 4x4x1 mosaic (12:00); ET imaging; 2x6 mosaic (17:30) |
| 230 | Jul 20 | | x | x | x | | | | | 4x4x1 (12:00 & 16:00); 3x6 mosaic (17:00); pit closeup |
| 240 | Jul 30 | RI-1 | x | x | x | x | x | | | Flat scoop push; 4x4x1 mosaic (16:00) |

| Sol | Date | Label | | | | | | | Notes | Activity |
|---|---|---|---|---|---|---|---|---|---|---|
| 243 | Aug 03 | | x | x | x | x | x | | | 1 scoop tip chop; 2x2x1 mosaic (14:40)<br>1 scoop tip chop; 2x2x1 mosaic (16:00) |
| 246 | Aug 05 | | x | x | x | x | x | | | 1 scoop tip chop - *no ground contact*; 2x2x1 mosaic (14:40)<br>1 scoop tip chop - *no ground contact*; 2x2x1 mosaic (16:00) |
| 250 | Aug 08 | | x | x | x | x | x | | | 1 scoop tip chop; 2x2x1 mosaic (14:40)<br>1 scoop tip chop; 2x2x1 mosaic (16:00) |
| 253 | Aug 12 | | x | x | x | x | x | | | 1 scoop tip chop; 2x2x1 mosaic (14:40)<br>1 flat scoop push; 2x2x1 mosaic (16:00) |
| 254 | Aug 13 | | x | x | x | x | x | | | pit imaging |
| 257 | Aug 15 | | x | x | x | x | x | | | 4x4x1 mosaic (16:00) – context |
| **Conjunction: 264 – 288** ||||||||||||
| 298 | Sep 28 | | x | x | x | | | | | 2x2x1 mosaic (16:00) |
| 302 | Oct 02 | P1 | x | x | x | x | x | | | Horizontal pinning |
| 305 | Oct 05 | | x | x | x | x | x | | | Vertical pinning |
| 308 | Oct 09 | | | x | x | x | x | | 20 | Hammer |
| 311 | Oct 12 | | x | x | x | x | x | | 101 | Vertical pin (5 mm) + hammer |
| 315 | Oct 16 | | x | x | x | x | x | | 101 | Vertical pin (5 mm) + hammer |
| 318 | Oct 19 | | | x | x | x | x | | 152 | Hammer |
| 322 | Oct 22 | REV1 | x | x | x | x | x | | 50 | Slight retract IDA, horizontal move, regolith push + hammer |
| 325 | Oct 26 | | x | x | x | x | x | | 304 | Hammer (150) + retract & push + hammer (150) → **Reversal** |
| 329 | Oct 29 | | x | x | x | | x | | | IDA retract (2 cm horizontal; 2 cm vertical) |
| 332 | Nov 02 | | x | x | x | | | | | 4x4x1 (13:00); 2x2x1 (16:00); pit closeup |
| 339 | Nov 09 | P2 | x | x | x | | x | | | Horizontal pinning |
| 342 | Nov 12 | | x | x | x | | x | | | Vertical pinning (1.5 cm down) |
| 346 | Nov 17 | | | x | x | x | x | | 40 | Hammer |
| 349 | Nov 19 | | x | x | x | x | x | | 50 | Vertical pinning (4 cm down) + hammer |
| 366 | Dec 07 | | | x | x | x | x | | 19 | Hammer |
| 373 | Dec 14 | | | x | x | x | x | | 127 | Hammer |
| 380 | Dec 21 | | | x | x | x | x | x | 126 | Hammer; Conductivity: 24 hours @ 2.0 W |
| | **2020** ||||||||||
| 400 | Jan 11 | REV2 | x | x | x | x | x | | | Vertical retract (1 cm up) + repining (3 cm down) |
| 407 | Jan 18 | | | x | x | x | x | | 151 | Hammer (150) → **Reversal** |
| 414 | Jan 25 | RI-2 | x | x | x | | x | | | Horizontal offload (37.5 cm) |
| 417 | Jan 29 | | | x | x | x | | x | | Vertical retract; 2x2x1 (15:00); scrape test; 2x2x1 (16:00) |
| 420 | Feb 01 | | | x | x | x | x | x | | Chop test; image; 2x2x1 mosaic (16:00) |

| Sol | Date | Phase | IDA | IDC | ICC | SEIS | STATIL | TEM-A | HAM | Activity |
|---|---|---|---|---|---|---|---|---|---|---|
| 454 | Mar 07 | BCP-H | x | x | x | ? | x | | | Back cap preload – horizontal scoop (BCP-H) 01; 28 mm |
| 458 | Mar 11 | | | x | x | x | x | | 24 | Back cap hammer – horizontal scoop (BCH-H) 01 |
| 461 | Mar 14 | | x | x | x | | | | | Re-applied preload on back cap |
| 468 | Mar 21 | | x | x | x | x | x | | | BCP-H 02 28 mm |
| 472 | Mar 25 | | | x | x | x | x | | 24 | BCH-H 02 |
| 482 | Apr 04 | | x | x | x | x | x | | | BCP-H 03 18 mm |
| 489 | Apr 11 | | | x | x | x | x | | 50 | BCH-H 03 |
| 502 | Apr 25 | | x | x | x | x | x | | | BCP-H 04 18 mm |
| 509 | May 02 | | | x | x | x | x | | 100 | BCH-H 04 |
| 516 | May 09 | | x | x | x | x | x | | | BCP-H 05 18 mm |
| 523 | May 16 | | | x | x | x | x | | 100 | BCH-H 05 |
| 530 | May 23 | | x | x | x | x | x | | | BCP-H 06 28 mm |
| 536 | May 30 | | | x | x | x | x | x | 151 | BCH-H 06; Conductivity: 24 hours @ 2.0 W |
| 543 | Jun 06 | | x | x | x | x | x | | 100 | BCP-H 07 32 mm; BCH-H 07 |
| 550 | Jun 13 | | x | x | x | x | x | | 126 | BCP-H 08 32 mm; BCH-H 08 |
| 557 | Jun 20 | | x | x | x | x | x | | 151 | BCP-H 09 32 mm; BCH-H 09 |
| 577 | Jul 11 | | x | x | x | x | x | | | Retract and 4x4x1 mosaic (15:00) |
| 598 | Aug 01 | RI-3 | x | x | x | x | x | | | 12 cm scrape (far-to-near) regolith into pit; mole obscured |
| 599 | Aug 02 | | x | x | x | | | | | 4x4x1 mosaic (15:30) |
| 611 | Aug 15 | BCP-I | x | x | x | x | x | | | Back Cap Preload – inclined scoop (BCP-I) 01 30 mm |
| 618 | Aug 22 | | | x | x | x | x | | 101 | Back Cap Hammer – inclined scoop (BCH-I) 01 |
| 632 | Sep 05 | | x | x | x | x | x | | 101 | BCH-I 02 |
| 645 | Sep 19 | | x | x | x | x | x | | 252 | BCH-I 03 |
| 659 | Oct 03 | | x | x | x | x | | | | Retract IDA, single images |
| 660 | Oct 04 | | x | x | x | | | | | 4x4x1 mosaic (16:00) |
| 673 | Oct 17 | RI-4 | x | x | x | | | | | Two 12 cm scrapes (far-to-near); stereo image |
| 680 | Oct 24 | | | | | | | x | | Conductivity: 24 hours @ 2.0 W |
| 686 | Oct 31 | | x | x | x | x | | | | Tamp regolith; 4x4x1 mosaic (15:30) |
| 700 | Nov 14 | | | x | x | x | x | | | One 12 cm scrape (far-to-near); single and stereo images |
| 734 | Dec 19 | | x | x | x | | | | | Reposition IDA; regolith preload |
| | 2021 | | | | | | | | | |
| 754 | Jan 09 | FFMT | x | x | x | x | x | | 506 | Final Free Mole Test |
| 775 | Jan 30 | | x | x | x | | | | | Retract IDA; 2x2x1 mosaic (15:30) |
| 795 | Feb 13 | | | | | | | x | | Conductivity: 24 hours @ 2.0 W |

Supplementary Table. Events during the mole recovery actions between Sols 253 and 472 on Mars. Listed are the sols on Mars, the terrestrial dates and the phases as given in Table 2. The activities of the various components and instruments are indicated. IDA: instrument deployment arm, IDC: instrument deployment camera, ICC: instrument context camera, SEIS: seismometer, STATIL: static tilt sensors, TEM-A: thermal conductivity measurement sensors, HAM gives the number of hammer strokes.

Ancillary movements of the IDA (coarse and fine positioning, grapple stow / release / restow, and such have been left out of the table. Pinning motions and imaging-only activities have been retained



\end{document}